\documentclass[
    superscriptaddress,
    reprint, 
    nofootinbib, 
    amsmath, 
    amssymb, 
    aps, 
    prd, 
    floatfix
]{revtex4-2}

\usepackage{graphicx}
\usepackage[
    colorlinks = true,
    linkcolor = magenta,
    citecolor = magenta,
    urlcolor = magenta
]{hyperref}

\usepackage{microtype}

\usepackage{physics}
\usepackage{eucal}
\usepackage{mathtools}
\usepackage{nccmath}
\usepackage{bbm}
\usepackage{color}
\usepackage[normalem]{ulem}
\usepackage{array}
\usepackage{ascmac}
\usepackage{fancybox}

\usepackage{tikz}

\allowdisplaybreaks[4]

\numberwithin{equation}{section}
\renewcommand{\theequation}{\arabic{section}.\arabic{equation}}

\makeatletter
    \def\frontmatter@affiliationfont{%
        \itshape
        \baselineskip=1.05\baselineskip
    }
\makeatother

\makeatletter
    \def\frontmatter@RRAP@format{%
        \addvspace{8pt}
        \small
        \centering
        \everypar{\hbox\bgroup(\@gobble@leavemode@uppercase}%
        \def\par{\@ifvmode{}{\unskip)\egroup\@@par}}%
    }
\makeatother

\begin{document}

\title{%
    Multifield stochastic inflation: Relevance of number of fields in statistical moments  
}%

\author{Tomo Takahashi}
\email{tomot@cc.saga-u.ac.jp}
\affiliation{%
	Department of Physics, Saga University, 1 Honjomachi, Saga 840-8502, Japan
}%
\author{Koki Tokeshi}
\email{koki.tokeshi@phys.ens.fr}
\affiliation{%
    Laboratoire de Physique de l’\'{E}cole Normale Sup\'{e}rieure, ENS, CNRS, 
    Universit\'{e} PSL, 
    Sorbonne Universit\'{e}, Universit\'{e} Paris Cit\'{e}, 75005 Paris, France
}

\date{\today}

\begin{abstract}
    \noindent
    In multifield inflation driven by $d$ scalar fields, $O(d)$ symmetry renders the number of fields irrelevant at classical level. 
    This ceases to be the case once stochastic effects are accommodated. 
    The statistical quantities such as the mean number and the variance of $e$-folds as well as the primordial power spectrum and its scale dependence 
    are perturbatively calculated in a small-noise regime. 
    In particular, a general formula is derived for arbitrary higher-order statistical moments of the stochastic number of $e$-folds at all perturbative orders,
    keeping the dependence on the number of fields fully analytical. 
    It is also discussed that the requirement for inflation to be successfully terminated 
    puts a theoretical bound on the number of fields from above. 
    Those general results are demonstrated for several $O (d)$-symmetric models. 
\end{abstract}

\maketitle 

\section{Introduction}
\label{sec:intro}

Over forty years have passed since it was first proposed that our Universe underwent a quasi-de-Sitter era shortly after its creation, known as cosmic inflation~\cite{Starobinsky:1980te, Sato:1980yn, Guth:1980zm, Linde:1981mu, Albrecht:1982wi, Linde:1983gd}. 
In addition to its original motivation to avoid the fine-tuning issues in the standard hot-universe scenario, it is equipped with the mechanism to generate primordial quantum fluctuations~\cite{Mukhanov:1981xt, Starobinsky:1982ee, Guth:1982ec, Bardeen:1983qw, Starobinsky:1979ty}. 
Those fluctuations are stretched and then considered to be classicalised when the length scale crosses out the Hubble horizon~\cite{Polarski:1995jg, Lesgourgues:1996jc}, and are later amplified due to gravitational instability to give rise to the seed of all the present cosmological structure such as stars and galaxies. 
Precise large-scale observations have confirmed the nearly scale-invariant and almost Gaussian primordial curvature perturbation~\cite{Planck:2018jri, Planck:2019kim}, as theoretically predicted from a wide class of inflationary models. 
The vanilla model of inflation consists of a single scalar field minimally coupled to gravity, which is undoubtedly the simplest choice amongst all the possible and consistent models that have been proposed until today. 

Extensions, or embeddings from the viewpoint of high-energy physics, of the vanilla model do not only include single-field models with an exotic potential, but also models with multiple scalar fields that drive inflation altogether, as well as models containing additional scalar fields that do not contribute to inflation, such as the curvaton model~\cite{Enqvist:2001zp, Lyth:2001nq, Moroi:2001ct}. 
Indeed, there are a lot of motivations to consider multifield models both theoretically and phenomenologically. 
What is attractive from a theoretical perspective is their compatibility with high-energy physics model constructions in which inflationary dynamics may be embedded~\cite{Wands:2007bd, Baumann:2014nda}. 
In addition to this, richer phenomenological consequences have been predicted in multifield models, such as the generation of non-adiabatic fluctuations as well as non-Gaussianities~\cite{Gordon:2000hv, Wands:2002bn, Bassett:2005xm, Peterson:2010np, Peterson:2011yt}. 
A particular motivation that the present article focusses on is that there exists a remarkable role of quantum fluctuations, distinctive to multifield models where the number of inflaton fields is much larger than unity. 

In a wide class of models with a single scalar field that slowly rolls on its potential, an inflationary solution exists at the classical level. 
There, the scalar field is deterministically dragged by the potential gradient, and it rolls down on the potential. 
The inflationary evolution is terminated when the vacuum energy of the inflaton field ceases to dominate the universe. 
From a given initial condition, the elapsed number of $e$-folds $N$ is measured until the termination of the inflation phase, and is deterministic. 

The classical treatment suffices to realise the inflationary universe, but the origin of the large-scale structure is encoded as quantum fluctuations that are superposed on the classical evolution. 
Various scales of quantum fluctuations are, because of the inflationary expansion, stretched and cross out the Hubble horizon. 
The large-scale field component undergoes the continuous but random inflow from small-scale quantum fluctuations, resulting in the stochastic evolution of the universe. 
In addition to the classical force that drives the scalar field down the potential, the stochastic random force originating from quantum fluctuations kicks it towards both the lower and higher regions of the potential. 
Whilst it evolves along the classical solution on average, the inflationary evolution slightly varies from region to region in space once quantum fluctuations are accounted for. 
This promotes the elapsed number of $e$-folds to a stochastic variable $\mathcal{N}$. 
When a single-field model is fixed, the deviation of $\mathcal{N}$ from the classical value $N$ is always due to the stochastic noise. 
That is, no deterministic source other than the gradient force does exist that affects the elapsed number of $e$-folds. 

As such, the stochastic effect plays an important role already in single-field models. 
For instance, it gives rise to non-Gaussian tails~\cite{Pattison:2017mbe, Prokopec:2019srf, Ezquiaga:2019ftu, Pattison:2021oen, Rigopoulos:2021nhv, Tada:2021zzj, Hooshangi:2021ubn, Achucarro:2021pdh, Animali:2022otk, Tomberg:2023kli, Vennin:2024yzl, Murata:2025onc} that cannot be found in the conventional Gaussian distribution. 
When rare fluctuations arise that contribute to the tail of the distribution of the stochastic duration of inflation, the large curvature perturbation may be realised that could lead to the formation of primordial black holes~\cite{Zeldovich:1967lct, Hawking:1971ei, Carr:1974nx, Carr:1975qj}. 
Though the stochastic effect tends to extend the duration of inflation~\cite{Vennin:2015hra}, inflation is terminated in almost all the regions unless the contribution of quantum fluctuations is too large. 
In other words, eternal inflation~\cite{Linde:1982ur, Vilenkin:1983xq, Winitzki:2008zz} can occur only as a rare stochastic event, whose probability is exponentially suppressed. 

Such stochastic effects also profoundly affect multifield models of inflation. 
A notable demonstration in the multifield scenario was recently presented in Ref.~\cite{Takahashi:2025hqt}, where the mean number of $e$-folds during inflation diverges once the number of fields exceeds a critical threshold. This phenomenon originates from the collective accumulation of inflaton fields, which generates a noise-induced deterministic force that compensates for the classical drift arising from the potential gradient. 
This noise-induced force cannot be found in any single-field model and is thus regarded as a collective force. 
The affected quantity is not restricted to the mean number of $e$-folds, but also its variance and higher-order statistical moments as well, which are directly related to the curvature perturbation $\zeta$ through the $\delta N$-formalism~\cite{Salopek:1990jq, Sasaki:1995aw, Sasaki:1998ug, Lyth:2004gb, Abolhasani:2019cqw} or the stochastic-$\delta \mathcal{N}$ formalism~\cite{Fujita:2013cna, Fujita:2014tja, Vennin:2015hra, Ando:2020fjm} in the context of stochastic inflation. 
It would allow for a further scrutiny of how stochastic effects in multifield models can shed light on the underlying inflationary scenario. 

It is thus important to find out how the number of fields $d$ affects the statistical quantities, which is a purely stochastic effect, but, as will be confirmed, affects the inflationary evolution in a deterministic way. 
This article scrutinises the stochastic effects in generic $O (d)$-symmetric multifield models, based on the stochastic formalism of inflation~\cite{Starobinsky:1986fx, Starobinsky:1994bd} (see Ref.~\cite{Cruces:2022imf} for a recent review), an effective field-theoretical treatment for the large-scale field configurations. 
The formalism has been applied to several multifield models in previous articles, see \textit{e.g.}~Ref.~\cite{Mollerach:1990zf}, and was systematically formulated and classified in Ref.~\cite{Assadullahi:2016gkk}. 
There, for $O (d)$-symmetric models, it was found at the nonzero leading order that the statistical quantities such as the mean number of $e$-folds, its variance, and the higher-order moments get corrections from the stochastic effects, explicitly depending on the number of fields. 

This article goes beyond the lowest nonzero order, fully generalising the previous studies. 
By restricting itself to a ``small-noise regime'', all the calculations are performed analytically to find the $d$-dependence of the stochastic correction terms at an arbitrary order. 
In addition to this, the concrete expressions of those terms are listed up to third order, emphasising the number-of-field-dependence of the statistical quantities in the small-noise regime. 
Both the mean number of $e$-folds and its variance are elongated under the stochastic effects in an arbitrary order, corresponding to the fact that the stochastic kick superposed on the classical trajectory in field space always realises an extra duration of inflation. 
In general, a $k$-th nonzero stochastic correction contains $d^{k}$ term. 
Those calculations of the statistical moments enable us to also derive the stochastic corrections in the observables through the stochastic $\delta \mathcal{N}$-formalism. 
The power spectrum and its tilt are therefore also derived, again to find their explicit dependences on the number of fields. 

The rest of this article is organised as follows. 
Section~\ref{sec:mfsi_rev} reviews the stochastic formalism applied to multifield inflationary models, as well as the necessary equations in this article. 
Section~\ref{sec:smex} focuses on a regime where the noise is small but cannot be neglected. 
The general expressions for the statistical moments of the stochastic number of $e$-folds in the small-noise regime are presented, which enable one to find the stochastic correction to the statistical quantities up to an arbitrary order in a systematic way. 
Section~\ref{sec:statistical_moments} gives the concrete expressions of those statistical quantities for a general $O (d)$-symmetric potential, demonstrating the relevance of the number of fields in the presence of the stochastic effects. 
Observable quantities such as the power spectrum and its tilt are also to be discussed. 
Section~\ref{sec:implications} discusses some explicit models such as monomial, $R^2$-type, and power-law models. 
Section~\ref{sec:conclusion} is devoted to summarising the article and as well as future possible directions that deserve to be further scrutinised. 

Throughout this article, natural units are used and $M_{\rm P} \simeq 2.4 \times 10^{18} \, \mathrm{GeV}$ denotes the reduced Planck mass. 

\section{Multifield stochastic inflation}
\label{sec:mfsi_rev}

A generic treatment of multifield models in the stochastic formalism of inflation is reviewed in this section, mainly based on Refs.~\cite{Assadullahi:2016gkk, Vennin:2016wnk}. 
The statistical properties of the stochastic duration of inflation are focussed on in particular, since they are related to the observables such as curvature perturbations through the stochastic-$\delta \mathcal{N}$ formalism, in which the dependence of the number of fields will be captured in Section~\ref{sec:statistical_moments}. 

\subsection{Stochastic evolution of inflaton fields}

In this section, a class of multifield inflationary models described by the action 
\begin{equation}
    S 
    = \int \dd^{4} x \, \sqrt{- g} \, \qty[ 
        \frac{M_{\rm P}^{2}}{2} R - \frac{g^{\mu \nu}}{2} \qty( \partial_{\mu} \vb*{\phi} ) \cdot \qty( \partial_{\nu} \vb*{\phi} ) 
        - V (\vb*{\phi}) 
    ] 
    \,\, , 
    \label{eq:pre_action}
\end{equation}
is considered. 
The background metric is assumed to be homogeneous and isotropic, so that the spacetime is characterised by the line element $\dd s^{2} = - \dd t^{2} + a^{2} (t) \, \dd \vb*{x}^{2}$, where $a = a (t)$ is the scale factor. 
The potential $V (\vb*{\phi})$ in Eq.~(\ref{eq:pre_action}) remains unspecified until the end of this section. 

The field-space is assumed to be flat with dimension $d$, denoting the inflaton fields by $\vb*{\phi} \equiv (\phi_{1}, \, \dots, \, \phi_{d})^{\mathsf{T}}$. 
The main idea to construct the stochastic formalism is to integrate out the sub-horizon (ultraviolet) modes in the entire Klein--Gordon equation, $\ddot{\vb*{\phi}} + 3 H \dot{ \vb*{\phi} } - \nabla_{\vb*{x}}^{2} \vb*{\phi} / a^{2} + \nabla V (\vb*{\phi}) = \vb*{0}$, decomposing the field configurations into the super-horizon (infrared) and sub-horizon modes as 
\begin{equation}
    \vb*{\phi} (N, \, \vb*{x}) 
    = \vb*{\phi}_{<} (N, \, \vb*{x}) + \vb*{\phi}_{>} (N, \, \vb*{x}) 
    \,\, , 
\end{equation}
where 
\begin{equation}
    \vb*{\phi}_{\gtrless} (N, \, \vb*{x}) 
    \equiv 
    \int \frac{\dd^{3} k}{(2 \pi)^{3}} \, \widetilde{\mathcal{W}}_{\pm} \qty[ \frac{k}{k_{\sigma} (N)} ] \widetilde{\vb*{\phi}} (N, \, \vb*{k}) e^{i \vb*{k} \cdot \vb*{x}} 
    \,\, . 
\end{equation}
Here and hereafter, time is labelled by the number of $e$-folds, $N \propto \ln a$, instead of the cosmic time $t$~\cite{Finelli:2008zg, Finelli:2010sh, Vennin:2015hra, Pattison:2019hef}. 
For a numerical constant $0 < \sigma \ll 1$, the time-dependent cutoff $k_{\sigma} (N) = \sigma \cdot aH$ separates the two scales. 
The most popular choice of the window function in the literature is $\widetilde{\mathcal{W}}_{\pm} (z) = \Theta (\pm z \mp 1)$, where $\Theta$ is the Heaviside function. 

The resultant effective equation of motion for $\vb*{\phi}_{<}$ describes the stochastic evolution of the inflaton fields, 
\begin{equation}
    \dv{\vb*{\phi}}{N} = - \frac{\nabla V (\vb*{\phi})}{3 H^{2} (\vb*{\phi})} + \frac{H (\vb*{\phi})}{2 \pi} \vb*{\xi} (N) 
    \,\, , 
    \label{eq:pre_lan}
\end{equation}
where the Hubble parameter $H \equiv \dd \ln a / \dd t$ is related to the inflationary potential by $H^{2} (\vb*{\phi}) \simeq V (\vb*{\phi}) / 3 M_{\rm P}^{2}$ since the slow-roll condition has been assumed. 
In Eq.~(\ref{eq:pre_lan}), $\nabla = \partial / \partial \vb*{\phi}$ is defined in the field space. 
In what follows, the large-scale part of the fields, $\vb*{\phi}_{<}$, is denoted simply by $\vb*{\phi}$. 
The normalised noise $\vb*{\xi}$ appearing in Eq.~(\ref{eq:pre_lan}) comes from the continuous but random transition from the sub-horizon to super-horizon modes. 
This noise is assumed to be uncoloured, though depending on the form of the window function coloured noise should follow as discussed in Refs.~\cite{Casini:1998wr, Winitzki:1999ve, Matarrese:2003ye, Liguori:2004fa, Andersen:2021lii, Mahbub:2022osb, Brahma:2024yor} (see also Refs.~\cite{Figueroa:2020jkf, Cruces:2024pni, Ahmadi:2025oon, Kawasaki:2026hnx} for recent studies on non-Markovianity in stochastic inflation). 
It is then characterised by the statistical properties, $\expval*{\vb*{\xi} (N, \, \vb*{x})} = \vb*{0}$ and 
\begin{align}
    &\quad 
    \expval*{\vb*{\xi} (N_{1}, \, \vb*{x}_{1}) \otimes \vb*{\xi} (N_{2}, \, \vb*{x}_{2})} 
    \notag \\ 
    &= \frac{ \sin \qty( k_{\sigma} \abs{ \vb*{x}_{1} - \vb*{x}_{2} } ) }{ k_{\sigma} \abs{ \vb*{x}_{1} - \vb*{x}_{2} } } \delta_{\rm D} (N_{1} - N_{2}) \mathbbm{1}_{d \times d} 
    \,\, . 
    \label{eq:pre_noise}
\end{align}
The Bunch--Davies vacuum initial condition ensures the Gaussianity of the noise with the vanishing mean, while our choice for the window function results in the $\delta$-function correlations with respect to time in Eq.~(\ref{eq:pre_noise}). 

In the limit $\vb*{x}_{1} = \vb*{x}_{2}$ in Eq.~(\ref{eq:pre_noise}), one has $\expval*{ \vb*{\xi} (N_{1}) \otimes \vb*{\xi} (N_{2}) } = \delta_{\rm D} (N_{1} - N_{2}) \mathbbm{1}_{d \times d}$ and the correlation between $\vb*{x}_{1}$ and $\vb*{x}_{2}$ vanishes, while the stochastic formalism is based on (the leading order of) gradient expansion. 
Therefore, each causal region evolves independently, and each stochastic realisation generated by Eq.~(\ref{eq:pre_lan}) corresponds to the evolution of each causal region. 
Starting from an initial condition $\vb*{\phi} = \vb*{\phi}_{0}$ at $N = N_{0} = 0$, the inflaton fields $\vb*{\phi}$ slowly roll on the potential $V (\vb*{\phi})$ until the slow-roll condition is violated. 
The terminating surface is in practice defined by $\mathcal{C}_{-} \equiv \qty{ \vb*{\phi}_{-} \mid \varepsilon (\vb*{\phi}_{-}) = 1}$, which can be regarded as the absorbing boundary since inflation is assumed to be terminated once $\vb*{\phi}$ hits $\mathcal{C}_{-}$, where $\varepsilon$ is the slow-roll parameter. 
In terms of the number of $e$-folds or as a function of the inflaton fields, $\varepsilon$ is given by 
\begin{equation}
	\varepsilon (\vb*{\phi}) 
	= - \dv{ \ln H (\vb*{\phi}) }{N} 
	\simeq \frac{M_{\rm P}^{2}}{2} \qty[ \frac{ \nabla V ( \vb*{\phi} ) }{ V (\vb*{\phi}) } ]^{2} 
	\,\, . 
    \label{eq:pre_srparam}
\end{equation}
For later use, the Hubble-flow parameters~\cite{Schwarz:2001vv, Schwarz:2004tz} are introduced here, 
\begin{equation}
	\varepsilon_{i + 1} (\vb*{\phi}) 
	\equiv \dv{ \ln \varepsilon_{i} (\vb*{\phi}) }{N} 
	\,\, , 
	\qquad 
	i \geq 0 
	\,\, . 
    \label{eq:pre_flowp}
\end{equation}
where $\varepsilon_{0} (\vb*{\phi}) \equiv H (\vb*{\phi}_{0}) / H (\vb*{\phi})$, $\varepsilon_{1} (\vb*{\phi}) = \varepsilon (\vb*{\phi})$, and so forth. 

It is convenient to nondimensionalise all the quantities. 
For the nondimensionalised fields $\vb*{r} \equiv \vb*{\phi} / M_{\rm P}$, the nondimensionalised potential is introduced by\footnote{
    Our definition of $v (\vb*{x})$ is the same as that of Ref.~\cite{Tokeshi:2023swe}, and is different by a factor of two from that in Ref.~\cite{Assadullahi:2016gkk}. 
    The replacement $v (\vb*{\phi})_{\text{\cite{Assadullahi:2016gkk}}} \to v (\vb*{\phi}) / 2$ reproduces the corresponding equations in the present article. 
} 
\begin{equation}
    v (\vb*{r}) 
    \equiv \frac{V (\vb*{\phi})}{12 \pi^{2} M_{\rm P}^{4}} 
    \,\, . 
\end{equation}
and is exclusively used instead of $V (\vb*{\phi})$ throughout. 
A differentiation $\nabla$ that acts on such nondimensionalised quantities is defined with respect to $\vb*{r}$ hereafter, so that $\nabla v (\vb*{r}) = \partial v (\vb*{r}) / \partial \vb*{r} = M_{\rm P} \cdot \partial \qty[ V (\vb*{\phi}) / 12 \pi^{2} M_{\rm P}^{4} ] / \partial \vb*{\phi}$ for instance. 
With those nondimensionalised quantities and derivatives, the stochastic equation (\ref{eq:pre_lan}) is recast into 
\begin{equation}
	\dv{\vb*{r}}{N} 
	= - \frac{ \nabla  v (\vb*{r}) }{v (\vb*{r}) } + \sqrt{ v (\vb*{r}) } \, \vb*{\xi} (N) 
	\,\, . 
	\label{eq:pre_lan_dimless}
\end{equation}
That is, the stochastic process $\vb*{r}$ is controlled by the deterministic drift vector $\mathsf{A} (\vb*{r}) \equiv - \nabla v (\vb*{r}) / v (\vb*{r})$ and the diffusion matrix $\mathsf{B} (\vb*{r}) \equiv \sqrt{ v (\vb*{r}) } \, \mathbbm{1}_{d \times d}$, for which Eq.~(\ref{eq:pre_lan_dimless}) can be written as $\dd \vb*{r} / \dd N = \mathsf{A} (\vb*{r}) + \mathsf{B} (\vb*{r}) \vb*{\xi} (N)$. 
This is a standard form of the multivariate stochastic process, in which neither the drift vector nor the (diagonal) diffusion matrix depends explicitly on the time variable. 

\subsection{Distribution functions}
\label{subsec:distfn}

The fact that the inflaton fields stochastically evolve under Eq.~(\ref{eq:pre_lan_dimless}) motivates us to consider the distribution function of them. 
For the infinitesimal generator of the stochastic process, $\mathcal{L}_{\rm FP}^{\dagger} \equiv - \nabla \cdot \mathsf{A} + \nabla \otimes \nabla : \mathsf{B}^{2} / 2$ (where the Frobenius inner product is used), the distribution function of $\vb*{r}$ evolves under the Fokker--Planck equation,  
\begin{equation}
    \pdv{f (\vb*{r}, \, N)}{N} 
    = \qty[ 
        \nabla \cdot \frac{ \nabla v (\vb*{r}) }{ v (\vb*{r}) } + \nabla^{2} \frac{v (\vb*{r})}{2} 
    ] f (\vb*{r}, \, N) 
    \,\, . 
    \label{eq:pre_fp}
\end{equation}
The distribution function (or the transition probability density to be more precise), $f = f (\vb*{r}, \, N) = f (\vb*{r}, \, N \mid \vb*{r}_{0}, \, N_{0})$, enables us to know the probability to find the field at the location $\vb*{r}$ in the field space, at a given time~$N$, provided that it started from an initial state $(\vb*{r}_{0}, \, N_{0})$. 

Since Eq.~(\ref{eq:pre_fp}) is of second order in $\vb*{r}$, two boundary conditions must be specified to write down the solution. 
One is the absorbing boundary, introduced above Eq.~(\ref{eq:pre_srparam}). 
It specifies a hypersurface in the field space that accumulates all the points where the slow-roll condition is violated, and $\mathcal{C}_{-} = \qty{ \vb*{r} \mid \varepsilon (\vb*{r}) = 1 }$ in terms of the nondimensionalised fields. 
It is here assumed that inflation is terminated once $\vb*{r}$ hits the absorbing boundary. 
Though a statistically rare random realisation of the noise could occur near $\mathcal{C}_{-}$ before (and even after) $\vb*{r}$ hits $\mathcal{C}_{-}$, such a possibility may safely be neglected since near the end of inflation effects of quantum fluctuations are very small in most models. 
For the other boundary condition, a reflective boundary $\mathcal{C}_{+}$ is introduced. 
It is placed in a high-potential region, and, when the effect of quantum fluctuations is very large, the existence of $\mathcal{C}_{+}$ prevents $\vb*{r}$ from its excursion into trans-Planckian regimes where quantum gravity may be relevant. 
In the small-noise regime, however, the exact location of $\mathcal{C}_{+}$ (as well as its presence itself) becomes irrelevant since $\vb*{r}$ can hardly ascend the potential, except that it is located near the reflective boundary initially, which is not the case of our interest. 

As mentioned in Introduction, the duration of inflation is also promoted to a stochastic number in the presence of stochastic effects. 
Suppose that, from an initial location $\vb*{r}$, it rolls on the potential under Eq.~(\ref{eq:pre_lan_dimless}). 
Without the stochastic noise $\vb*{\xi}$, it rolls down the potential until it hits $\mathcal{C}_{-}$, dragged by the gradient force, or equivalently the drift term. 
In this classical limit, the evolution of $\vb*{r}$ follows a uniquely determined classical trajectory, and the elapsed number of $e$-folds is a deterministic quantity. 

When the stochastic noise is accounted for, on the other hand, one stochastic realisation generated by Eq.~(\ref{eq:pre_lan_dimless}) and another one in general record different numbers of $e$-folds, due to the randomness of the stochastic noise $\vb*{\xi}$. 
Therefore, the number of $e$-folds elapsed from an initial location $\vb*{r}$ to the end of inflation occurring on $\mathcal{C}_{-}$ must be treated as a stochastic number, denoted by $\mathcal{N}$ and often called the \textit{first-passage time}. 
Hereafter, both the classical $e$-fold and the stochastic $e$-fold $\mathcal{N}$ will be referred to simply as the number of $e$-folds.
The distribution of $\mathcal{N}$, $f_{\rm FPT} (\vb*{r}, \, \mathcal{N})$, satisfies the adjoint Fokker--Planck equation, 
\begin{equation}
    \pdv{f_{\rm FPT} (\vb*{r}, \, \mathcal{N})}{\mathcal{N}} 
    = \qty[ 
        - \frac{ \nabla v (\vb*{r}) }{ v (\vb*{r}) } \cdot \nabla + \frac{v (\vb*{r})}{2} \nabla^{2} 
    ] f_{\rm FPT} (\vb*{r}, \, \mathcal{N}) 
    \,\, . 
    \label{eq:pre_fpt}
\end{equation}
Since the infinitesimal generator for $f_{\rm FPT}$ is the adjoint of $\mathcal{L}_{\rm FP}^{\dagger}$, Eq.~(\ref{eq:pre_fpt}) can also be written as $\partial f_{\rm FPT} / \partial \mathcal{N} = \mathcal{L}_{\rm FP} f_{\rm FPT}$. 
The ``initial`` condition to solve Eq.~(\ref{eq:pre_fpt}) is given by $f_{\rm FPT} (\vb*{r} \notin \mathcal{C}_{-}, \, \mathcal{N} = 0) = 0$. 
This is because every realisation needs a finite time to hit the absorbing boundary unless $\vb*{r} \in \mathcal{C}_{-}$ initially. 
On the other hand, nonzero duration of inflation cannot be elapsed if $\vb*{r} \in \mathcal{C}_{-}$ initially, so that $f_{\rm FPT} (\vb*{r} \in \mathcal{C}_{-}, \, \mathcal{N}) = \delta_{\rm D} (\mathcal{N})$ is required together. 

The first raw moment of the first-passage time, $\expval*{ \mathcal{N} } (\vb*{r})$, measures the mean duration of inflation starting from an initial location $\vb*{r}$ until it hits $\mathcal{C}_{-}$. 
More generally, the recurrence and differential equation for the statistical moments of the first-passage time, $\expval*{ \mathcal{N}^{n} } (\vb*{r})$, can be derived from Eq.~(\ref{eq:pre_fpt}). 
It is given, for $n \geq 1$, by~\cite{Assadullahi:2016gkk} 
\begin{equation}
    \qty[ \frac{v (\vb*{r})}{2} \nabla^{2} - \frac{\nabla v (\vb*{r})}{v (\vb*{r})} \cdot \nabla ] \expval*{ \mathcal{N}^{n} } (\vb*{r}) 
    + n \expval*{ \mathcal{N}^{n-1} } (\vb*{r}) 
    = 0 
    \,\, . 
    \label{eq:pre_dfeq_statmom}
\end{equation}
This in principle enables us, starting from $\expval*{ \mathcal{N}^{0} } (\vb*{r}) \equiv 1$, to obtain the higher moments recursively. 
However, it is in practice notoriously difficult to solve a multi-dimensional partial differential equation such as Eq.~(\ref{eq:pre_dfeq_statmom}), even numerically. 
This is one of the reasons why a class of $O(d)$-symmetric models is focused on in the following, through which the stochastic effects on the statistical moments of $\mathcal{N}$, in particular effects coming from the prevalence of a large number of fields $d$, will be derived 
analytically. 
Another reason is that our analysis serves as a benchmark when one goes beyond the $O (d)$-symmetric case to study more complicated but realistic models, which is beyond the scope of the present article. 

\subsection{\texorpdfstring{$O (d)$}{O(d)}-symmetric models}

A particular interest lies in a class of $O (d)$-symmetric models (see \textit{e.g.}~Refs.~\cite{Adshead:2020ijf, Tada:2023fvd} for studies under the symmetry), where $r \equiv \norm{ \vb*{r} } = \norm{ \vb*{\phi} } / M_{\rm P}$ is the only effective degree of freedom. 
In such cases, a set of stochastic processes (\ref{eq:pre_lan_dimless}) can be recast into a single process of $r$. 

In general, for 
\begin{equation}
	\mathbb{R}^{d} \ni (r_{1}, \, \dots, \, r_{d})^{\mathsf{T}} 
	\mapsto 
	r = \sqrt{ r_{1}^{2} + \cdots + r_{d}^{2} } \in \mathbb{R}^{1} 
	\,\, , 
\end{equation}
It\^{o} formula (It\^{o} lemma) derives the stochastic process for~$r$~\cite{gardiner2009stochastic} 
\begin{align}
	\dv{r}{N} 
	&= - \nabla r \cdot \frac{ \nabla v (\vb*{r}) }{v (\vb*{r})} 
	+ \frac{v (\vb*{r})}{2} \cdot \tr \qty( \nabla \otimes \nabla r ) 
	\notag \\ 
	&\quad 
	+ \sqrt{ v (\vb*{r}) } \, \qty( \nabla r ) \cdot \qty[ \vb*{\xi} (N) ]  
	\,\, . 
\end{align}
The second term in right-hand side is the so-called It\^{o} correction term. 
Let us now assume $O (d)$ symmetry, so that the dependence on all the variables but $r$ vanishes in all the quantities, for instance $v (\vb*{r}) = v (r)$. 
One then obtains $\tr \qty[ \nabla \otimes \nabla r ] = (d-1)/r$ for the It\^{o} correction term. 
For the stochastic noise, let us define the projected noise, $\xi \equiv ( \nabla r ) \cdot \vb*{\xi} (N)$. 
It should be noted here that $\nabla r$ is the unit vector in the $r$ direction. 
The statistical properties of this newly-introduced projected noise are given by 
\begin{subequations}
	\begin{align}
		\expval*{ \xi (N) } 
		&= \qty( \nabla r ) \cdot \expval*{ \vb*{\xi} (N) } 
		= 0 
		\,\, , 
		\\ 
		\expval*{ \xi (N_{1}) \xi (N_{2}) } 
		&= \frac{ \vb*{r}^{\mathsf{T}} \expval*{ \vb*{\xi} (N_{1}) \otimes \vb*{\xi} (N_{2}) } \vb*{r} }{ r^{2} } 
		= \delta_{\rm D} (N_{1} - N_{2}) 
		\,\, . 
	\end{align}
\end{subequations}
Therefore, $\xi (N)$ is again the uncoloured and standardised Gaussian noise. 
By combining the ingredients until here, the resultant stochastic evolution of $r$ is described by 
\begin{equation}
	\dv{r}{N} 
	= - \frac{v' (r)}{v (r)} + \frac{v (r)}{2} \frac{d-1}{r} 
	+ \sqrt{ v (r) } \, \xi (N) 
	\,\, . 
	\label{eq:pre_lan_r}
\end{equation}
This demonstrates a derivation of the effective stochastic process by $r$, directly from the original set of the Brownian motions by $\vb*{r}$ in the presence of (internal) spherical symmetry. 

Continuing the discussion about $O(d)$-symmetric cases, all the relevant quantities such as the distribution functions and statistical moments are independent of the angular variables. 
This enables us to remove $\nabla$'s from Eqs.~(\ref{eq:pre_fp}), (\ref{eq:pre_fpt}), and (\ref{eq:pre_dfeq_statmom}). 
For the Fokker--Planck equation (\ref{eq:pre_fp}), the distribution function in the polar-coordinate variables is introduced through the conservation of the probabilities, $f (\vb*{r}, \, N) \, \dd^{d} r = f (r, \, N) \, r^{d-1} \, \dd r \, \dd \Omega_{d-1}$, where the same symbol $f$ is used for the Cartesian and polar variables. 
There is no angular dependence in $f$ in the right-hand side due to $O (d)$ symmetry, and one then has $f (\vb*{r}, \, N) \propto f (r, \, N) / r^{d-1}$. 
This leads to the evolution equation of $f = f (r, \, N)$, which is given by 
\begin{equation}
	\pdv{f}{N} 
	= 
	- \pdv{r} \qty{ \qty[ - \frac{ v' (r) }{ v (r) } + \frac{v (r)}{2} \frac{d-1}{r} ] f } 
	+ 
	\pdv[2]{r} \qty[ \frac{v (r)}{2} f ] 
	\,\, . 
	\label{eq:pre_fp_r}
\end{equation}

The adjoint of Eq.~(\ref{eq:pre_fp_r}) reads, for $f_{\rm FPT} (r, \, \mathcal{N})$ defined in the same manner as $f$, 
\begin{equation}
	\pdv{f_{\rm FPT}}{\mathcal{N}} 
	= 
	\qty[ - \frac{ v' (r) }{ v (r) } + \frac{v (r)}{2} \frac{d-1}{r} ] \pdv{f_{\rm FPT}}{r} + \frac{v (r)}{2} \pdv[2]{f_{\rm FPT}}{r} 
	\,\, . 
    \label{eq:pre_fpt_r}
\end{equation}
The ordinary differential equation for the statistical moments of $\mathcal{N}$, for $n \geq 1$, follows from Eq.~(\ref{eq:pre_fpt_r}) that~\cite{Assadullahi:2016gkk}
\begin{equation}
    \frac{v}{2} \qty{ \dv[2]{r} + \qty[ \frac{d-1}{r} - 2 \frac{v'}{v^{2}} ] \dv{r} } \expval*{ \mathcal{N}^{n} } 
    = - n \expval*{ \mathcal{N}^{n-1} } 
    \,\, . 
    \label{eq:pre_dfeq_statmom_vr}
\end{equation}
Here, $v' (r) = \dd v (r) / \dd r$. 
Under the two boundary conditions introduced in Section~\ref{subsec:distfn}, the analytical but formal solution to Eq.~(\ref{eq:pre_dfeq_statmom_vr}) can be written down, given by~\cite{Assadullahi:2016gkk} 
\begin{align}
    \expval*{ \mathcal{N}^{n} } (r) 
    &= n 
    \int_{r_{-}}^{r} \frac{\dd x}{x^{d-1}} \, \exp \qty[ - \frac{2}{v (x)} ] 
    \notag \\ 
    &\quad \times 
    \int_{x}^{r_{+}} \dd y \, y^{d-1} \frac{2}{v (y)} \exp \qty[ \frac{2}{v (y)} ] 
    \expval*{ \mathcal{N}^{n-1} } (y) 
    \,\, . 
    \label{eq:d_stmm_master}
\end{align}
It should be noted that, in Eq.~(\ref{eq:d_stmm_master}), $x, \, y, \, r_{-}$, and $r_{+}$ are all the nondimensionalised fields. 
The analytical formula~(\ref{eq:d_stmm_master}) enables us to recursively derive the statistical moments of $\mathcal{N}$, starting from $\expval*{ \mathcal{N}^{0} } (r) = \expval*{ 1 } = 1$ and $n = 1$, however in most cases, they should be evaluated numerically unfortunately. 

\begin{figure}
	\centering
	\includegraphics[width = 0.995\linewidth]{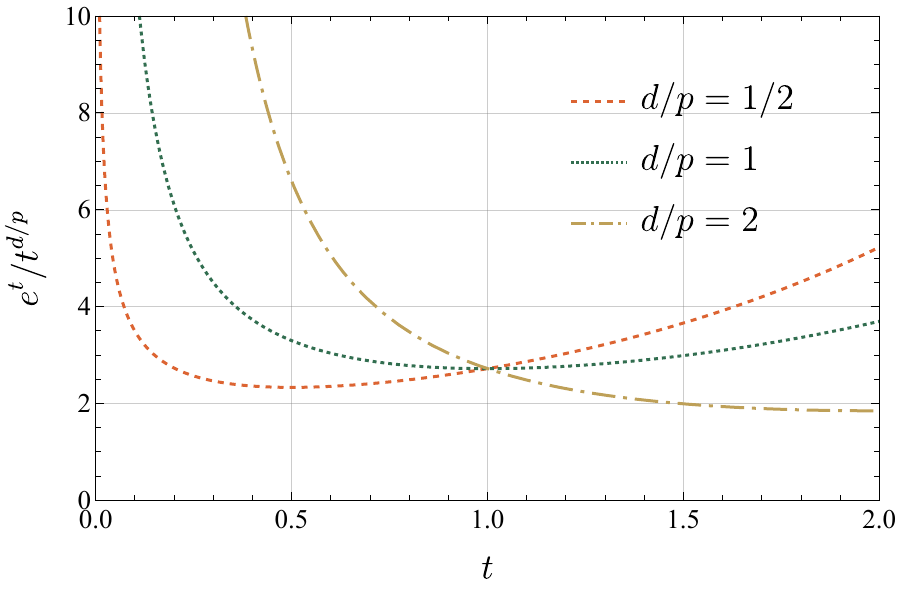}
	\caption{
		The integrand of Eq.~(\ref{eq:pre_infinf}). 
		It diverges as $t \to 0$ regardless of $d/p$, but the integral remains finite only if $d/p < 1$ when the integration region is extended to $t = 0$. 
	} 
	\label{fig:div}
\end{figure}

\vspace{1.0\baselineskip}
\paragraph*{\textit{\textbf{Monomial potential.}}}

When stochastic effects are accounted for, it occurs in a wide class of models that the inflationary evolution becomes eternal, by chance or in a deterministic manner. 
There are several mechanisms that lead to eternal evolution, and one of the simplest models to handle is reviewed here. 

To demonstrate it, let us focus on the mean number of $e$-folds ($n = 1$), and on the monomial potential, $v (r) = v_{0} r^{p}$. 
The inner integral over $y$ in Eq.~(\ref{eq:d_stmm_master}) behaves as 
\begin{equation}
	\int_{x}^{r_{+}} \dd y \, 
	y^{d-1} 
	\exp \qty[ 
		\frac{2}{v_{0} y^{p}} 
	] 
	\frac{2}{v_{0} y^{p}} 
	\propto 
	\int_{2 / v_{0} r_{+}^{p}}^{2 / v_{0} x^{p}} 
	\dd t \, 
	\frac{e^{t}}{t^{d/p}} 
	\,\, , 
	\label{eq:pre_infinf}
\end{equation}
where the irrelevant proportional constant is $(2 / v_{0})^{d/p} / p$. 
Figure~\ref{fig:div} shows the integrand in Eq.~(\ref{eq:pre_infinf}), the behaviour of which is determined by the ratio $d/p$. 
An interesting behaviour in $\expval*{ \mathcal{N} }$ arises in the limit where the reflective boundary is sent to infinity. 
In this limit, Eq.~(\ref{eq:pre_infinf}) converges to give a finite number of the mean number of $e$-folds if and only if $d / p < 1$. 
When $d / p = 1$, on the other hand, it diverges logarithmically, and the divergence becomes worse for $d / p > 1$. 
In both cases, the divergence of Eq.~(\ref{eq:pre_infinf}) gives rise to $\expval*{ \mathcal{N} } = \infty$. 
The number of $e$-folds in the limit $r_{+} \to \infty$ is therefore summarised as follows:
\begin{equation}
	\lim_{r_{+} \to \infty} \expval*{ \mathcal{N} } (r) 
	\quad 
	\begin{cases}
		< \infty &\quad (d / p < 1) 
		\,\, , 
		\\[2.0ex]
		= \infty &\quad (d / p \geq 1) 
		\,\, .  
	\end{cases}
	\label{eq:pre_rctinf}
\end{equation}
The emergence of an infinite number of $e$-folds in the latter case is called \textit{infinite inflation} in Ref.~\cite{Vennin:2016wnk} (see also Refs.~\cite{Noorbala:2019kdd} for a relevant literature). 
It should be noted that this divergence is controlled by the reflective boundary, and $\expval*{ \mathcal{N} }$ remains finite as long as $r_{+}$ is finite even if it is placed extremely far away. 
That $\expval*{ \mathcal{N} } = \infty$ in this context can thus be found if and only if $r_{+} = \infty$, in addition to $d / p \geq 1$. 
Whatever all the other parameters are, such as $v_{0}$ (\textit{e.g.}~the mass of the inflatons) and $r$ (the initial location), inflation lasts forever on average if $d / p \geq 1$ in the limit $r_{+} \to \infty$. 
It should also be noted that this way of realisation of an infinite number of $e$-folds is different from \textit{eternal inflation}~\cite{Linde:1982ur, Vilenkin:1983xq, Winitzki:2008zz} occurred by the rare realisations of the stochastic noise, as mentioned in Ref.~\cite{Vennin:2016wnk}. 

{
\makeatletter
\renewcommand{\arraystretch}{1.8}
\begin{table}
    \caption{
        The presence or absence of the reflective boundary in relation to the mean number of $e$-folds. 
        ``Small noise'' means that the presence or absence of $r_{+}$ is irrelevant in the small-noise regime, whereas in the other two columns it is relevant. 
    }
	\begin{tabular}{c||c|c|c}
		~$r_{+}$~ & ~$r_{+} < \infty$~ & ~$r_{+} = \infty$~ & ~small noise~ 
		\\ \hline 
		~$\expval*{ \mathcal{N} }$~ & ~$\expval*{ \mathcal{N} } < \infty$~ & ~Eq.~(\ref{eq:pre_rctinf})~ & ~Eq.~(\ref{eq:rvd_mean_master})~  
	\end{tabular}
    \label{tab:situ}
\end{table}
\makeatother
}%

As will be confirmed, the exact location of $r_{+}$ becomes irrelevant in the small-noise regime. 
This means that an infinite number of $e$-folds \textit{controlled by the reflective boundary} is absent in the small-noise regime, as 
the contribution from $r_{+}$ is exponentially suppressed. 
As a result, the formula (\ref{eq:d_stmm_master}) allows us to analytically compute the statistical moment $\expval*{ \mathcal{N}^{n} }$
within a perturbative framework, yielding results consistent with those obtained non-perturbatively from the stochastic process (\ref{eq:pre_lan_r}). 
This means that the stochastic process with and without the reflective boundary, assuming $r_{-} < r \ll r_{+}$, are in practice identical in the small-noise regime. 
When the stochastic noise is not necessarily small, on the other hand, $r$ tends to enter the high-potential region close to $r_{+}$, and the presence of the reflective boundary may matter. 
See also Table~\ref{tab:situ}. 

\subsection{Discretisation scheme}

Though It\^{o} discretisation scheme is exclusively used following the literature~\cite{Pinol:2020cdp, Tokuda:2017fdh, Tokuda:2018eqs}, there is another famous discretisation scheme known as Stratonovich discretisation scheme. 
While the ambiguity of the discretisation scheme is irrelevant since this article focusses, from Section~\ref{sec:smex}, on the small-noise regime, the two are not exactly the same to be precise. 
However, there is a fact between the two discretisation schemes~\cite{gardiner2009stochastic} that states that an It\^{o} stochastic process of the form 
\begin{equation}
    \dv{x}{N} 
    = \mathsf{A} (x) + \mathsf{B} (x) \xi (N) \,,
\end{equation}
where $\mathsf{A} (x)$ and $\mathsf{B} (x)$ respectively represent the deterministic drift and the diffusion coefficient, 
is the same as the Stratonovich stochastic process of the form 
\begin{equation}
    \dv{x}{N} 
    = \mathsf{A} + \frac{ \mathsf{B} (x)}{2} \dv{ \mathsf{B} (x) }{x} 
    + \mathsf{B} (x) \xi (N) 
    \,\, . 
\end{equation}

For the stochastic process (\ref{eq:pre_lan_r}), the Stratonovich-discretised version therefore reads 
\begin{equation}
	\dv{r}{N} 
	= - \frac{v' (r)}{v (r)} + \frac{v (r)}{2} \qty[ \frac{d-1}{r} + \frac{v' (r)}{2 v (r)} ] + \sqrt{v (r)} \, \xi (N) 
	\,\, . 
	\label{eq:disc_lan2}
\end{equation}
The term $v' (r) / 2 v (r)$ arises in addition to the noise-induced centrifugal-force term, multiplied by $v(r)$. 
For the monomial model, $v (r) = v_{0} r^{p}$, Eq.~(\ref{eq:disc_lan2}) reduces to 
\begin{equation}
	\dv{r}{N} 
	= - \frac{v' (r)}{v (r)} + \frac{v (r)}{2} \frac{d - 1 + p/2}{r} + \sqrt{v (r)} \, \xi (N)  
	\,\, . 
\end{equation}
This means that, the difference between the two discretisation schemes can be absorbed as the increment of the number of fields, from $d$ to $d + p/2$. 
In particular, it becomes irrelevant when $d$ is very large while keeping $p = \mathcal{O} (1)$, which is the situation of our interest, 
and the difference between the two schemes is therefore negligible for $d \gg p$.
As long as one focusses on the slow-roll regime, the same statement can be justified even for models that do not respect $O (d)$ symmetry, since the additional term in the Stratonovich scheme, $v' (r) / 2 v (r)$, is proportional to~$\sqrt{ \varepsilon }$, where $\varepsilon$ is the slow-roll parameter that satisfies $\varepsilon\ll 1$ during inflation. 

\section{Statistical moments \texorpdfstring{\protect\\}{} in small-noise regime}
\label{sec:smex}

As was mentioned at the end of the previous section, a regime of particular interest concerns the statistical moments of $\mathcal{N}$ under small noises.
At zeroth order of the noise, the trajectory of the $r$-field is deterministically realised. 
In such situations, the classically-elapsed number of $e$-folds is a function of the initial location $r$, where no dependence on $d$ can be found. 
At first order of the noise, a stochastic correction that depends on $d$ linearly extends the mean number of $e$-folds. 
The quadratic dependence on $d$ appears at second order, and higher-order terms follow accordingly. 
Without specifying the potential with $O (d)$ symmetry, this section derives the general-order expression for the statistical moments of the stochastic number of $e$-folds.  

Given that the strength of each stochastic kick is determined by $v$, the small-noise regime refers to situations where $v \ll 1$. 
The perturbative expansion is therefore performed with $v$ itself being the expansion parameter~\cite{Vennin:2015hra}. 

\subsection{Classical formula}
\label{subsec:cl}

Upon first inspection, it is not so trivial how the well-known formulas for the classically-realised number of $e$-folds can be revalidated from Eq.~(\ref{eq:d_stmm_master}). 
Before going to higher-order calculations, this subsection derives the classical formula of the mean number of $e$-folds from Eq.~(\ref{eq:d_stmm_master}). 
The content in this subsection is primarily a review of Refs.~\cite{Vennin:2015hra, Assadullahi:2016gkk}, though it is provided in more detail as calculations themselves will be useful for our later general derivations. 

The mean number of $e$-folds, \textit{i.e.}, the first statistical moment of $\mathcal{N}$ that can in principle be calculated by setting $n = 1$ in Eq.~(\ref{eq:d_stmm_master}), reads 
\begin{align}
    \expval*{ \mathcal{N} } (r) 
    &= 
    \int_{r_{-}}^{r} \frac{\dd x}{x^{d-1}} \, \exp \qty[ - \frac{2}{v (x)} ] 
    \notag \\ 
    &\quad \times 
    \int_{x}^{r_{+}} \dd y \, y^{d-1} \frac{2}{v (y)} \exp \qty[ \frac{2}{v (y)} ] 
    \,\, . 
    \label{eq:sncl_mean}
\end{align}
For a fixed $x$ such that $r_{-} \leq x \leq r$, the $y$-integral in Eq.~(\ref{eq:d_stmm_master}) is performed in the domain $x \leq y \leq r_{+}$. 
More specifically, $x$ always comes inside the region between the terminating surface and the initial location, which can be realised at the classical level. 
This is why, in the small-noise regime, the main contribution in the $y$-integral mainly comes from the vicinity of $y = x$. 
This motivates us to perform a Taylor expansion of the integrand at $y = x$, as 
\begin{equation}
	\frac{v (x)}{v (y)} 
	= 
	1 - f_{1} (x) (y - x) + \mathcal{O} ( (y - x)^{2} ) 
	\,\, , 
	\label{eq:sncl_vT}
\end{equation}
where $f_{1} (x) \equiv v' (x) / v (x)$. 
The exponential factor is also expanded in such a way that 
\begin{equation}
	\exp \qty[ \frac{2}{v (y)} ] 
	\simeq \exp \qty[ \frac{2}{v (x)} ] \exp \qty[ 
		- \frac{2 f_{1} (x)}{v (x)} (y - x) 
	] 
	\,\, . 
	\label{eq:sncl_evT}
\end{equation}
The second exponential function in Eq.~(\ref{eq:sncl_evT}) plays the role of the saddle point, and the higher-order corrections can be neglected here in deriving the leading-order result. 
Substitution of Eqs.~(\ref{eq:sncl_vT}) and (\ref{eq:sncl_evT}) into Eq.~(\ref{eq:sncl_mean}) gives 
\begin{align}
	\expval*{ \mathcal{N} } (r) 
	&\simeq \int_{r_{-}}^{r} \frac{\dd x}{x^{d-1}} 
	\frac{2}{v (x)} 
	\notag \\ 
	&\quad \times 
	\int_{x}^{r_{+}} \dd y \, y^{d-1} 
	\exp \qty[ - \frac{2 f_{1} (x)}{v (x)} (y - x) ] 
	\,\, . 
	\label{eq:sncl_n1}
\end{align}
The second line in Eq.~(\ref{eq:sncl_n1}) can be performed analytically in terms of the incomplete gamma function, defined by 
\begin{equation}
	\Gamma (s, \, z) 
	\equiv \int_{z}^{\infty} \dd t \, t^{s-1} e^{-t} 
	\,\, , 
\end{equation}
as 
\begin{align}
    &\quad 
    \int \dd y \, y^{d-1} 
	\exp \qty[ - \frac{2 f_{1} (x)}{v (x)} (y - x) ] 
    \notag \\ 
    &= 
	- \exp \qty[ \frac{2 f_{1} (x)}{v (x)} x ] 
	\qty[ \frac{v (x)}{2 f_{1} (x)} ]^{d} 
	\Gamma \qty[ d, \, \frac{2 f_{1} (x)}{v (x)} y ] 
	\,\, .  
    \label{eq:sncl_int}
\end{align}
Let us now assume that the second argument of the incomplete gamma function is positive. 
That is to say, it is assumed that the inflationary potential $v (x) > 0$ is a monotonically increasing function, provided that $y$ is always positive. 
In the small-noise regime where $v \ll 1$, the asymptotic formula of the incomplete gamma function,  
\begin{equation}
	\Gamma (d, \, z) 
	\approx  z^{d-1} e^{-z} 
	\,\, , 
	\qquad 
	z \to \infty 
	\,\, , 
    \label{eq:incg_ep1}
\end{equation}
can be used. 
This leads to 
\begin{equation}
	\expval*{ \mathcal{N} } (r)  
	\simeq 
	\int_{r_{-}}^{r} 
	\frac{ \dd x }{ f_{1} (x) } 
	\qty( \frac{y}{x} )^{d-1} 
	\eval{ 
		\exp [ \frac{ 2 f_{1} (x) }{v (x)} (x - y) ] 
	}_{y = r_{+}}^{y = x} 
    \,\, . 
\end{equation}
Now, as announced previously, it can be confirmed that the contribution from $y = r_{+}$ is exponentially suppressed for a sufficiently large $r_{+}$ because of $r_{-} \leq x \leq r$. 
Therefore, in the small-noise regime, it turns out that the reflective boundary has no effect and is just formally introduced 
as a boundary condition to Eq.~(\ref{eq:pre_dfeq_statmom_vr}), in addition to the absorbing boundary condition. 
The result is given by 
\begin{equation}
	\expval*{ \mathcal{N} } (r) 
	\simeq 
	\int_{r_{-}}^{r} \frac{ \dd x }{ f_{1} (x) } 
	= 
	\int_{r_{-}}^{r} \dd x \, \frac{v (x)}{v' (x)} 
	\,\, . 
	\label{eq:sncl_clef}
\end{equation}

This is an $\mathcal{O} (v^{0})$ quantity as it should be, and matches the well-known classical formula for the number of $e$-folds. 
Indeed, it reduces to $\expval*{ \mathcal{N} } (r) = (r^{2} - r_{-}^{2}) / 2 p$ for the monomial model, $v (r) = v_{0} r^{p}$, which will be scrutinised in Section~\ref{sec:implications}. 
Since no stochastic effect is incorporated at the leading-order result (\ref{eq:sncl_clef}) in $v$, neither $\mathcal{O} (v)$ term nor $d$-dependence can be found. 
It contains the effect coming from the classical gradient force only, by which $r$ deterministically rolls down on the potential. 
Though a similar calculation can be performed for $\expval*{ \mathcal{N}^{2} }$, to derive the zeroth-order term of the variance $\expval*{ \mathcal{N}^{2} } - \expval*{ \mathcal{N} }^{2}$ [$= 0$ at that order since $\mathcal{N}$ is deterministic], it is postponed to Section~\ref{subsec:var} after the general formulas are derived. 

In the presence of the stochastic effects, on the other hand, the motion driven by the stochastic kicks orthogonal to the classical trajectory affects the mean number of $e$-folds, by which $r$ loiters and $\expval*{ \mathcal{N} }$ is thus extended, as will be confirmed in Section~\ref{sec:statistical_moments}. 
There, it will be seen that from the next orders $d$-dependent terms arise in $\expval*{ \mathcal{N} }$. 
Though the perturbative calculations can be performed order by order, the next subsection presents the general expression from which one can derive the correction term at an arbitrary order systematically. 
Besides $\expval*{ \mathcal{N} }$, the derived result applies to arbitrary higher statistical moments of $\mathcal{N}$ thanks to its recursive structure.

\subsection{General formula}
\label{subsec:deriv}

The small-noise calculations reviewed in Section~\ref{subsec:cl} can be extended to derive the statistical moments, $\expval*{ \mathcal{N}^{n} }$, for an arbitrary $n$. 
The aim of this subsection is to derive the general perturbative result, Eq.~(\ref{eq:sngen_main}), in which perturbative terms at all orders in the noise are included. 

In general, the $n$-th moment of the number of $e$-folds is expanded in the small-noise regime as a series of $v$, 
\begin{equation}
    \expval*{ \mathcal{N}^{n} } (r) 
    = \sum_{\ell = 0}^{\infty} \expval*{ \mathcal{N}^{n} }^{(\ell)} (r)  
    \,\, . 
    \label{eq:d_stmm_oex}
\end{equation}
In Eq.~(\ref{eq:d_stmm_oex}), $\expval*{ \mathcal{N}^{n} }^{(\ell)} (r) = \mathcal{O} (v^{\ell})$ stands for the $\ell^{\text{th}}$-order correction term that accounts for the stochastic effect. 
The first ($n = 1$) and the second ($n = 2$) moments are of particular interest since they are directly related to the most important observables, namely the power spectrum and its spectral index. 
In addition to this, having the third moment enables us to calculate the local non-Gaussianity parametrised by $f_{\rm NL}$, which is however beyond the scope of the present article. 
 
Let us first rewrite the master formula (\ref{eq:d_stmm_master}) in the differentiated form, 
\begin{align}
    \frac{1}{n+1} 
    \dv{\expval*{ \mathcal{N}^{n+1} } (x)}{x}  
    &= \frac{1}{x^{d-1}} \int_{x}^{r_{+}} \dd y \, y^{d-1} \frac{2}{v (y)}  
    \notag \\ 
    &\quad \times 
    \exp \qty[ - \frac{2}{v (x)} + \frac{2}{v (y)} ] 
    \expval*{ \mathcal{N}^{n} } (y) 
    \,\, ,
    \label{eq:sngen_master}
\end{align}
since $x$-integral becomes relevant only after all the perturbative calculations are performed, as will be discussed below. 

For the reason described below Eq.~(\ref{eq:sncl_mean}), the $n$-th moment of $\mathcal{N}$ in the integrand in Eq.~(\ref{eq:sngen_master}) is Taylor expanded around $y = x$, 
\begin{align}
	\expval*{ \mathcal{N}^{n} } (y) 
	&= 
	\expval*{ \mathcal{N}^{n} } (x) 
	\sum_{i = 0}^{\infty} \frac{g_{n, \, i} (x)}{i!} (y - x)^{i} 
	\,\, , 
	\label{eq:sngen_nyT}
	\\[2.0ex]  
	g_{n, \, i} (x) 
	&\equiv \frac{1}{\expval*{ \mathcal{N}^{n} } (x)} \dv[i]{\expval*{\mathcal{N}^{n}} (x)}{x} 
	\,\, . 
	\label{eq:sngen_gnT}
\end{align}
Provided that the two $\expval*{ \mathcal{N}^{n} } (x)$ in Eq.~(\ref{eq:sngen_gnT}) are expanded in powers of $v$, later $g_{n, \, i} (x)$ is expanded in the same manner, see Eq.~(\ref{eq:d_stmm_oex}). 
The factor $2 / v (y)$ in the integrand is also Taylor expanded around $y = x$, 
\begin{equation}
	\frac{2}{v (y)} 
	= \frac{2}{v (x)} \sum_{i = 0}^{\infty} \frac{ (-)^{i} f_{i} (x) }{i!} (y - x)^{i} 
	\,\, . 
	\label{eq:sngen_vT}
\end{equation}
This is the generalisation of Eq.~(\ref{eq:sncl_vT}). 
Here, $f_{i} (x)$ is defined as the $i$-th derivative of $1/v (x)$:
\begin{align}
	f_{i} (x) 
	&\equiv 
	(-)^{i} v (x) \dv[i]{x} \qty[ \frac{1}{v (x)} ] 
	\label{eq:sngen_fnT} 
    \\ 
	&= \sum_{j = 0}^{i} 
	\frac{ (-)^{i + j} j! }{ [ v (x) ]^{j} } 
	B_{i, \, j} [ v' (x), v'' (x), \cdots, v^{(i - j + 1)} (x) ] 
	\,\, . 
	\notag 
\end{align}
The second line follows due to Fa\`{a} di Bruno's formula~\cite{faa}, in which $B_{i, \, j} (x_{1}, \, x_{2}, \, \cdots, \, x_{i - j + 1})$ is the (partial or incomplete exponential) Bell polynomial~\cite{5ff4f149-c808-3dd9-a33e-1cf21fe301dc, Louis:1974}. 
It is defined through the generating function,
\begin{equation}
    \frac{1}{j!}
    \qty( 
        \sum_{i = 1}^{\infty} x_{i} \frac{t^{i}}{i!}
    )^{j}
    =
    \sum_{i = j}^{\infty}
    B_{i, \, j} 
    (x_1, \, \dots, \, x_{i - j + 1}) 
    \frac{t^{i}}{i!} 
    \,\, . 
    \label{eq:def_Bell}
\end{equation}

The same expansion as Eq.~(\ref{eq:sngen_vT}) is performed to the $2  / v (y)$ factor in the exponential. 
When the terms are split into the leading saddle and the remainders, it reads 
\begin{align}
    &\exp \qty[ - \frac{2}{v (x)} + \frac{2}{v (y)} ] 
	\notag \\
    &= \exp \qty[ - \frac{2 f_{1} (x)}{v (x)} (y - x) ] 
	\notag \\
	&\quad \times 
	\exp \qty[ \frac{2}{v (x)} \sum_{i = 2}^{\infty} \frac{ (-)^{i} f_{i} (x) }{ i! } (y-x)^{i} ] 
    \vphantom{\int_{\displaystyle\frac{a}{b}}^{\displaystyle\frac{a}{b}}} 
	\,\, , 
	\label{eq:sngen_evT} 
\end{align}
which is a generalisation of Eq.~\eqref{eq:sncl_evT}. 
Though the expression 
will be reorganised later in Eq.~(\ref{eq:sngen_expT_3}), a straightforward manipulation is to expand the second line in Eq.~(\ref{eq:sngen_evT}) into the power series of $(y-x)$, 
\begin{widetext}
	\begin{subequations}
	\begin{align}
		&\quad \exp \qty[ \frac{2}{v (x)} \sum_{i = 2}^{\infty} \frac{ (-)^{i} f_{i} (x) }{ i! } ( y - x )^{i} ] 
		= \sum_{j = 0}^{\infty} \frac{1}{j!} \qty[ \frac{2}{v (x)} \sum_{i = 2}^{\infty} \frac{ (-)^{i} f_{i} (x) }{ i! } ( y - x )^{i} ]^{j} 
		= \sum_{j = 0}^{\infty} 
		\raisebox{-0.34ex}{\fbox{$v^{j}$}} \, \times 
		\qty[ \sum_{i = 2}^{\infty} 
		\tikz[baseline={(0, -0.1)}]{
			\fill (0, 0) circle (2.5pt);
  			\node at (0, -0.3) {\scriptsize {$i$}};
		} 
		\times ( y - x )^{i} ]^{j} 
		\label{eq:sngen_expT_2a}
		\\ 
		&= 1 
            	+ 
            	\raisebox{-0.34ex}{\fbox{$v^{1}$}} \, \times 
		\tikz[baseline={(0, -0.1)}]{
			\fill (0, 0) circle (2.5pt);
  			\node at (0, -0.3) {\scriptsize {$2$}};
		} 
		\times (y - x)^{2} 
		+ 
		\raisebox{-0.34ex}{\fbox{$v^{1}$}} \, \times 
		\tikz[baseline={(0, -0.1)}]{
			\fill (0, 0) circle (2.5pt);
  			\node at (0, -0.3) {\scriptsize {$3$}};
		} 
		\times (y - x)^{3} 
		+ 
		\qty( 
			\raisebox{-0.34ex}{\fbox{$v^{1}$}} \, \times 
			\tikz[baseline={(0, -0.1)}]{
				\fill (0, 0) circle (2.5pt);
  				\node at (0, -0.3) {\scriptsize {$4$}};
			} 
			+ 
			\raisebox{-0.34ex}{\fbox{$v^{2}$}} \, \times 
			\tikz[baseline={(0, -0.1)}]{
				\fill (0, 0) circle (2.5pt);
  				\node at (0, -0.3) {\scriptsize {$2$}};
				\fill (1, 0) circle (2.5pt);
  				\node at (1, -0.3) {\scriptsize {$2$}};
				\draw[thick] (0, 0) -- (1, 0);
			}
		) \times (y - x)^{4} 
		\notag \\[2.0ex] 
		&\quad~~\, + 
		\qty[ \, 
			\raisebox{-0.34ex}{\fbox{$v^{1}$}} \, \times 
			\tikz[baseline={(0, -0.1)}]{
				\fill (0, 0) circle (2.5pt);
  				\node at (0, -0.3) {\scriptsize {$5$}};
			} 
			+ 
			\raisebox{-0.34ex}{\fbox{$v^{2}$}} \, \times 
			\qty( 
				\tikz[baseline={(0, -0.1)}]{
					\fill (0, 0) circle (2.5pt);
  					\node at (0, -0.3) {\scriptsize {$2$}};
					\fill (1, 0) circle (2.5pt);
  					\node at (1, -0.3) {\scriptsize {$3$}};
					\draw[thick] (0, 0) -- (1, 0.0);
				} 
				+ 
				\tikz[baseline={(0, -0.1)}]{
					\fill (0, 0) circle (2.5pt);
  					\node at (0, -0.3) {\scriptsize {$3$}};
					\fill (1, 0) circle (2.5pt);
  					\node at (1, -0.3) {\scriptsize {$2$}};
					\draw[thick] (0, 0) -- (1, 0);
				} 
			) 
		] \times (y - x)^{5} 
		\notag \\[2.0ex] 
		&\quad~~\, + \qty{ 
			\raisebox{-0.34ex}{\fbox{$v^{1}$}} \, \times 
			\tikz[baseline={(0, -0.1)}]{
				\fill (0, 0) circle (2.5pt);
  				\node at (0, -0.3) {\scriptsize $6$};
			} 
			+ 
			\raisebox{-0.34ex}{\fbox{$v^{2}$}} \, \times 
			\qty[ \, 
				\tikz[baseline={(0, -0.1)}]{
					\fill (0, 0) circle (2.5pt);
  					\node at (0, -0.3) {\scriptsize {$2$}};
					\fill (1, 0) circle (2.5pt);
  					\node at (1, -0.3) {\scriptsize {$2$}};
					\draw[thick] (0, 0) -- (1, 0);
				} 
				+ \qty( 
					\tikz[baseline={(0, -0.1)}]{
						\fill (0, 0) circle (2.5pt);
  						\node at (0, -0.3) {\scriptsize $2$};
						\fill (1, 0) circle (2.5pt);
  						\node at (1, -0.3) {\scriptsize $4$};
						\draw[thick] (0, 0) -- (1, 0);
					} 
					+ 
					\tikz[baseline={(0, -0.1)}]{
						\fill (0, 0) circle (2.5pt);
  						\node at (0, -0.3) {\scriptsize $4$};
						\fill (1, 0) circle (2.5pt);
  						\node at (1, -0.3) {\scriptsize $2$};
						\draw[thick] (0, 0) -- (1, 0);
					} 
				) 
			] 
			+ 
			\raisebox{-0.34ex}{\fbox{$v^{3}$}} \, \times 
			\tikz[baseline={(0, -0.1)}]{
				\fill (0, 0) circle (2.5pt);
  				\node at (0, -0.3) {\scriptsize $2$};
				\fill (1, 0) circle (2.5pt);
  				\node at (1, -0.3) {\scriptsize $2$};
				\fill (2, 0) circle (2.5pt);
  				\node at (2, -0.3) {\scriptsize $2$};
				\draw[thick] (0, 0) -- (1, 0) -- (2, 0);
			} 
		} \times (y - x)^{6} 
		+ \cdots 
        \,\, . 
		\label{eq:sngen_expT_2}
        	\end{align}
	\end{subequations}
\end{widetext}

Here, the diagrammatic and symbolic notations, 
\begin{equation}
	\tikz[baseline={(0, -0.1)}]{ 
		\fill (0, 0) circle (2.5pt); 
		\node at (0, -0.3) {\scriptsize {$i$}};
	} 
	\, 
	\longleftrightarrow 
	\, 
	\frac{ (-)^{i} f_{i} (x) }{ i! } 
	\quad\text{and}\quad   
	\raisebox{-0.34ex}{\fbox{$v^{j}$}} \, 
	\, 
	\longleftrightarrow 
	\, 
	\frac{1}{j!} \frac{ 2^{j} }{ \qty[ v (x) ]^{j} } 
	\,\, , 
    \label{eq:def_diag}
\end{equation}
have been introduced to reduce clutter. 
The connected dots mean the product of the factors allocated to each dot, and it should be noted that $\raisebox{-0.34ex}{\fbox{$v^{j}$}} = \mathcal{O} (v^{-j})$. 
In Eq.~(\ref{eq:sngen_expT_2}), the terms are organised in powers of $y - x$. 
The coefficient of $(y - x)^{p}$-term for $p \geq 2$ consists of 
\begin{equation}
	\qty{ \raisebox{-0.34ex}{\fbox{$v^{q}$}} }_{q = 1, \, \dots, \, [p/2]} 
	\,\, , 
\end{equation}
each of which is multiplied by one or more diagrams ($[ z ]$ denotes the floor function). 
For each fixed $p \geq 2$ and $1 \leq q \leq [p/2]$, the number of diagrams, including a single dot, can be counted to give 
\begin{equation}
	N_{p, \, q} 
	= \frac{ \Gamma (p - q) }{ \Gamma (q) \Gamma (p - 2 q + 1) } 
	\,\, . 
	\label{eq:sngen_npq}
\end{equation}
When the factor multiplied by $\raisebox{-0.34ex}{\fbox{$v^{j}$}}$ in Eq.~(\ref{eq:sngen_expT_2a}) is expanded, for each fixed $j$, the lowest term is proportional to $(y - x)^{2 j}$ (recall that the index $i$ starts from $2$), and for $p \geq 2 j$, the number of the terms proportional to $(y - x)^{p}$ is given by $N_{p, \, q}$. 
One therefore arrives at Eq.~(\ref{eq:sngen_npq}). 
Table~\ref{tab:numterms} summarises the number of relevant terms, $N_{p, \, q}$. 

\begin{table*}
    \caption{
        The list of the numbers $N_{p, \, q}$ given in Eq.~(\ref{eq:sngen_npq}). 
        The reorganisation of the terms, Eq.~(\ref{eq:sngen_dijsum}) as illustrated in Eq.~(\ref{eq:sngen_expT_3}), is performed in the diagonal directions along each coloured entry. 
    } 
    \includegraphics[width = 0.85\linewidth]{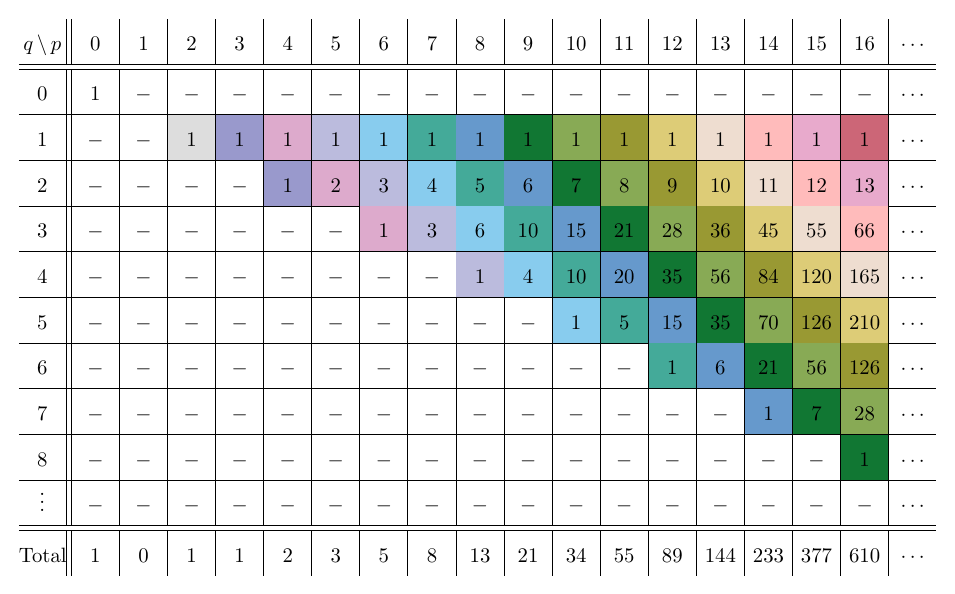}
    \label{tab:numterms}
\end{table*}

As an illustration, let us see the coefficient of $(y-x)^{5}$ in Eq.~(\ref{eq:sngen_expT_2}), and one finds one diagram and two diagrams associated with $\raisebox{-0.34ex}{\fbox{$v^{1}$}}$ and $\raisebox{-0.34ex}{\fbox{$v^{2}$}}$ respectively. 
These correspond to $N_{5, \, 1} = 1$ and $N_{5, \, 2} = 2$ (see also Table~\ref{tab:numterms}). 
For the $(y-x)^{6}$ term in the third line in Eq.~(\ref{eq:sngen_expT_2}), the numbers of the diagrams are $N_{6, \, 1} = 1$, $N_{6, \, 2} = 3$, and $N_{6, \, 3} = 1$. 
This way of arrangement of Eq.~(\ref{eq:sngen_expT_2}), that is, the result of organising the terms in each vertical direction in Table~\ref{tab:numterms}, can be summarised into 
\begin{equation}
	1 
	+ 
	\sum_{p=2}^{\infty} 
	\qty[ 
		\sum_{q = 1}^{[p/2]} 
		\, \raisebox{-0.34ex}{\fbox{$v^{q}$}} \times (\text{diagrams}) 
	] {} (y-x)^{p} 
	\,\, . 
	\label{eq:sngen_orgs1}
\end{equation}
In Eq.~(\ref{eq:sngen_orgs1}), the number of ``(diagrams)'' in the summation over $q$ is equal to $N_{p, \, q}$ for each fixed $p$ and $q$. 

When one adds all the nonzero numbers in the table in the vertical axis (namely over all the possible $q$ for each fixed $p$), it results in 
\begin{equation}
	\sum_{q=1}^{[p/2]} N_{p, \, q} = F_{p - 1} 
	\,\, , 
	\qquad 
	p \geq 1 
	\,\, , 
\end{equation}
where $F_{i \geq 0} = 0, \, 1, \, 1, \, 2, \, 3, \, \dots$ are the Fibonacci numbers, see the bottom line in Table~\ref{tab:numterms}. 

The way of organisation (\ref{eq:sngen_orgs1}) follows straightforwardly from Eq.~(\ref{eq:sngen_expT_2a}), 
but turns out not to be the most convenient manner to proceed further, provided that our aim is to obtain all the statistical moments \textit{in powers of} $v$. 
What should be done here is rather to reorganise Eq.~(\ref{eq:sngen_expT_2}) in powers of $v$. 
To do so, let us use the fact that terms proportional to $(y-x)^{p}$ give rise to the terms of order of $v^{p}$, or higher order, after the integration over $y$ is performed. 
This can be confirmed soon, but let us just accept it for a while. 

Then, it leads to the statement that, if a term is proportional to $(y -x)^{a} / v^{b}$, it starts from the term of order of $v^{a - b}$. 
This observation motivates us to reorganise Eq.~(\ref{eq:sngen_expT_2}) in the most convenient manner, resumming the terms along with the diagonal directions in Table~\ref{tab:numterms}. 
In the table, the same coloured entries that form Pascal's triangle are now organised to give rise to the same order of $v$. 
Reorganising the terms according to this rule reads 
\begin{widetext}
	\begin{align}
		&\quad \exp \qty[ \frac{2}{v (x)} \sum_{k = 2}^{\infty} \frac{ (-)^{k} f_{k} (x) }{ k! } (y-x)^{k} ] 
		\notag \\[2.0ex] 
		&= 
		\underbrace{ 
		1 
		\vphantom{
			\fbox{$v^{1}$} \, \times 
			\tikz[baseline={(0, -0.1)}]{
				\fill (0, 0) circle (2.5pt);
  				\node at (0, -0.3) {\scriptsize {$2$}};
			} 
		} 
		}_{\mathrel{\leadsto} \mathcal{O} (v^{0})} 
		+ 
		\underbrace{ 
		\fbox{$v^{1}$} \, \times 
		\tikz[baseline={(0, -0.1)}]{
			\fill (0, 0) circle (2.5pt);
  			\node at (0, -0.3) {\scriptsize {$2$}};
		} 
		\times (y - x)^{2} 
		}_{\mathrel{\leadsto} \mathcal{O} (v^{1})}
		+ 
		\underbrace{
		\fbox{$v^{1}$} \, \times 
		\tikz[baseline={(0, -0.1)}]{
			\fill (0, 0) circle (2.5pt);
  			\node at (0, -0.3) {\scriptsize {$3$}};
		} 
		\times (y - x)^{3} 
		+ 
		\fbox{$v^{2}$} \, \times 
		\tikz[baseline={(0, -0.1)}]{
			\fill (0, 0) circle (2.5pt);
  			\node at (0, -0.3) {\scriptsize {$2$}};
			\fill (1, 0) circle (2.5pt);
  			\node at (1, -0.3) {\scriptsize {$2$}};
			\draw[thick] (0, 0) -- (1, 0);
		}
		\times (y - x)^{4} 
		}_{\mathrel{\leadsto} \mathcal{O} (v^{2})}
		\notag \\[2.0ex] 
		&\quad~~\, 
		+ 
		\underbrace{ 
		\fbox{$v^{1}$} \, \times 
		\tikz[baseline={(0, -0.1)}]{
			\fill (0, 0) circle (2.5pt);
  			\node at (0, -0.3) {\scriptsize {$4$}};
		} 
		\times (y - x)^{4} 
		+ 
		\fbox{$v^{2}$} \, \times \qty( 
			\tikz[baseline={(0, -0.1)}]{
			\fill (0, 0) circle (2.5pt);
  			\node at (0, -0.3) {\scriptsize {$2$}};
			\fill (1, 0) circle (2.5pt);
  			\node at (1, -0.3) {\scriptsize {$3$}};
			\draw[thick] (0, 0) -- (1, 0);
		} 
		+ \tikz[baseline={(0, -0.1)}]{
			\fill (0, 0) circle (2.5pt);
  			\node at (0, -0.3) {\scriptsize {$3$}};
			\fill (1, 0) circle (2.5pt);
  			\node at (1, -0.3) {\scriptsize {$2$}};
			\draw[thick] (0, 0) -- (1, 0);
		} 
		) 
		\times (y - x)^{5} 
		+ \fbox{$v^{3}$} \, \times 
		\tikz[baseline={(0, -0.1)}]{
			\fill (0, 0) circle (2.5pt);
  			\node at (0, -0.3) {\scriptsize {$2$}};
			\fill (1, 0) circle (2.5pt);
  			\node at (1, -0.3) {\scriptsize {$2$}};
			\fill (2, 0) circle (2.5pt);
  			\node at (2, -0.3) {\scriptsize {$2$}};
			\draw[thick] (0, 0) -- (1, 0) -- (2, 0);
		} 
		\times (y - x)^{6} 
		}_{\mathrel{\leadsto} \mathcal{O} (v^{3})}
		+ \cdots 
	\label{eq:sngen_expT_3}
\end{align}
In Eq.~(\ref{eq:sngen_expT_3}), ``$\leadsto \mathcal{O} (v^{n})$'' indicates that each set of the terms gives rise to the $\mathcal{O} (v^{n})$ term after the integration over $y$ is performed. 
Equation~(\ref{eq:sngen_expT_3}) can be summarised into  
\begin{align}
	\exp \qty[ \frac{2}{v (x)} \sum_{i = 2}^{\infty} \frac{ (-)^{i} f_{i} (x) }{ i! } (y-x)^{i} ] 
    = \sum_{i = 0}^{\infty} \sum_{j = 0}^{i} 
	\fbox{$v^{j} (x)$} \, \times 
	D_{i, \, j} (x) \times (y - x)^{i + j} 
	\,\, . 
	\label{eq:sngen_dijsum}
\end{align}
\end{widetext}
Here, the factor $D_{i, j} (x)$ has been introduced, which collects all the relevant diagrams. 
For $1 \leq j \leq i$, each $D_{i, \, j} (x)$ consists of one or more dots, where the sum of all the numbers attached to the dots is equal to $i + j$. 
When either $i$ or $j$ is zero, one defines 
\begin{subequations}
	\label{eq:sngen_dsp}
	\begin{align}
		D_{i = 0, \, j = 0} (x) 
		&\equiv 1 
		\,\, , 
		\\ 
		D_{i = 0, \, j \geq 1} (x) 
		= D_{i \geq 1, \, j = 0} (x) 
		&\equiv 0 
		\,\, . 
	\end{align}
\end{subequations}
For instance, in the last line in Eq.~(\ref{eq:sngen_expT_3}), one observes 
\begin{subequations}
	\begin{align}
		D_{3, \, 0} 
		&= 0 
		\,\, , 
		\\ 
		D_{3, \, 1} 
		&= 
		\tikz[baseline={(0, -0.1)}]{
			\fill (0, 0) circle (2.5pt);
  			\node at (0, -0.3) {\scriptsize {$4$}};
		} 
		= \frac{ (-)^{4} f_{4} (x) }{4!} 
		= \frac{f_{4} (x)}{24} 
		\,\, , 
		\\ 
		D_{3, \, 2} 
		&= 
		\tikz[baseline={(0, -0.1)}]{
			\fill (0, 0) circle (2.5pt);
  			\node at (0, -0.3) {\scriptsize {$2$}};
			\fill (1, 0) circle (2.5pt);
  			\node at (1, -0.3) {\scriptsize {$3$}};
			\draw[thick] (0, 0) -- (1, 0);
		} 
		+ \tikz[baseline={(0, -0.1)}]{
			\fill (0, 0) circle (2.5pt);
  			\node at (0, -0.3) {\scriptsize {$3$}};
			\fill (1, 0) circle (2.5pt);
  			\node at (1, -0.3) {\scriptsize {$2$}};
			\draw[thick] (0, 0) -- (1, 0);
		} 
		= - \frac{ f_{2} (x) f_{3} (x) }{6} 
		\notag \\ 
		&= \frac{ (-)^{2} f_{2} (x) }{2!} \frac{ (-)^{3} f_{3} (x) }{3!} + \frac{ (-)^{3} f_{3} (x) }{3!} \frac{ (-)^{2} f_{2} (x) }{2!} 
		\,\, , 
		\\ 
		D_{3, \, 3} 
		&= 
		\tikz[baseline={(0, -0.1)}]{
			\fill (0, 0) circle (2.5pt);
  			\node at (0, -0.3) {\scriptsize {$2$}};
			\fill (1, 0) circle (2.5pt);
  			\node at (1, -0.3) {\scriptsize {$2$}};
			\fill (2, 0) circle (2.5pt);
  			\node at (2, -0.3) {\scriptsize {$2$}};
			\draw[thick] (0, 0) -- (1, 0) -- (2, 0);
		} 
		= \qty[ \frac{ (-)^{2} f_{2} (x) }{2!} ]^{3} 
		= \frac{ [ f_{2} (x) ]^{3} }{8} 
		\,\, . 
	\end{align}
\end{subequations}
In general, all the relevant diagrams can be collected to calculate the factor $D_{i, \, j} (x)$ by summing over combinations of the integers that satisfy certain conditions as follows: 
\begin{align}
    D_{i, \, j} (x) 
    &= \sum_{\qty{ p_{1}, \, \dots, \, p_{j}} \in P} 
    \tikz[baseline={(0, -0.1)}]{
			\fill (0, 0) circle (2.5pt);
  			\node at (0, -0.3) {\scriptsize {$p_{1}$}};
			\fill (1, 0) circle (2.5pt);
  			\node at (1, -0.3) {\scriptsize {$p_{2}$}};
			\fill (3, 0) circle (2.5pt);
  			\node at (3, -0.3) {\scriptsize {$p_{j}$}};
			\draw[thick] (0, 0) -- (1, 0) -- (1.5, 0); 
            \draw[thick] (2.5, 0) -- (3, 0); 
            \node at (2, 0) {\scriptsize {$\cdots$}};
		} 
    \rule{0pt}{5ex} 
    \notag \\ 
    &= \sum_{\qty{ p_{1}, \, \dots, \, p_{j}} \in P} 
    \qty( \prod_{i = 1}^{j} 
    \tikz[baseline={(0, -0.1)}]{
			\fill (0, 0) circle (2.5pt);
  			\node at (0, -0.3) {\scriptsize {$p_{i}$}};
		} 
    ) 
    \,\, , 
    \label{eq:fd_gen}
\end{align}
where, for $\vb*{p} \equiv (p_{1}, \, p_{2}, \, \dots, \, p_{j})^{\mathsf{T}}$, the set $P$ is defined by 
\begin{equation}
    P 
    \equiv \qty{ 
        ~ \vb*{p} 
        ~~\middle|~~ 
        \sum_{k = 1}^{j} p_{k} = i + j 
        ~~\text{and}~~ 
        p_{k} \geq 2 ~ 
    } 
    \,\, . 
\end{equation}
The fact that 
\begin{equation}
    D_{i, \, 1} (x) 
    = \tikz[baseline={(0, -0.1)}]{
			\fill (0, 0) circle (2.5pt);
  			\node at (0, -0.3) {\scriptsize {$i + 1$}};
		} 
    \label{eq:sdot}
\end{equation}
implies that any $D_{i, \, j} (x)$ can be expressed in terms of $D_{1 \leq k \leq i - j + 1, \, 1} (x)$, where the second subscript is unity. 
This is because connected dots are just a product of several dots. 
It turns out that the Bell polynomial used in Eq.~(\ref{eq:sngen_fnT}) enables to reduce Eq.~(\ref{eq:fd_gen}) to a closed form:
\begin{align}
    D_{i, \, j} (x) 
    &= \frac{j!}{i!} 
    B_{i, \, j} \left[ 
        1! D_{1, \, 1} (x), \, 
        2! D_{2, \, 1} (x), \, 
    \right. 
    \label{eq:fnD_Bell}
    \\ 
    &\quad \qquad\quad~~ \left. 
        \dots, \, 
        (i - j + 1)! D_{i - j + 1, \, 1} (x) 
    \right] 
    \,\, . 
    \notag 
\end{align}

For each $i$ in the above equation, the $i$-th order terms in $v$ are exhausted. 
This is because, as will be seen in later, the integration of $(y - x)^{i + j}$ over $y$ yields the contribution of order $v^{i + j}$, while the squared $v^{j} (x)$ contributes as the term of order $v^{- j}$, see Eq.~(\ref{eq:def_diag}).  
One therefore obtains the terms of order $v^{(i + j) - j} = v^{i}$ in total for each fixed $i$. 

Substitution of Eqs.~(\ref{eq:sngen_dijsum}), (\ref{eq:sngen_evT}), (\ref{eq:sngen_vT}), and (\ref{eq:sngen_nyT}) into Eq.~(\ref{eq:sngen_master}) leads to the following expression, which is now ready for being performed the integration over $y$, 
\begin{widetext}
	\begin{align}
		\frac{1}{n+1} \dv{\expval*{ \mathcal{N}^{n+1} } (x)}{x}  
		&= 
		\frac{1}{x^{d-1}} 
		\int_{x}^{r_{+}} \dd y \, y^{d-1} 
		\qty[ 
			\frac{2}{v (x)} \sum_{k_{1} = 0}^{\infty} \frac{ (-)^{k_{1}} f_{k_{1}} (x) }{k_{1}!} (y - x)^{k_{1}} 
		] 
		\exp \qty[ - \frac{2 f_{1} (x)}{v (x)} (y - x) ] 
		\notag \\ 
		&\quad \times 
		\sum_{k_{2} = 0}^{\infty} \sum_{k_{3} = 0}^{k_{2}} 
		\fbox{$v^{k_{3}} (x)$} \, \times 
		D_{k_{2}, \, k_{3}} (x) \times (y - x)^{k_{2} + k_{3}} 
		\expval*{ \mathcal{N}^{n} } (x) 
		\sum_{k_{4} = 0}^{\infty} \frac{g_{n, \, k_{4}} (x)}{k_{4}!} (y - x)^{k_{4}} 
		\notag \\ 
		&= 
		\sum_{k_{1} = 0}^{\infty} 
		\sum_{k_{2} = 0}^{\infty} 
		\sum_{k_{3} = 0}^{k_{2}} 
		\sum_{k_{4} = 0}^{\infty} 
		\qty[ 
			\tikz[baseline={(0, -0.1)}]{
				\fill (0, 0) circle (2.5pt);
  				\node at (0, -0.3) {\scriptsize {$k_{1}$}};
			} \times 
			D_{k_{2}, \, k_{3}} (x) 
		] 
		\frac{ \fbox{$v^{k_{3}} (x)$} }{k_{4}!} 
		\times \expval*{ \mathcal{N}^{n} } (x) g_{n, \, k_{4}} (x) 
		\notag \\ 
		&\quad \times 
		\frac{1}{x^{d-1}} 
		\frac{2}{v (x)} 
		\int_{x}^{r_{+}} \dd y \, y^{d-1} 
		\exp \qty[ - \frac{2 f_{1} (x)}{v (x)} (y - x) ] {} 
		(y-x)^{ k_{1} + k_{2} + k_{3} + k_{4} } 
		\notag \\ 
		&= 
		\sum_{k_{1} = 0}^{\infty} 
		\sum_{k_{2} = 0}^{\infty} 
		\sum_{k_{3} = 0}^{k_{2}} 
		\sum_{k_{4} = 0}^{\infty} 
		\sum_{k_{5} = 0}^{\infty} 
		\qty[ 
			\tikz[baseline={(0, -0.1)}]{
				\fill (0, 0) circle (2.5pt);
  				\node at (0, -0.3) {\scriptsize {$k_{1}$}};
			} \times 
			D_{k_{2}, \, k_{3}} (x) 
		] 
		\frac{ \fbox{$v^{k_{3}} (x)$} }{k_{4}!} \times 
		\dv[k_{4}]{\expval*{ \mathcal{N}^{n} }^{(k_{5})} (x)}{x} 
		\notag \\ 
		&\quad \times 
		\frac{1}{x^{d-1}} 
		\frac{2}{v (x)} 
		\int_{x}^{r_{+}} \dd y \, y^{d-1} 
		\exp \qty[ - \frac{2 f_{1} (x)}{v (x)} (y - x) ] {} 
		(y-x)^{ k_{1} + k_{2} + k_{3} + k_{4} } 
		\,\, . 
    		\label{eq:sngen_T4}
	\end{align}
\end{widetext}
Going from the second to the third line, the following expansion from Eqs.~(\ref{eq:sngen_gnT}) and (\ref{eq:d_stmm_oex}) has been used:
\begin{equation}
	\expval*{ \mathcal{N}^{n} } (x) g_{n, \, i} (x) 
	= \dv[i]{\expval*{ \mathcal{N}^{n} } (x)}{x} 
	= \sum_{j=1}^{\infty} \dv[i]{\expval*{ \mathcal{N}^{n} }^{(j)} (x)}{x} 
	\,\, . 
\end{equation}

The remaining $y$-integral is evaluated as follows. 
The binomial expansion is applied to the factor $(y - x)^{k}$, and it is then integrated over $y$, resulting in 
\begin{align}
	G_{i} (x) 
	&\equiv \frac{1}{x^{d-1}} \frac{2}{v (x)} 
	\int_{x}^{r_{+}} \dd y \, y^{d-1} 
	\notag \\ 
	&\quad \times 
	\exp \qty[ - \frac{2 f_{1} (x)}{v (x)} (y - x) ] {} 
	(y - x)^{i} 
	\notag \\ 
	&= \frac{1}{x^{d-1}} \frac{2}{v (x)} 
	\exp \qty[ \frac{2 f_{1} (x)}{v (x)} x ] 
	\sum_{j = 0}^{i} C (i, \, j) (- x)^{i - j} 
	\notag \\ 
	&\quad \times 
	\qty[ \frac{v (x)}{2 f_{1} (x)} ]^{d + j} 
	\eval{ 
		\Gamma \qty[ d + j, \, \frac{2 f_{1} (x)}{v (x)} y ] 
	}_{y = r_{+}}^{y = x} 
	\,\, , 
	\label{eq:sngen_gx}
\end{align}
where $C (i, \, j)$ is the binomial coefficient. 
What should be done next is to expand $G_{i} (x)$ in powers of $v$. 
As was done in Eq.~(\ref{eq:sncl_int}) using Eq.~(\ref{eq:incg_ep1}), the asymptotic expansion of the incomplete gamma function~\cite{NIST:DLMF}, 
\begin{equation}
	\Gamma (s, \, z) 
	= z^{s - 1} e^{- z} \sum_{k = 0}^{\infty} \frac{\Gamma (s)}{\Gamma (s - k)} \frac{1}{z^{k}} 
	\,\, , 
	\label{eq:sngen_asicg}
\end{equation}
is used for this purpose. 
Substitution of Eq.~(\ref{eq:sngen_asicg}) into Eq.~(\ref{eq:sngen_gx}), one obtains 
\begin{align}
    G_{i} (x) 
    &= \qty( \frac{y}{x} )^{d-1} 
	\exp \qty[ \frac{2 f_{1} (x)}{v (x)} (x-y) ] 
    \notag \\ 
    &\quad \times 
	\sum_{j = 0}^{\infty} 
	\frac{(-x)^i}{y^{j}}
	\qty[ \frac{v (x)}{2} ]^{j} 
	\frac{1}{[ f_{1} (x) ]^{j+1}} 
    \sum_{k = 0}^{i} 
	\frac{(-)^{k}}{k!} 
	\qty( \frac{y}{x} )^{k} 
	\notag \\ 
	&\qquad \times 
	\eval{ 
	\frac{ \Gamma (i+1) }{ \Gamma (i - k + 1) } 
	\frac{ \Gamma (d + k) }{ \Gamma (d + k - j) } 
	}_{y = r_{+}}^{y = x} 
    \,\, , 
    \label{eq:sngen_gx2}
\end{align}
where the summation over $0 \leq j \leq i$ in Eq.~(\ref{eq:sngen_gx}) and $k \geq 0$ in Eq.~(\ref{eq:sngen_asicg}) respectively correspond to the summations over $0 \leq k \leq i$ and $j \geq 0$ in Eq.~(\ref{eq:sngen_gx2}). 
That is, the summation indices $j$ and $k$ have been interchanged. 

The second summation over $k$ in Eq.~(\ref{eq:sngen_gx2}) can be performed and expressed in terms of a Gauss' hypergeometric function, defined by 
\begin{equation}
	{}_{2} F_{1} \qty( \begin{matrix} a, \, b \\ c \end{matrix} ~\middle|~ z ) 
	\equiv 
	\sum_{k=0}^{\infty} 
    \frac{ 
        (a)_{k} (b)_{k}
    }{
        (c)_{k} 
    } 
    \frac{z^{k}}{k!} 
    \,\, , 
    \label{eq:def_2F1}
\end{equation}
where $(p)_{k} \equiv \Gamma (p+ k) / \Gamma (p)$ is the Pochhammer symbol. 
For $(a, \, b, \, c) = (d, \, -i, \, d - j)$ and $z = y/x$, the infinite sum in Eq.~(\ref{eq:def_2F1}) is terminated due to the non-positive integer $- i \leq 0$. 
The definition of ${}_{2} F_{1}$ (\ref{eq:def_2F1}) for those arguments can be cast into the form  
\begin{align}
    {}_{2} F_{1} \qty( \begin{matrix} d, \, -i \\ d-j \end{matrix} ~\middle|~ z ) 
    &= 
    \frac{ \Gamma (d-j) }{ \Gamma (d) } 
    \sum_{k=0}^{\infty} 
    \frac{z^{k}}{k!} 
    \notag \\ 
    &\quad \times 
	\frac{ \Gamma (-i + k) }{ \Gamma (-i) } 
	\frac{ \Gamma (d + k) }{ \Gamma (d-j+k) } 
    \notag \\ 
    &= 
    \frac{ \Gamma (d-j) }{ \Gamma (d) } 
    \sum_{k=0}^{\infty} 
    \frac{ (-)^{k} }{k!} z^{k} 
    \notag \\ 
    &\quad \times 
	\frac{ \Gamma (i + 1) }{ \Gamma [ i - (k-1) ] } 
	\frac{\Gamma (d + k)}{\Gamma (d - j + k)} 
    \,\, . 
    \label{eq:arg_2F1}
\end{align}
The following reexpression for $k \geq 0$ has been used in the second equality in Eq.~(\ref{eq:arg_2F1}), 
\begin{equation}
    \frac{ \Gamma (-i + k) }{ \Gamma (-i) } 
	= (-)^{k} \frac{ \Gamma (i + 1) }{ \Gamma [ i - (k-1) ] } 
    \,\, . 
\end{equation}
That $j$ is an order of $v$ in Eq.~(\ref{eq:sngen_gx2}), and that $d \gg 1$ is of interest here, ensure that $d - j + k > 0$ in most practical situations, and there is no singular behaviour in the last gamma function in Eq.~(\ref{eq:arg_2F1}). 
On the other hand, $\Gamma [ i - (k - 1) ]$ diverges for $k \geq i + 1$, by which the summation over $k$ in Eq.~(\ref{eq:arg_2F1}) is indeed restricted to $0 \leq k \leq i$. 
Together with Eqs.~(\ref{eq:sngen_gx2}) and (\ref{eq:arg_2F1}), this leads to 
\begin{align}
	G_{i} (x) 
	&= \qty(\frac{y}{x})^{d-1} 
	\exp \qty[ \frac{2 f_{1} (x)}{ v (x) } (x - y) ] 
	\notag \\ 
	&\quad \times 
	\sum_{j = 0}^{\infty} 
	\frac{1}{y^{j}} \qty[ \frac{v (x)}{2} ]^{j} 
	\frac{1}{\qty[ f_{1} (x) ]^{j+1}}
	\notag \\ 
	&\qquad 
	\times 
	\eval{ 
		\frac{(-x)^{i} \Gamma (d)}{\Gamma (d - j)} 
		{}_{2} F_{1} \qty( 
			\begin{matrix} d, \, -i \\ d - j \end{matrix} ~\middle| ~ \frac{y}{x} 
		) 
	}^{y = x}_{y = r_{+}} 
	\,\, . 
\end{align}

The contribution from $y = r_{+}$ is again neglected since the small-noise regime is of interest, then the ${}_{2} F_{1}$ is to be evaluated at $z = 1$. 
Here, the Gauss' summation formula~\cite{gauss1813disquisitiones}, 
\begin{equation}
	{}_{2} F_{1} \qty( \begin{matrix} a, \, b \\ c \end{matrix} ~\middle|~ 1 ) 
	= \frac{ \Gamma (c) \Gamma (c - a - b) }{ \Gamma (c - a) \Gamma (c - b) } 
	\,\, , 
	\label{eq:sngen_gsumf}
\end{equation}
for $\Re (c - a - b) > 0$, may be used. 
Precisely speaking, since now $(d - j) - d - (- i) = i - j \leq 0$, the direct use of the summation formula may not be validated, however, due to the cancellation between the gamma functions in Eq.~(\ref{eq:sngen_gsumf}), the formal use of it turns out to be justified. 
This results in  
\begin{equation}
	\frac{\Gamma (d)}{\Gamma (d - i)} 
	\eval{ {}_{2} F_{1} \qty( 
		\begin{matrix} d, \, - i \\ d - j \end{matrix} ~\middle| ~ \frac{y}{x} 
	) }_{y = x} 
	= (-)^{i} \frac{\Gamma (j + 1)}{ \Gamma ( j - i + 1 )} 
	\,\, . 
\end{equation}
It should be noted that the denominator diverges to give zero for $0 \leq j \leq i - 1$. 
The resultant expansion of the function $G_{i} (x)$ in powers of $v$ then reads 
\begin{align}
	G_{i} (x) 
	&\simeq \sum_{j = i}^{\infty} \frac{1}{x^{j - i}} 
	\qty[ \frac{v (x)}{2} ]^{j} 
	\notag \\ 
	&\quad \times 
	\frac{ \Gamma (d) }{ \Gamma (d - j + i) } 
	\frac{ \Gamma (j + 1) }{ \Gamma (j - i + 1) } 
	\frac{1}{\qty[ f_{1} (x) ]^{j + 1} } 
	\,\, . 
	\label{eq:sngen_gT}
\end{align}
It is now clear that, as announced previously, the integral over $y$ involving $(y - x)^{i}$ indeed gives the terms of order of $v^{i}$ and higher. 
Finally, substitution of Eq.~(\ref{eq:sngen_gT}) into Eq.~(\ref{eq:sngen_T4}) leads to the all-order recursive formula for our perturbative expansion for the statistical moments of $\mathcal{N}$, given by 
\begin{widetext}
	\begin{align}
		\frac{1}{n+1} \dv{\expval*{ \mathcal{N}^{n+1} } (x)}{x}  
		&= 
		\sum_{k_{1} = 0}^{\infty} 
		\sum_{k_{2} = 0}^{\infty} 
		\sum_{k_{3} = 0}^{k_{2}} 
		\sum_{k_{4} = 0}^{\infty} 
		\sum_{k_{5} = 0}^{\infty} 
		\sum_{k_{6} = k_{1} + k_{2} + k_{3} + k_{4}}^{\infty} 
		\frac{
			\tikz[baseline={(0, -0.1)}]{
				\fill (0, 0) circle (2.5pt);
  				\node at (0, -0.3) {\scriptsize $k_{1}$};
			} \times 
			D_{k_{2}, \, k_{3}} (x) 
		}{k_{3}! k_{4}!} 
		\qty[ \frac{v (x)}{2} ]^{ - k_{3} + k_{6} } 
		\dv[k_{4}]{\expval*{ \mathcal{N}^{n} }^{(k_{5})} (x)}{x} 
		\notag \\[2.0ex] 
		&\quad \times 
		\frac{1}{x^{k_{6} - (k_{1} + k_{2} + k_{3} + k_{4})}} 
		\frac{ \Gamma (d) }{ \Gamma (d - k_{6} + k_{1} + k_{2} + k_{3} + k_{4}) } 
		\frac{ \Gamma (k_{6} + 1) }{ \Gamma \qty[ k_{6} - ( k_{1} + k_{2} + k_{3} + k_{4} ) + 1 ] } 
		\frac{1}{ \qty[ f_{1} (x) ]^{k_{6} + 1} } 
		\,\, . 
		\label{eq:sngen_main}
	\end{align}
\end{widetext}

This is the main result of the present section. 
In Eq.~(\ref{eq:sngen_main}), the relevant factors in counting the order of $v$ are $[ v (x) / 2 ]^{- k_{3} + k_{6}}$ and $\dd^{k_{4}} \expval*{ \mathcal{N}^{n} }^{(k_{5})} (x) / \dd x^{k_{4}}$. 
Those factors give rise to the term of order of $v^{- k_{3} + k_{5} + k_{6}}$. 
In other words, the terms of order of $v^{m}$ in $\dd \expval*{ \mathcal{N}^{n+1} } (x) / \dd x$, or equivalently $\expval*{ \mathcal{N}^{n+1} } (x)$, arise from all the combinations of the six integers that satisfy $- k_{3} + k_{5} + k_{6} = m$. 

Let us set $n = 0$ in Eq.~(\ref{eq:sngen_main}) and confirm that it correctly gives the leading-order result of the mean number of $e$-folds, Eq.~(\ref{eq:sncl_clef}), which consists of one term of order of $v^{0}$ only. 
To extract the term, the relevant combination of the dummy indices in Eq.~(\ref{eq:sngen_main}) is the one that satisfies $- k_{3} + k_{5} + k_{6} = 0$, which implies that $k_{1} = \cdots = k_{6} = 0$. 
This gives rise to $\dd \expval*{ \mathcal{N} } (x) / \dd x = 1 / f_{1} (x) = v (x) / v' (x)$, and the well-known formula for the classical number of $e$-folds is therefore revalidated. 

In the same manner, higher-order terms of $\expval*{ \mathcal{N} }$ can also be systematically obtained by exhausting all the relevant combinations of integers. 
What follows from those terms of the mean number of $e$-folds is the second statistical moment, $\expval*{ \mathcal{N}^{2} }$, and so forth. 
The derived result (\ref{eq:sngen_main}) will fully be used to derive the perturbative expansion of the relevant statistical moments in the next section. 

The mean number of $e$-folds at an arbitrary order in $v$ can be calculated from Eq.~(\ref{eq:sngen_main}) with $n = 0$, as will be later seen in Eq.~(\ref{eq:rvd_mean_master}), or equivalently in Eq.~(\ref{eq:rvd_mean_k_c1}). 
Higher-order calculation of the mean does not require the lower-order terms. 
On the other hand, lower-order statistical moments are needed in calculation of the higher-order statistical moments, as can be seen in Eq.~(\ref{eq:sngen_main}). 
To obtain $\expval*{ \mathcal{N}^{2} }^{(1)}$, one needs $\expval*{ \mathcal{N} }^{(0)}$ and $\expval*{ \mathcal{N} }^{(1)}$, and to calculate $\expval*{ \mathcal{N}^{2} }^{(2)}$, one needs $\expval*{ \mathcal{N} }^{(0)}$, $\expval*{ \mathcal{N} }^{(1)}$, and $\expval*{ \mathcal{N} }^{(2)}$, etc. 

\section{Emergence of number of fields \texorpdfstring{\protect\\}{} in statistical moments}
\label{sec:statistical_moments}

Having derived the general formula that allows us to calculate $\expval*{ \mathcal{N}^{n} }$ for arbitrary $n$ and at arbitrary order, the perturbative terms of the first and second moments of $\mathcal{N}$, up to several orders are derived for demonstration in this section. 
The irrelevance of the number of fields $d$ can be observed for the classical number of $e$-folds, 
\textit{e.g.}~in Eq.~(\ref{eq:sncl_clef}), 
which is at leading order in $v$. 
Once the stochastic effect is accounted for,  by going to the next order in $v$, however, the trajectories illustrated in the field space loiter in general, deviating from the classical straight line, 
and the extension of the mean number of $e$-folds 
explicitly depends on the number of fields $d$. 
This is the very phenomenon that the present article is interested in. 
The dependence on $d$ at each order of the perturbative expansion of the statistical moments of $\mathcal{N}$ in the small-noise regime will be explicitly identified. 
The results presented in this section not only correctly revalidate the results in the literature, but also go to higher orders. 

\subsection{Mean}
\label{subsec:mean}

Let us first derive the perturbative series of the mean number of $e$-folds, the first statistical moment of $\mathcal{N}$, in which the dependence on the number of fields $d$ is observed from the first-order correction term in $v$. 
What will be done here is to derive each term in Eq.~(\ref{eq:d_stmm_oex}) with $n = 1$, namely each $\expval*{ \mathcal{N} }^{(k)} (r) = \mathcal{O} (v^k)$ in the expansion, 
\begin{equation}
	\expval*{ \mathcal{N} } (r) 
	= \sum_{k = 0}^{\infty} \expval*{ \mathcal{N} }^{(k)} (r) 
	\,\, . 
	\label{eq:rvd_mean_sum}
\end{equation}

In the calculation of $\expval*{ \mathcal{N} }$, one sets $n = 0$ in Eq.~(\ref{eq:sngen_main}), so that only $k_{4} = 0$ (and $k_{5} = 0$ consequently) term survives. 
This leads, by relabelling the dummy indices, to 
\begin{align}
    \dv{\expval*{ \mathcal{N} } (x)}{x}  
	&\simeq 
		\sum_{k_{1} = 0}^{\infty} 
		\sum_{k_{2} = 0}^{\infty} 
		\sum_{k_{3} = 0}^{k_{2}} 
		\sum_{k_{4} = k_{1} + k_{2} + k_{3}}^{\infty} 
		\frac{
			\tikz[baseline={(0, -0.1)}]{
				\fill (0, 0) circle (2.5pt);
  				\node at (0, -0.3) {\scriptsize $k_{1}$};
			} \times 
			D_{k_{2}, \, k_{3}} (x) 
		}{k_{3}!} 
	\notag \\[0.5ex]
    &\quad \times 
    \qty[ \frac{v (x)}{2} ]^{ - k_{3} + k_{4} } 
		\frac{1}{x^{k_{4} - (k_{1} + k_{2} + k_{3})}} 
		\frac{1}{ \qty[ f_{1} (x) ]^{k_{4} + 1} } 
    \notag \\[0.5ex]
    &\quad \times 
		\frac{ \Gamma (d) }{ \Gamma (d - k_{4} + k_{1} + k_{2} + k_{3}) } 
    \notag \\[0.5ex]
    &\quad \times 
		\frac{ \Gamma (k_{4} + 1) }{ \Gamma \qty[ k_{4} - ( k_{1} + k_{2} + k_{3} ) + 1 ] } 
		\,\, . 
    \label{eq:rvd_mean_master}
\end{align}
For each fixed $-k_{3} + k_{4} = k$, the number of the relevant combinations of $(k_{1}, \, k_{2}, \, k_{3}, \, k_{4})$ is given by $(k+1) (k+2) (k+3) / 6$. 
However, $D_{i, \, j} = 0$ for some of the combinations, \textit{i.e.}, $D_{i \geq 1, \, j = 0}$ or $D_{i = 0, \, j \geq 1}$, see Eq.~(\ref{eq:sngen_dsp}). 
The net number of the nonzero combinations is then counted to be 
\begin{equation}
	N_{1} (k) 
	= \frac{ (k + 1) (k^{2} + 2 k + 6) }{ 6 } 
	\,\, . 
\end{equation}
Given that $G_{k} (x)$ starts at $\mathcal{O} (v^{k})$ term, all the relevant terms for each order in $v$ can systematically be exhausted. 

Below, the perturbative terms in $\expval*{ \mathcal{N} }$ are derived up to third order using Eq.~(\ref{eq:rvd_mean_master}), to find their number-of-field-dependence. 
The general-order formula will then be followed. 

\subsubsection{Zeroth order}

As was already confirmed near the end of Section~\ref{subsec:deriv}, at zeroth order, only the combination of the dummy indices $k_{1} = k_{2} = k_{3} = k_{4} = 0$ is relevant, as can also be seen from $N_{1} (0) = 1$. 
It then follows from Eq.~(\ref{eq:rvd_mean_master}) that $\dd \expval*{ \mathcal{N} }^{(0)} (x) / \dd x \simeq 1 / f_{1} (x)$. 
This appropriately reproduces the classical number of $e$-folds, given by Eq.~(\ref{eq:sncl_clef}). 

\subsubsection{First order}

At first order in $v$, the exponent of $v (x)$ in Eq.~(\ref{eq:rvd_mean_master}) must be $- k_{3} + k_{4} = 1$. 
There exist $N_{1} (1) = 3$ terms in Eq.~(\ref{eq:rvd_mean_master}), and the combinations of integers are exhausted by 
\begin{equation}
	(k_{1}, \, k_{2}, \, k_{3}, \, k_{4}) 
	= 
	\begin{cases}
		(0, \, 0, \, 0, \, 1) \,\, , \\[1.0ex] 
		(0, \, 1, \, 1, \, 2) \,\, , \\[1.0ex]
		(1, \, 0, \, 0, \, 1) \,\, . 
	\end{cases}
\end{equation}\\
The corresponding diagrammatic factors are $D_{0, \, 0} (x) = 1$ and $D_{1, \, 1} (x) = (-)^{2} f_{2} (x) / 2!$ only. 
The result then reads 
\begin{align}
	\dv{\expval*{ \mathcal{N} }^{(1)} (x)}{x} 
	&= v (x) \qty[ \frac{v (x)}{v' (x)} ]^{2} \left[ 
            \frac{d-1}{2 x} 
	\right. 
        \notag \\[0.5ex]  
        &\quad~~ \left. {} 
        + \frac{1}{2} \frac{v' (x)}{v (x)} - \frac{1}{2} \frac{v (x)}{v' (x)} \frac{v'' (x)}{v (x)} 
        \right] 
        \,\, . 
        \label{eq:mean_o1}
\end{align}
It is clear that there exists a $d$-dependent contribution linear in $d$, which is absent in the leading-order term. 
When $d = 1$, the term vanishes and the single-field result in Ref.~\cite{Vennin:2015hra} is correctly reproduced. 
For the monomial potential, $v (r) = v_{0} r^{p}$, all the three terms in the second bracket in Eq.~(\ref{eq:mean_o1}) are proportional to $1/x$. 

This implies that, even when the stochastic noise is small, the mean number of $e$-folds can significantly deviate from the classical prediction in the presence of many fields. 
While the perturbative calculations may break down if Eq.~(\ref{eq:mean_o1}) dominates over the leading term, this nonetheless indicates that the dynamics of the inflatons can be altered when the number of fields $d$ is extremely large. 

\subsubsection{Second order}

In deriving the second-order perturbative term in $\expval*{\mathcal{N}}$, there are $N_{1} (2) = 7$ relevant terms that must be included in Eq.~(\ref{eq:rvd_mean_master}). 
The corresponding combinations of the integers are summarised in Appendix. 
Summing up all those contributions, the result is given by 
\begin{align}
	\dv{ \expval*{ \mathcal{N} }^{(2)} (x) }{x} 
	&= 
	\qty[ v (x) ]^{2} \qty[ \frac{v (x)}{v' (x)} ]^{3} 
	\left\{ 
		\frac{(d-1)(d-2)}{4 x^{2}} 
	\right. 
	\notag \\[0.5ex] 
	&\quad 
		+ \frac{d-1}{x} \qty[
                \frac{v' (x)}{v (x)} - \frac{3}{4} \frac{v (x)}{v' (x)} \frac{v'' (x)}{v (x)} 
            ] 
	\notag \\[0.5ex] 
	&\quad 
        + \frac{1}{2} \qty[ \frac{v' (x)}{v (x)} ]^{2}  
		- \frac{v'' (x)}{v (x)} 
    \notag \\[0.5ex] 
    &\quad 
		+ \frac{3}{4} \qty[ \frac{v (x)}{v' (x)} ]^{2} \qty[ \frac{v'' (x)}{v (x)} ]^{2} 
    \notag \\[0.5ex] 
    &\quad 
    \left. 
		- \, \frac{1}{4} \frac{ v (x) }{v' (x)} \frac{v''' (x)}{v (x)} 
    \right\} 
	\,\, . 
	\label{eq:d_mean_o2_res}
\end{align}
Similarly to the first-order result, Eq.~(\ref{eq:mean_o1}), the $d$-dependence can again be observed. 
In addition to the linear term, at second order the quadratic dependence on $d$ is present. 
The linear term in $d$ vanishes at $d = 1$, whereas the quadratic term in $d$ vanishes at both $d = 1$ and $d = 2$. 

\subsubsection{Third order}

The third-order calculation includes $N_{1} (3) = 14$ terms (again see Appendix), and a careful calculation leads to Eq.~(\ref{eq:mean_o3}) below. 
As the pattern can be read off from Eqs.~(\ref{eq:mean_o1}) and (\ref{eq:d_mean_o2_res}), the third-order result (\ref{eq:mean_o3}) involves the term cubic in the number of fields, which vanish at $d = 1$, $2$, and also $3$. 

\begin{widetext}
\begin{align}
    \dv{\expval*{ \mathcal{N} }^{(3)} (x) }{x} 
    &= 
    \qty[ v (x) ]^{3} \qty[ \frac{v (x)}{v' (x)} ]^{4} 
    \Biggl( 
        \frac{(d-1)(d-2)(d-3)}{8 x^{3}} 
        + \frac{3 (d-1) (d-2)}{8 x^{2}} \qty[ 
            3 \frac{v' (x)}{v (x)} - 2 \frac{ v (x) }{v' (x)} \frac{v'' (x)}{v (x)} 
        ] 
    \Biggr. 
    \notag \\[2.0ex] 
    &\hspace{-1.5cm} \Biggl. 
    +\, \frac{(d-1)}{8 x} \qty{ 
        18 \qty[ \frac{v' (x)}{v (x)} ]^{2} - 27 \frac{v'' (x)}{v (x)} + 15 \qty[ \frac{v (x)}{v' (x)} ]^{2} \qty[ \frac{v'' (x)}{v (x)} ]^{2} - 4 \frac{v (x)}{v' (x)} \frac{v''' (x)}{v (x)} 
    } 
    + \frac{3}{4} \qty[ \frac{v' (x)}{v (x)} ]^{3} 
    - \frac{9}{4} \frac{v' (x)}{v (x)} \frac{v'' (x)}{v (x)} 
    \Biggr. 
    \notag \\[2.0ex] 
    &\hspace{-1.5cm} \Biggl. 
    +\, \frac{27}{8} \frac{v (x)}{v' (x)} \qty[ \frac{v'' (x)}{v (x)} ]^{2} 
    - \frac{15}{8} \qty[ \frac{v (x)}{v' (x)} ]^{3} \qty[ \frac{v'' (x)}{v (x)} ]^{3} 
    - \frac{9}{8} \frac{v''' (x)}{v (x)} 
	+ \frac{5}{4} \qty[ \frac{v (x)}{v' (x)} ]^{2} \frac{ v'' (x)}{v (x)} \frac{v''' (x)}{v (x)} 
    - \frac{1}{8} \frac{v (x)}{v' (x)} \frac{v'''' (x)}{v (x)} 
    \Biggr) 
    \,\, . 
    \label{eq:mean_o3}
\end{align}
\end{widetext}

\subsubsection{General order}

In principle, the procedure employed until here applies to calculate $\mathcal{O} (v^{k})$ term in the mean number of $e$-folds for arbitrary $k \geq 0$, collecting the terms corresponding to $(k_{1}, \, k_{2}, \, k_{3}, \, k_{4})$ that give rise to the desired order of $v$. 
However, writing down the explicit forms of all the factors becomes more and more complicated when $k$ gets larger. 
For instance, at fourth order one has $N_{1} (4) = 25$, which is already unwieldy. 
Therefore, instead of going to higher-order calculations, let us observe the general structure of the $k$-th order term, to see how the $d$-dependent terms arise from the general formula (\ref{eq:rvd_mean_master}). 

The order in $v$ of each term in Eq.~(\ref{eq:rvd_mean_master}) is determined by the combination $- k_{3} + k_{4}$, as was mentioned previously. 
In other words, a term proportional to $v^{k}$ only comes from the terms with $k_{4} = k + k_{3}$. 
Then, the gamma functions in the denominators restrict the possible range of $k_{1}$ and $k_{2}$ to $\mathrm{max} (0, \, k - (d - 1) ) \leq k_{1} + k_{2} \leq k$, where the left-hand side of the inequality can be set to $0$ if a very large $d$ is in mind. 
Consequently, Eq.~(\ref{eq:rvd_mean_master}) reduces to 
\begin{align}
		\dv{\expval*{ \mathcal{N} }^{(k)} (x)}{x}  
		&\simeq 
		\sum_{k_{1} = 0}^{k} 
		\sum_{k_{2} = \max (0, \, k - k_{1} - (d - 1))}^{k - k_{1}} 
		\sum_{k_{3} = 0}^{k_{2}} 
		\qty[ \frac{v (x)}{2} ]^{ k } 
		\notag \\[0.5ex]
		&\quad \times 
        \frac{
			\tikz[baseline={(0, -0.1)}]{
				\fill (0, 0) circle (2.5pt);
  				\node at (0, -0.3) {\scriptsize $k_{1}$};
			} \times 
			D_{k_{2}, \, k_{3}} (x) 
		}{x^{k - (k_{1} + k_{2})} k_{3}!} 
		\frac{ \Gamma (d) }{ \Gamma (d - k + k_{1} + k_{2}) } 
        \notag \\[0.5ex]
        &\quad \times 
		\frac{ \Gamma (k + k_{3} + 1) }{ \Gamma \qty[ k - ( k_{1} + k_{2} ) + 1 ] } 
		\frac{1}{ \qty[ f_{1} (x) ]^{k + k_{3} + 1} } 
		\,\, . 
		\label{eq:rvd_mean_k_c1}
\end{align}

\begin{figure}
    \centering
    \includegraphics[width=0.8\linewidth]{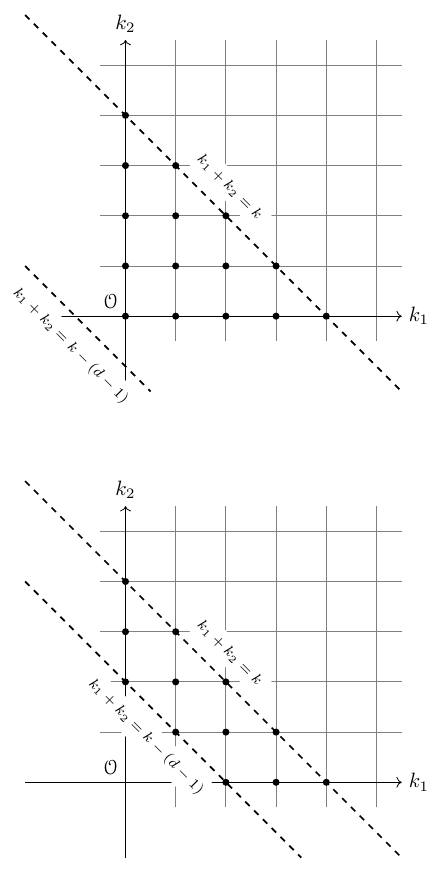}
    \caption{
        The summation over $k_{1}$ and $k_{2}$ in Eq.~(\ref{eq:rvd_mean_k_c1}). 
        Depending on the ordering between $0$ and $k - k_{1} - (d-1)$, the origin $\mathcal{O}$, \textit{i.e.}, $(k_{1}, \, k_{2}) = (0, \, 0)$ may or may not be included. 
    }
    \label{fig:lat}
\end{figure}

With $(k, \, d) = (4, \, 6)$ for instance, the lower bound of the summation over $k_{2}$ is $k - (d - 1) = - 1 < 0$. 
Hence, $k_{2}$ runs from $0$ to $k - k_{1}$ for each $0 \leq k_{1} \leq k$, and in the summation over $k_{1}$ and $k_{2}$ the lattice points listed below are collected as shown in the top panel in Figure~\ref{fig:lat}. 
With $(k, \, d) = (4, \, 3)$ as another illustration, on the other hand, $k_{2}$ start from $k - (d - 1) = 2 > 0$ instead of zero. 
The summation over $k_{1}$ and $k_{2}$ now collects the lattice points listed in the bottom panel in Figure~\ref{fig:lat}. 

\subsection{Variance}
\label{subsec:var}

In the previous subsection, the first statistical moment of $\mathcal{N}$, \textit{i.e.}, the mean number of $e$-folds, is focussed on. 
In order to relate it to the observables, the variance of $\mathcal{N}$ must also be calculated.
Each perturbative term in the variance can be obtained as well from the general formula (\ref{eq:sngen_main}) for arbitrary $n \geq 0$. 
With the variance in addition to the mean in hand, the power spectrum of the curvature perturbation together with its spectral index can be derived by virtue of the stochastic-$\delta \mathcal{N}$ formalism. 
In this subsection, let us therefore turn our attention to the derivation of the variance of the number of $e$-folds. 

The variance of $\mathcal{N}$ is defined by  
\begin{equation}
    \delta \mathcal{N}^{2} (r) 
    \equiv 
    \expval*{ 
        ( \mathcal{N} - \expval*{ \mathcal{N} } )^{2} 
    } (r) 
    = 
    \expval*{ \mathcal{N}^{2} } (r) - \expval*{ \mathcal{N} }^{2} (r) 
    \,\, . 
    \label{eq:rvd_var_def}
\end{equation}
As in Eq.~(\ref{eq:rvd_mean_sum}), the second statistical moment, $\expval*{ \mathcal{N}^{2} } (r)$, can also be expanded in a series in $v$, 
\begin{equation}
	\expval*{ \mathcal{N}^{2} } (r) 
	= \sum_{k = 0}^{\infty} \expval*{ \mathcal{N}^{2} }^{(k)} (r) 
	\,\, . 
	\label{eq:rvd_var_sum}
\end{equation}
Substitution of Eqs.~(\ref{eq:rvd_mean_sum}) and~(\ref{eq:rvd_var_sum}) into Eq.~(\ref{eq:rvd_var_def}) gives the perturbative expansion of $\delta \mathcal{N}^{2}$, 
\begin{equation}
	\delta \mathcal{N}^{2} (r) 
	= \sum_{k = 0}^{\infty} \qty[ \delta \mathcal{N}^{2} ]^{(k)} (r) 
	\,\, , 
\end{equation}
where the $k$-th order term is of order $v^{k}$ and is given by 
\begin{equation}
	\qty[ \delta \mathcal{N}^{2} ]^{(k)} (r) 
	= \expval*{ \mathcal{N}^{2} }^{(k)} (r) 
	- \sum_{i = 0}^{k} \expval*{ \mathcal{N} }^{(i)} (r) \expval*{ \mathcal{N} }^{(k-i)} (r) 
	\,\, . 
	\label{eq:rvd_var_k}
\end{equation}
Each factor of $\expval*{ \mathcal{N} }^{(i)}$ in Eq.~(\ref{eq:rvd_var_k}) can be calculated from the master formula (\ref{eq:rvd_mean_master}), or Eq.~(\ref{eq:rvd_mean_k_c1}), and has already been derived in the previous subsection up to third order. 

What remains is thus to derive each order of the second moment, and it can also be calculated from the general formula~(\ref{eq:sngen_main}) with $n = 1$. 
One difference from the calculation of $n = 0$ case is the presence of the $\expval*{ \mathcal{N}^{n} } (x)$ factor on the right-hand side in Eq.~(\ref{eq:sngen_main}), which is absent in the calculations of the mean number of $e$-folds. 
However, the general formula is already prepared and can be applied to the second moment, which reads 
	\begin{align}
		&\quad 
        \frac{1}{2} \dv{\expval*{ \mathcal{N}^{2} } (x)}{x}  
		\notag \\[0.5ex]
        &\simeq 
		\sum_{k_{1} = 0}^{\infty} 
		\sum_{k_{2} = 0}^{\infty} 
		\sum_{k_{3} = 0}^{k_{2}} 
		\sum_{k_{4} = 0}^{\infty} 
		\sum_{k_{5} = 0}^{\infty} 
		\sum_{k_{6} = k_{1} + k_{2} + k_{3} + k_{4}}^{\infty} 
        \notag \\[0.5ex] 
        &\quad\quad 
		\frac{
			\tikz[baseline={(0, -0.1)}]{
				\fill (0, 0) circle (2.5pt);
  				\node at (0, -0.3) {\scriptsize $k_{1}$};
			} \times 
			D_{k_{2}, \, k_{3}} (x) 
		}{k_{3}! k_{4}!} 
		\dv[k_{4}]{\expval*{ \mathcal{N} }^{(k_{5})} (x)}{x} 
		\notag \\[0.5ex] 
		&\quad\quad \times 
        \qty[ \frac{v (x)}{2} ]^{ - k_{3} + k_{6} } 
		\frac{1}{x^{k_{6} - (k_{1} + k_{2} + k_{3} + k_{4})}} 
        \frac{1}{ \qty[ f_{1} (x) ]^{k_{6} + 1} } 
		\notag \\[0.5ex] 
        &\quad\quad \times 
		\frac{ \Gamma (d) }{ \Gamma (d - k_{6} + k_{1} + k_{2} + k_{3} + k_{4}) } 
        \notag \\[0.5ex] 
        &\quad\quad \times 
		\frac{ \Gamma (k_{6} + 1) }{ \Gamma \qty[ k_{6} - ( k_{1} + k_{2} + k_{3} + k_{4} ) + 1 ] } 
		\,\, . 
		\label{eq:gen_main}
	\end{align}

All the combinations that satisfy $- k_{3} + k_{5} + k_{6} = k$ give rise to the contribution of order $v^{k}$. 
Among those combinations that yield $(k+1) (k+2) (k+3) (k+4) (k+5) / 120$ terms, after removing the ones that vanish due to $D_{i = 0, \, j \geq 1} (x) = D_{i \geq 0, \, j = 0} (x) = 0$, the net number of the nonzero combinations is counted to be 
\begin{equation}
	N_{2} (k) 
	= \frac{ (k+1) (k+2) (k+3) (k^{2} + 4 k + 20) }{ 120 } 
	\,\, . 
    \label{eq:nts_s2}
\end{equation}
Let us now collect the relevant terms at each order. 

\subsubsection{Zeroth order}

At zeroth order in $v$, the evolution of the inflaton fields is deterministic, so that there is no fluctuations in the number of $e$-folds and its variance must vanish. 
The combination of the indices in Eq.~(\ref{eq:gen_main}) at this order is only $k_{1} = \cdots = k_{6} = 0$, corresponding to $N_{2} (0) = 1$. 
This gives rise to 
\begin{equation}
	\frac{1}{2} \dv{ \expval*{ \mathcal{N}^{2} }^{(0)} (x) }{x} 
	\simeq \frac{ \expval*{ \mathcal{N} }^{(0)} (x) }{ f_{1} (x) } 
	= \frac{1}{2} \dv{x} \qty[ \expval*{ \mathcal{N} }^{(0)} (x) ]^{2} 
	\,\, . 
	\label{eq:rvd_var_d0}
\end{equation}
Thus, as expected, the variance is found to vanish from Eq.~(\ref{eq:rvd_var_k}), 
\begin{equation}
	\qty[ \delta \mathcal{N}^{2} ]^{(0)} (r) 
	= \expval*{ \mathcal{N}^{2} }^{(0)} (r) - \qty[ \expval*{ \mathcal{N} }^{(0)} (r) ]^{2} 
	= 0 
	\,\, , 
	\label{eq:rvd_var_r0}
\end{equation}
Upon integrating Eq.~(\ref{eq:rvd_var_d0}) over the domain $x \in [r_{-}, \, r]$, the contribution from $r_{-}$ vanishes by definition of the absorbing boundary, see also below Eq.~(\ref{eq:pre_fpt}). 

\subsubsection{First order}

Though Eq.~(\ref{eq:nts_s2}) tells us that the number of the terms increases as $k^{5}$, at first order, there are only 5 relevant combinations, \textit{i.e.}, $N_{2} (1) = 5$, 
\begin{equation}
	(k_{1}, \, k_{2}, \, k_{3}, \, k_{4}, \, k_{5}, \, k_{6}) 
	= 
	\begin{cases}
		(0, \, 0, \, 0, \, 0, \, 0, \, 1) \,\, , 
		\\[1.0ex] 
		(0, \, 0, \, 0, \, 0, \, 1, \, 0) \,\, , 
		\\[1.0ex] 
		(0, \, 0, \, 0, \, 1, \, 0, \, 1) \,\, ,
		\\[1.0ex] 
		(0, \, 1, \, 1, \, 0, \, 0, \, 2) \,\, ,
		\\[1.0ex] 
		(1, \, 0, \, 0, \, 0, \, 0, \, 1) \,\, . 
	\end{cases}
\end{equation}
Summing up all the contributions, and then simplifying the expression further, one obtains 
\begin{align}
	\frac{1}{2} \dv{ \expval*{ \mathcal{N}^{2} }^{(1)} (x)}{x} 
	&= \dv{x} \qty[ \expval*{ \mathcal{N} }^{(0)} (x) \expval*{ \mathcal{N} }^{(1)} (x) ] 
	\notag \\[0.5ex] 
	&\quad 
	+ \frac{1}{2} \frac{v (x)}{ \qty[ f_{1} (x) ]^{2} } \dv{ \expval*{ \mathcal{N} }^{(0)} (x)}{x} 
	\,\, .  
	\label{eq:rvd_var_d1}
\end{align}
This leads to 
\begin{equation}
	\dv{x} \qty[ \delta \mathcal{N}^{2} ]^{(1)} (x) 
	= v (x) \qty[ \frac{v (x)}{v' (x)} ]^{3} 
	\,\, . 
	\label{eq:rvd_var_r1}
\end{equation}

In deriving the variance at first order, in Eq.~(\ref{eq:rvd_var_d1}), the derivative of $\expval*{ \mathcal{N}^{2} } (x)$ contains the total-derivative term and terms that are not reducible to a total derivative. 
The quantity inside the total derivative in the right-hand side is subtracted from the second moment, which is on the left-hand side, to define the variance. 
This is a common structure at arbitrary order in calculating $\expval*{ \mathcal{N}^{2} } (x)$. 
That is, the derivative $\dd \expval*{ \mathcal{N}^{2} } (x) / \dd x$ can always be decomposed into the total-derivative term and the remaining terms, and the latter directly gives the variance of the number of $e$-folds, such as Eq.~(\ref{eq:rvd_var_r1}). 
The only exception is at zeroth order, see Eq.~(\ref{eq:rvd_var_d0}), where non-total-derivative term is absent. 

No term that depends on $d$ can be found at first order in the variance
in Eq.~(\ref{eq:rvd_var_r1}). 
This is similar to the zeroth-order result of the mean number of $e$-folds, see Eq.~(\ref{eq:sncl_clef}). 
Both results are valid regardless of the number of fields. 
Indeed, Eq.~(\ref{eq:rvd_var_r1}) revalidates the result of Ref.~\cite{Vennin:2015hra}, 
which is of interest in a single-field case. 

\subsubsection{Second order}

The second-order calculation requires collecting 16 terms, \textit{i.e.}, $N_{2} (2) = 16$. 
All those combinations of the indices can be found in Appendix, and subtraction of the total-derivative terms from $\expval*{ \mathcal{N}^{2} } (r)$ as was done in Eq.~(\ref{eq:rvd_var_d1}) arrives at 
\begin{align}
	\dv{x} \qty[ \delta \mathcal{N}^{2} ]^{(2)} (x) 
	&= \qty[ v (x) ]^{2} \qty[ \frac{v (x)}{ v' (x) } ]^{4} 
	\left[ 
		\frac{3 (d-1)}{2 x} 
	\right. 
	\notag \\[0.5ex] 
	&\quad \left. 
		+\, 3 \frac{v' (x)}{v (x)} 
		- \frac{5}{2} \frac{v (x)}{v' (x)} \frac{v'' (x)}{v (x)} 
	\right] 
	\,\, . 
    \label{eq:rvd_var_o2}
\end{align}
As in the first-order calculation of the mean, the linear $d$-dependent term appears in the variance at second order. 
When $d = 1$, that term vanishes and the expression matches the previously-derived single-field result~\cite{Vennin:2015hra}. 

\subsubsection{Third order}

The third-order term includes
the $N_{2} (3) = 41$ combinations in Eq.~(\ref{eq:gen_main}).
The computation of this term is cumbersome, but after some algebra it results in 
\begin{align}
	\dv{x} \qty[ \delta \mathcal{N}^{2} ]^{(3)} (x) 
	&\simeq \qty[ v (x) ]^{3} \qty[ \frac{ v (x) }{ v' (x) } ]^{5} 
	\notag \\[0.5ex] 
    &\quad \times \left\{ 
        \frac{ (d-1) (6d - 11) }{ 4 x^{2} } 
    \right. 
    \notag \\[0.5ex] 
    &\qquad 
        + \frac{d-1}{2 x} \qty[ 
			17 \frac{v' (x)}{v (x)} - \frac{25}{2} \frac{v (x)}{v' (x)} \frac{v'' (x)}{v (x)} 
		] 
    \notag \\[0.5ex] 
    &\qquad 
        +\, \frac{35}{4} \qty[ \frac{v' (x)}{v (x)} ]^{2} 
		- \frac{29}{2} \frac{v'' (x)}{v (x)} 
    \notag \\[0.5ex] 
    &\qquad 
        + 8 \qty[ \frac{v (x)}{v' (x)} ]^{2} \qty[ \frac{v'' (x)}{v (x)} ]^{2} 
    \notag \\[0.5ex] 
    &\qquad \left. 
        - \, \frac{7}{4} \frac{v (x)}{v' (x)} \frac{ v''' (x)}{v (x)} 
    \right\} 
	\,\, . 
    \label{eq:rvd_var_o3}
\end{align}

In addition to the linearly dependent term on $d$, as is also present at second order, the quadratic term in $d$ is involved in the third-order term in the variance, Eq.~(\ref{eq:rvd_var_o3}). 
This outcome is similar to the quadratic term in $d$ found in the second-order term in the mean number of $e$-folds, Eq.~(\ref{eq:d_mean_o2_res}). 
Contrary to the mean number of $e$-folds, however, the quadratic term in $d$ in the variance does not vanish at $d = 2$, though it still vanishes together with the term proportional to $(d - 1) / x$ at $d = 1$. 

\vspace{0.8\baselineskip}
\subsubsection{General order}

The higher-order terms in the variance can also be enumerated order by order. 
However, at third order, a large number of terms already appears. 
Therefore, let us instead present the general-order formula for the variance, which comes from the total-derivative terms subtracted from $\expval*{ \mathcal{N}^{2} }$. 
Such total derivatives appear as the $k_{4} = 0$ term in Eq.~(\ref{eq:gen_main}). 
Therefore, similar to Eq.~(\ref{eq:rvd_mean_k_c1}), the general expression for the variance (instead of the second moment itself) reads 
\begin{align}
    \frac{1}{2} \dv{ \, [ \delta \mathcal{N}^{2} ]^{(k)} (x) }{ x } 
    &\simeq 
    \sum_{k_{1} = 0}^{k} 
    \sum_{k_{2} = 0}^{k - k_{1}} 
    \sum_{k_{3} = 0}^{k_{2}} 
    \sum_{k_{4} = 1}^{k - (k_{1} + k_{2})} 
    \sum_{k_{5} = 0}^{k - (k_{1} + k_{2} + k_{4})}  
    \notag \\[0.5ex] 
    &\quad \times 
    \frac{
			\tikz[baseline={(0, -0.1)}]{
				\fill (0, 0) circle (2.5pt);
  				\node at (0, -0.3) {\scriptsize $k_{1}$};
			} \times 
			D_{k_{2}, \, k_{3}} (x) 
	}{k_{3}! k_{4}!} 
    \qty[ \frac{v (x)}{2} ]^{k - k_{5}} 
    \notag \\[0.5ex] 
    &\quad \times 
    \, \dv[k_{4} - 1]{x} 
    \qty[ 
        \dv{ \expval*{ \mathcal{N} }^{(k_{5})} (x) }{ x } 
    ] 
    \notag \\[0.5ex] 
    &\quad \times 
    \frac{ 1 }{ x^{k - (k_{1} + k_{2} + k_{4} + k_{5})} } 
    \frac{1}{ [ f_{1} (x) ]^{k + k_{3} - k_{5} + 1} } 
    \notag \\[0.5ex] 
    &\quad \times 
    \frac{ \Gamma (d) }{ \Gamma [ d + (k_{1} + k_{2} + k_{4} + k_{5}) - k ] } 
    \notag \\[0.5ex] 
    &\quad \times 
    \frac{ \Gamma (k + k_{3} - k_{5} + 1 ) }{ \Gamma [ k - (k_{1} + k_{2} + k_{4} + k_{5}) + 1 ] } 
    \,\, . 
    \label{eq:var_ex_gen}
\end{align}

While in principle calculations can be extended to derive the higher-order statistical moments, this article restricts itself not to continue further since the strategy remains the same. 
Instead, let us now turn our attention to the observables, \textit{i.e.}, the power spectrum of the curvature perturbation and its spectral index. 

\subsection{Observables}

From the derived quantities until here, it is now possible to find how the basic observables depend on the number of fields $d$. 
This subsection restricts itself to the power spectrum $\mathcal{P}_{\zeta}$ of the curvature perturbation $\zeta$ and its spectral index $n_{\rm S}$. 
In the same manner as the previous sections, those are derived as perturbative series in $v$. 

\subsubsection{Power spectrum}

By virtue of the stochastic-$\delta \mathcal{N}$ formalism, it reads 
\begin{equation}
	\mathcal{P}_{\zeta} (x) 
	= \dv{ \, \delta \mathcal{N}^{2} (x) }{ \expval*{ \mathcal{N} } (x) } 
	= \qty[ \dv{ \expval*{ \mathcal{N} } (x) }{x} ]^{-1} \, 
    \dv{ \, \delta \mathcal{N}^{2} (x) }{x} 
    \,\, . 
    \label{eq:obs_psdef}
\end{equation}
In Eq.~(\ref{eq:obs_psdef}), it is understood that the field value $x$ is implicitly evaluated at the horizon crossing of a relevant mode. 
Provided that both factors $\expval*{ \mathcal{N} } (x)$ and $\delta \mathcal{N}^{2} (x)$ have been obtained in Sections~\ref{subsec:mean} and \ref{subsec:var} respectively, as power series in $v$, the power spectrum can also be expanded similarly. 
For brevity, the following shorthand notations are introduced, 
\begin{equation}
	\mu_{i} (x) 
	\equiv \dv{ \expval*{ \mathcal{N} }^{(i)} (x) }{x} 
	\,\, , 
	\qquad 
	\nu_{i} (x) 
	\equiv \dv{ \qty[ \delta \mathcal{N}^{2} ]^{(i)} (x) }{x} 
	\,\, . 
    \label{eq:obs_mntn}
\end{equation}
When $\mathcal{P}_{\zeta} (x)$ is written as a power series in $v$, 
\begin{equation}
	\mathcal{P}_{\zeta} (x) 
	= \sum_{k = 1}^{\infty} \mathcal{P}_{\zeta}^{(k)} (x) 
	\,\, , 
\end{equation}
terms in each order $\mathcal{P}_{\zeta}^{(k)} (x)$ can be expressed by the quantities introduced in Eq.~(\ref{eq:obs_mntn}). 
Since the power spectrum is defined for the fluctuations, it starts from $\mathcal{O} (v)$. 

The first-order contribution in Eq.~(\ref{eq:obs_psdef}) comes from the zeroth-order term in the mean and the first-order term in the variance. 
It results in 
\begin{equation}
	\mathcal{P}_{\zeta}^{(1)} (x) 
	\simeq \frac{\nu_{1} (x)}{\mu_{0} (x)} 
	\simeq v (x) \qty[ \frac{v (x)}{v' (x)} ]^{2} 
	\,\, . 
\end{equation}
This of course matches the standard slow-roll result, $\mathcal{P}_{\zeta}^{(1)} (x) \simeq [ (H / M_{\rm P}) / 2 \pi ]^{2} / 2 \varepsilon$, where $\varepsilon (x) = [ v' (x) / v (x) ]^{2} / 2$ is the slow-roll parameter defined in Eq.~(\ref{eq:pre_srparam}). 

The second-order term in Eq.~(\ref{eq:obs_psdef}) is given by  
\begin{align}
	\mathcal{P}_{\zeta}^{(2)} (x) 
	&\simeq \frac{ \mu_{0} (x) \nu_{2} (x) - \mu_{1} (x) \nu_{1} (x) }{ \qty[ \mu_{0} (x) ]^{2} } 
	\notag \\[0.5ex] 
	&\simeq 
	\qty[ v (x) ]^{2} \qty[ \frac{ v (x) }{ v' (x) } ]^{3} 
	\left[ 
		\frac{d-1}{x} 
    \right. 
    \notag \\[0.5ex] 
    &\quad \left. 
        + \, \frac{5}{2} \frac{v' (x)}{v (x)} - 2 \frac{v (x)}{v' (x)} \frac{v'' (x)}{v (x)} 
	\right] 
	\,\, . 
\end{align}
This matches with the result at $d = 1$ given in Ref.~\cite{Vennin:2015hra}. 
For $d \geq 1$, the result given in Ref.~\cite{Vennin:2016wnk} is reproduced. 

Since the perturbative correction terms have been derived up to third order, one can go to the next order of the power spectrum:
\begin{align}
	\mathcal{P}_{\zeta}^{(3)} (x) 
	&\simeq \frac{ \mu_{0}^{2} \nu_{3} - \mu_{0} \mu_{1} \nu_{2} + \qty( \mu_{1}^{2} - \mu_{0} \mu_{2} ) \nu_{1} }{ \mu_{0}^{3} } 
	\notag \\[0.5ex] 
	&\simeq 
	\qty[ v (x) ]^{3} \qty[ \frac{ v (x) }{ v' (x) } ]^{4} 
	\left\{ 
		\frac{(d-1) (3 d - 7)}{4 x^{2}} 
	\right. 
	\notag \\[0.5ex] 
	&\quad \left. 
	+\, \frac{d-1}{x} \qty[ 
			\frac{23}{4} \frac{v' (x)}{v (x)} - 4 \frac{v (x)}{v' (x)} \frac{v'' (x)}{v (x)} 
		] 
	\right. 
	\notag \\[0.5ex] 
	&\quad \left. 
		+ \, 7 \qty[ \frac{v' (x)}{v (x) } ]^{2} 
        - \frac{45}{4} \frac{v'' (x)}{v (x)} 
    \right. 
    \notag \\[0.5ex] 
    &\quad \left. 
		+ \, \frac{25}{4} \qty[ \frac{v (x)}{v' (x)} ]^{2} \qty[ \frac{v'' (x)}{v (x)} ]^{2} 
		- \frac{3}{2} \frac{v (x)}{v' (x)} \frac{v''' (x)}{v (x)} 
	\right\} 
	\,\, . 
\end{align}

\subsubsection{Tilt} 

From the power spectrum, the spectral index $n_{\rm S}$ is defined as 
\begin{align}
	n_{\rm S} - 1 
	&\equiv \eval{ \dv{ \ln \mathcal{P}_{\zeta} (k) }{\ln k} }_{k = aH} 
	\simeq - \dv{x}{\expval*{ \mathcal{N} } (x)} \cdot \dv{\ln \mathcal{P}_{\zeta}}{x} 
	\notag \\[0.5ex] 
    &= - \frac{
		\displaystyle \dv{\ln \mathcal{P}_{\zeta}}{x}  
	}{ 
		\displaystyle 
		\dv{\mathcal{N} (x)}{x}
	} 
	= - \frac{
		\displaystyle \frac{1}{\mathcal{P}_{\zeta} (x)} \dv{\mathcal{P}_{\zeta} (x)}{x}  
	}{ 
		\displaystyle 
		\dv{\mathcal{N} (x)}{x}
	} 
	\,\, . 
    \label{eq:obs_nsdef}
\end{align}
Provided that the power spectrum has been obtained up to third order, the spectral index can be derived up to second order.
To calculate $n_{\rm S}$, let us introduce 
\begin{equation}
	\lambda_{i} (x) 
	\equiv \mathcal{P}_{\zeta}^{(i)} (x) 
	\,\, , 
	\qquad 
	\sigma_{i} (x) 
	\equiv \dv{ \mathcal{P}_{\zeta}^{(i)} (x) }{x} 
	\,\, . 
\end{equation}

After one writes Eq.~(\ref{eq:obs_nsdef}) in powers of $v$, the spectral index at zeroth order can be obtained, 
\begin{equation}
	(n_{\rm S} - 1)^{(0)} 
	= - \frac{\sigma_{1} (x)}{\mu_{0} (x) \lambda_{1} (x)}
	= 2 \frac{v'' (x)}{v (x)} - 3 \qty[ \frac{v' (x)}{v (x)} ]^{2} 
	\,\, . 
\end{equation}
This can be expressed in terms of the Hubble flow parameters defined in Eq.~(\ref{eq:pre_flowp}), as $(n_{\rm S} - 1)^{(0)} \simeq - 2 \varepsilon_{1} - \varepsilon_{2}$, reproducing the known result~\cite{Schwarz:2001vv, Vennin:2015hra}. 
It is noted that the Hubble-flow parameters are defined at the classical level, that is to say, the noise-induced centrifugal force is not included, so one has 
\begin{equation}
    \varepsilon_{2} (x) 
    \simeq 2 \qty[ \frac{v' (x)}{v (x)} ]^{2} - 2 \frac{v'' (x)}{v (x)} 
    \,\, . 
\end{equation}

The spectral index at first order in $v$ is computed as 
\begin{align}
    (n_{\rm S} - 1)^{(1)} 
    &\simeq - \frac{ \mu_{0} \lambda_{1} \sigma_{2} - ( \mu_{1} \lambda_{1} + \mu_{0} \lambda_{2} ) \sigma_{1} }{ (\mu_{0} \lambda_{1} )^{2} } 
    \notag \\[0.5ex] 
    &\simeq v (x) \left\{ 
		\frac{d-1}{x^{2}} - \frac{d-1}{2 x} \frac{v' (x)}{v (x)} 
		- \qty[ \frac{v' (x)}{v (x)} ]^{2} 
	\right. 
	\notag \\[0.5ex] 
	&\quad \left. 
		+ \frac{3}{2} \frac{v'' (x)}{v (x)} 
		- 3 \qty[ \frac{v (x)}{v' (x)} ]^{2} \qty[ \frac{v'' (x)}{v (x)} ]^{2} 
    \right. 
    \notag \\[0.5ex] 
    &\quad \left. 
		+ 2 \frac{v (x)}{v' (x)} \frac{v''' (x)}{v (x)} 
	\right\} 
	\,\, . 
\end{align}

The next-order correction to $n_{\rm S}$ reads 
\begin{widetext}
\begin{align}
    ( n_{\rm S} - 1 )^{(2)} 
    &\simeq - \frac{  
    \left[ 
        \mu_{0}^{2} \lambda_{1}^{2} \sigma_{3} 
        - \mu_{0} \lambda_{1} ( \mu_{1} \lambda_{1} + \mu_{0} \lambda_{2} ) \sigma_{2} 
    + \left( 
        \mu_{0} \mu_{1} \lambda_{1} \lambda_{2} 
        - \mu_{0} \mu_{2} \lambda_{1}^{2} 
        + \mu_{1}^{2} \lambda_{1}^{2} 
        \mu_{0}^{2} \lambda_{1} \lambda_{3} 
        + \mu_{0}^{2} \lambda_{2}^{2} 
    \right) \sigma_{1} 
    \right] }{ (\mu_{0} \lambda_{1})^{3} } 
    \notag \\[2.0ex] 
    &\simeq [ v (x) ]^{2} 
    \left( 
        \frac{d-1}{x} 
        \qty{ 
            - 6 \frac{v' (x)}{v (x)} 
            + 6 \frac{v (x)}{v' (x)} \frac{v'' (x)}{v (x)} 
            - 3 \qty[ \frac{v (x)}{v' (x)} ]^{3} \qty[ \frac{v'' (x)}{v (x)} ]^{2} 
            + \qty[ \frac{v (x)}{v' (x)} ]^{2} \frac{v''' (x)}{v (x) }
        }
    \right. 
    \notag \\[2.0ex] 
    &\quad \qquad
    + \frac{d-1}{x^{2}} \qty{ 
        6 - 3 \qty[ \frac{v (x)}{v' (x)} ]^{2} \frac{v'' (x)}{v (x)} 
    } 
    - 2 \frac{d-1}{x^{3}} \frac{v (x)}{v' (x)} 
    - \frac{23}{4} \qty[ \frac{v' (x)}{v (x)} ]^{2} 
    + \frac{27}{2} \frac{v'' (x)}{v (x)} 
    - 23 \qty[ \frac{v (x)}{v' (x)} ]^{2} \qty[ \frac{v'' (x)}{v (x)} ]^{2} 
    \notag \\[2.0ex] 
    &\quad \qquad \left. 
    + \, \frac{21}{2} \frac{v (x)}{v' (x)} \frac{v''' (x)}{v (x)} 
    + 14 \qty[ \frac{v (x)}{v' (x)} ]^{4} \qty[ \frac{v'' (x)}{v (x)} ]^{3} 
    - \frac{23}{2} \qty[ \frac{v (x)}{v' (x)} ]^{3} \frac{v'' (x)}{v (x)} \frac{v''' (x)}{v (x)} 
    + \frac{3}{2} \qty[ \frac{v (x)}{v' (x)} ]^{2} \frac{v'''' (x)}{v (x)} 
    \right) 
    \,\, . 
\end{align}

The perturbative series of the statistical moments of the first-passage time, as well as the power spectrum and the spectral index, have been derived analytically until here. 
Those quantities are expected to describe the dynamics of multifield slow-roll over $O(d)$-symmetric potential, but their use is restricted to the small-noise regime. 
It should be emphasised that a non-negligible deviation from the classical prediction can be found depending on the number of fields $d$, even when $v_{0}$ is small. 
\end{widetext}

\subsubsection{\texorpdfstring{$f_{\rm NL}$}{fNL}}

The local-type non-Gaussianity conventionally parametrised by the $f_{\rm NL}$ parameter can also be calculated from the statistical moments of $\mathcal{N}$ as follows, 
\begin{align}
    f_{\rm NL} 
    &= \frac{5}{72} \dv[2]{ \, \delta \mathcal{N}^{3} (x) }{ [ \expval*{ \mathcal{N} } (x) ] } \qty[ \dv{ \, \delta \mathcal{N}^{2} (x) }{ \expval*{ \mathcal{N} } (x) } ]^{-2} 
    \notag \\[0.5ex] 
    &= \frac{5}{72} 
    \frac{
        \displaystyle 
        \dv{ \expval*{ \mathcal{N} } (x) }{ x } 
        \dv[2]{ \, \delta \mathcal{N}^{3} (x) }{ x } 
        - \dv[2]{ \expval*{ \mathcal{N} } (x) }{ x } 
        \dv{ \, \delta \mathcal{N}^{3} (x) }{ x } 
    }{
        \displaystyle 
        \dv{ \expval*{ \mathcal{N} } (x) }{ x } 
        \qty[ 
            \dv{ \, \delta \mathcal{N}^{2} (x) }{ x } 
        ]^{2} 
    } 
    \,\, ,
\end{align}
which needs the third moment $\expval*{ \mathcal{N}^{3} }$ in addition to the mean and variance of $\mathcal{N}$. 
Though the derivation of $\expval*{ \mathcal{N}^{3} }$ and the higher-order moments proceeds in the same manner as that for the mean $\expval*{ \mathcal{N} }$ and the variance $\expval*{ \mathcal{N}^{2} }$,
it is not pursued further here, and let us now turn to demonstrate the derived formulas and to extract the physical consequences. 

\section{Demonstrative examples}
\label{sec:implications}

Having derived the small-noise stochastic correction terms in the statistical moments as well as the observables, let us finally consider several concrete models to derive the implications. 
Our primary interest continues to lie in the effects that come from the number of fields. 

\subsection{Monomial model}
\label{subsec:demo_mon}

As the easiest model to handle, the monomial model, 
\begin{equation}
    v (r) = v_{0} r^{p} 
    \,\, , 
    \label{eq:m1_pot}
\end{equation}
is considered first. 
At each order, the correction terms to the mean number of $e$-folds derived in Section~\ref{subsec:mean} reduce to 
\begin{subequations}
    \begin{align}
        \expval*{ \mathcal{N} }^{(0)} (r) 
	    &= \frac{ (r^{2} - r_{-}^{2}) }{ 2 p }
	    \,\, , 
        \label{eq:rvd_mean0_mon}
        \\[0.5ex] 
        \expval*{ \mathcal{N} }^{(1)} (r) 
	    &= v_{0} \frac{ d }{ 2 p^{2} (p+2) } ( r^{p+2} - r_{-}^{p+2} ) 
    	\,\, , 
	    \label{eq:rvd_mean1_mon}
        \\[0.5ex] 
        \expval*{ \mathcal{N} }^{(2)} (r) 
	    &= v_{0}^{2} \frac{ d (d + p) }{ 4 p^{3} (2p+2) } 
	    ( r^{2p+2} - r_{-}^{2p+2} )
	    \,\, , 
        \label{eq:rvd_mean2_mon}
        \\[0.5ex] 
        \expval*{ \mathcal{N} }^{(3)} (r) 
	    &= v_{0}^{3} \frac{ d (d + p) (d + 2 p) }{ 8 p^{4} (3 p + 2) } 
	    ( r^{3p+2} - r_{-}^{3p+2} )
	    \,\, , 
	\label{eq:rvd_mean3_mon}    
    \end{align}
\end{subequations}
The contribution of those terms to the total number of $e$-folds can also be found in our previous article~\cite{Takahashi:2025hqt}, in which, from Eqs.~(\ref{eq:rvd_mean0_mon}), (\ref{eq:rvd_mean1_mon}), (\ref{eq:rvd_mean2_mon}), and (\ref{eq:rvd_mean3_mon}), the general-order term is deduced to be 
\begin{equation}
    \expval*{ \mathcal{N} }^{(k)} (r) 
    = \qty( \frac{v_{0}}{2} )^{k} \frac{ \Gamma (d/p + k) }{ p \Gamma (d/p) } \frac{r^{kp+2} - r_{-}^{kp+2}}{k p + 2} 
    \,\, , 
    \label{eq:d_mean_ogen}
\end{equation}
for $k = 0, \, 1, \, \cdots$. 

The variance is absent in the zeroth order and starts from first order in $v$, 
\begin{subequations}
    \begin{align}
    	\qty[ \delta \mathcal{N}^{2} ]^{(1)} (r) 
	    &\simeq v_{0} \frac{1}{p^{3} (p+4)} 
        ( r^{p+4} - r_{-}^{p+4} ) 
	    \,\, , 
        \\[0.5ex] 
        \qty[ \delta \mathcal{N}^{2} ]^{(2)} (r) 
	    &\simeq v_{0}^{2} \frac{p + 3 d + 2}{2 p^{4} (2 p + 4)} 
	    ( r^{2 p + 4} - r_{-}^{2 p + 4} ) 
	    \,\, , 
        \\[0.5ex] 
        \qty[ \delta \mathcal{N}^{2} ]^{(3)} (r) 
	    &\simeq v_{0}^{3} 
	    \frac{6 d^{2} + (9 p + 8) d + 2 (p + 1) (p + 2)}{ 4 p^{5} (3 p + 4) } 
        \notag \\[0.5ex] 
        &\quad \quad \times 
    	( r^{3 p + 4} - r_{-}^{3 p + 4} ) 
	    \,\, , 
    \end{align}
\end{subequations}
and so forth. 
Similarly, the power spectrum of the curvature perturbation can also be obtained order by order, as 
\begin{subequations}
	\begin{align}
		\mathcal{P}_{\zeta}^{(1)} (x) 
		&\simeq v_{0} \frac{1}{p^{2}} x^{p + 2} 
		\,\, , 
		\\[0.5ex] 
		\mathcal{P}_{\zeta}^{(2)} (x) 
		&\simeq v_{0}^{2} \frac{p + 2 d + 2}{2 p^{3}} x^{2 p + 2} 
		\,\, , 
		\\[0.5ex] 
        \mathcal{P}_{\zeta}^{(3)} (x) 
		&\simeq v_{0}^{3} \frac{ 4 + 3 d (2 + d) + 6 p + 7 d p + 2 p^2 }{ 4 p^{4} } x^{3 p + 2} 
		\,\, , 
	\end{align}
\end{subequations}
where one has in mind that $x$ is a scale of interest. 
The scale-dependence of the power spectrum, the spectral index, at each order reads 
\begin{subequations}
    \label{eq:mon_nS}
    \begin{align}
        ( n_{\rm S} - 1 )^{(0)} 
        &\simeq - \frac{p (p + 2)}{x^{2}} 
        \,\, , 
        \\[0.5ex] 
        ( n_{\rm S} - 1 )^{(1)} 
        &\simeq - v_{0} \frac{(p - 2) d + p^{2} + 2 p}{2 x^{2}} \, x^{p} 
        \,\, , 
        \\[0.5ex] 
        ( n_{\rm S} - 1 )^{(2)} 
        &\simeq - v_{0}^{2} \frac{ 8 p d + 3 p^{2} + 8 p + 4}{4 x^{2}} x^{2 p} 
        \,\, . 
    \end{align}
\end{subequations}
An immediate consequence is that, in the presence of the stochastic effects, the tilt becomes increasingly red as their strength grows. 

The $k$-th order result of the mean number of $e$-folds, Eq.~(\ref{eq:d_mean_ogen}), derives another important implication. 
Substitution of Eq.~(\ref{eq:d_mean_ogen}) into Eq.~(\ref{eq:d_stmm_oex}) gives the all-order perturbative formula of $\expval*{ \mathcal{N} } (r)$, in the small-noise regime where $v_{0} \ll 1$ is in mind. 
As usual, the perturbative series has a radius of convergence, inside which it approximates the exact behaviour and thus remains reliable. 
On the other hand, the series exhibits a badly divergent behaviour outside the radius of convergence, which implies the breakdown of the perturbative treatment and calls for a non-perturbative analysis. 

To identify the radius of convergence of the series, let us now examine the asymptotic behaviour of $\expval*{ \mathcal{N} }^{(k)}$. 
In the regime where $k \gg 1$ and $d \gg 1$, it approaches 
\begin{equation}
    k \expval*{ \mathcal{N} }^{(k)} (r) 
    \approx \qty( \frac{v_{0}}{2} \frac{d}{p} r^{p} )^{k} \qty( \frac{r}{p} )^{2} 
    \,\, , 
\end{equation}
so that the reliability of the perturbative series can be assessed by the criterion that 
\begin{equation}
    \frac{v_{0}}{2} \frac{d}{p} r^{p} \gtrless 1 
    \,\, . 
\end{equation}
At this point, one notices that, even though $v \ll 1$ was originally required to conduct the perturbative expansion, the relevant condition is actually stronger and is given by $v d / 2 p \ll 1$. 

The existence of this nontrivial radius of convergence can be understood as follows. 
Writing down Eq.~(\ref{eq:pre_lan_r}) for the monomial potential, it reads 
\begin{equation}
    \dv{r}{N} 
    = 
    - \frac{p}{r} + \frac{v_{0} r^{p}}{2} \frac{d-1}{r} + \sqrt{ v_{0} r^{p} } \, \xi (N) 
    \,\, . 
    \label{eq:d_mean_geno_lan}
\end{equation}
The deterministic trajectory follows from Eq.~(\ref{eq:d_mean_geno_lan}) with the noise term being neglected, in which the drift and stochastic-noise induced centrifugal forces compete. 
When one of those dominates over the other, the dynamics can be classified according to 
\begin{equation}
    \begin{cases}
        \displaystyle \frac{p}{r} > \frac{v_{0} r^{p}}{2} \frac{d-1}{r}  
        &~ \Longrightarrow ~~ \text{drift-dominated} 
        \,\, , \\[2.0ex] 
        \displaystyle \frac{p}{r} < \frac{v_{0} r^{p}}{2} \frac{d-1}{r}  
        &~ \Longrightarrow ~~ \text{centrifugal-force-dominated} 
        \,\, . 
    \end{cases}
    \label{eq:dm_mon_class}
\end{equation}
On the other hand, $r$ stays at rest when the two forces are balanced. 
The condition under which $r$ experiences no force but the stochastic noise is given by 
\begin{equation}
    \frac{p}{r} 
    = \frac{v_{0} r^{p}}{2} \frac{d-1}{r} 
    \,\, , 
\end{equation}
which can be approximated by $v d / 2 p \approx 1$ when $d \gg 1$. 
This is nothing but the radius of convergence that has been found through the perturbative calculation. 
In other words, the derived perturbative formula can be trusted as long as the drift force dominates the dynamics, although more and more perturbative terms are needed for the correct value of $\expval*{ \mathcal{N} }$ as the number of fields $d$ increases, see also Figure~1 in Ref.~\cite{Takahashi:2025hqt}. 

\begin{figure*}
    \centering
    \includegraphics[width=0.995\linewidth]{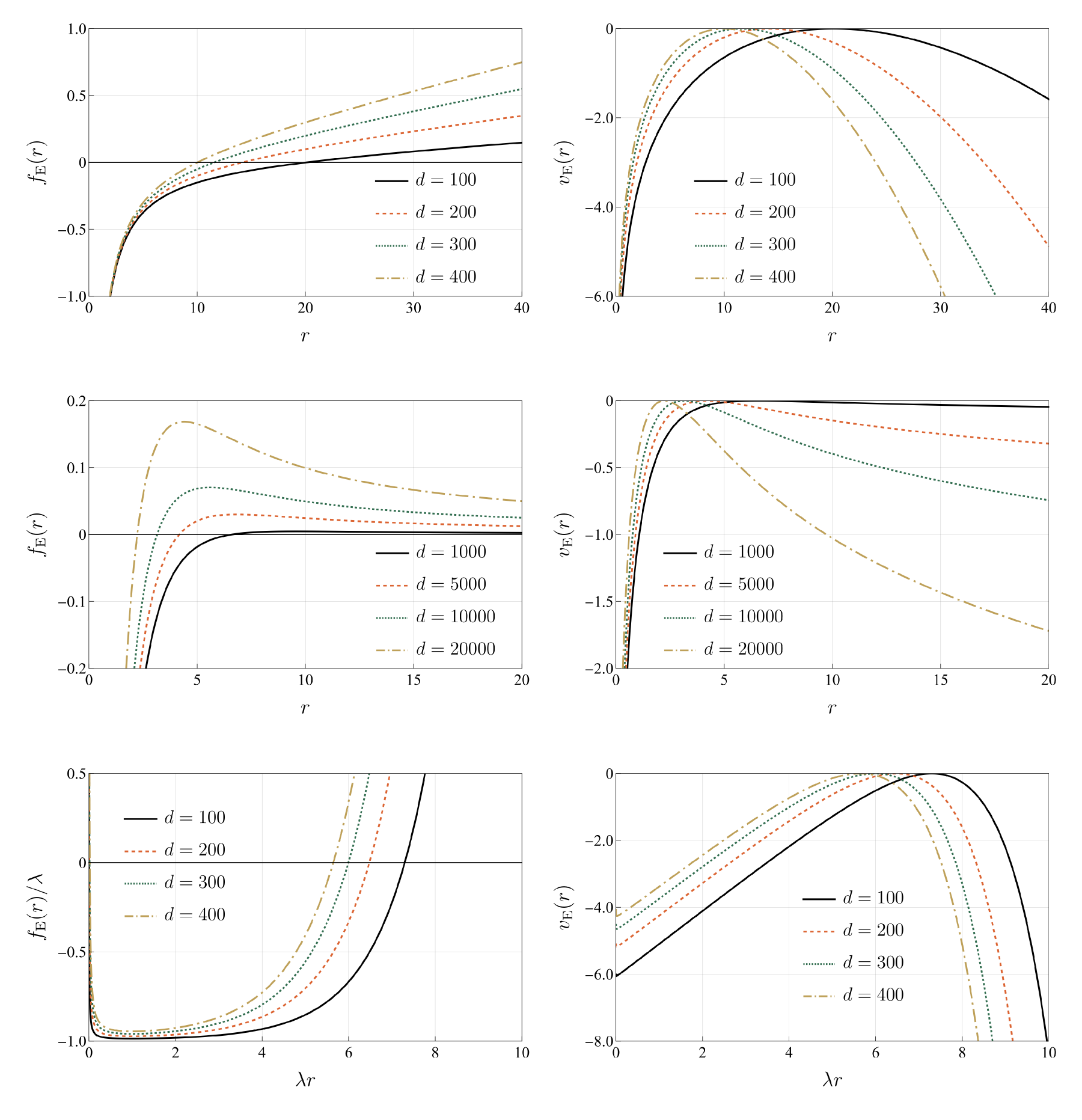}
    \caption{
        The effective force $f_{\rm E} (r)$ and the effective potential $v_{\rm E} (r)$ defined in Eq.~(\ref{eq:d_mean_veff}), for  the monomial model (\textit{top}), the $R^{2}$-type model (\textit{middle}), and the power-law model (\textit{bottom}). 
        In all the panels, $v_{0} = 10^{-4}$ is fixed and $v_{\rm E} (r)$ is normalised in such a way that the maximum becomes zero. 
    }
    \label{fig:fneff}
\end{figure*}

\begin{figure*}
    \centering
    \includegraphics[width=0.995\linewidth]{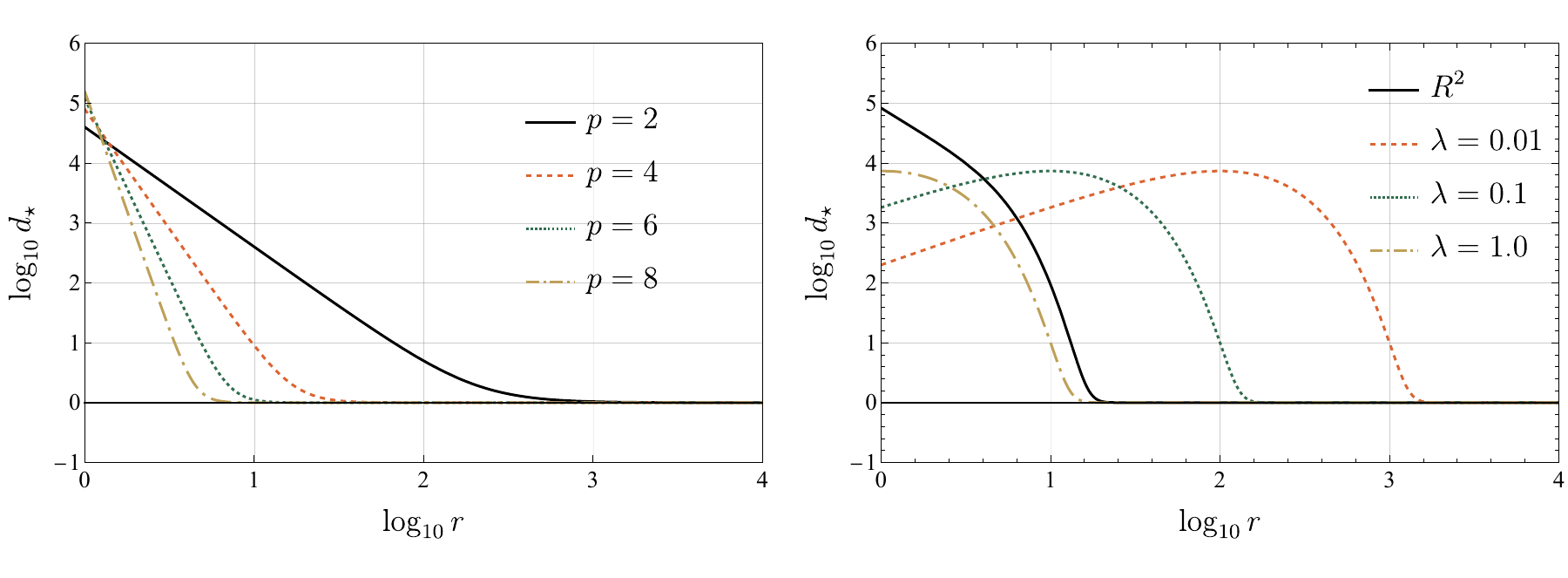}
    \caption{
        The bound on the number of fields as a function of $r$, Eq.~(\ref{eq:critd_gen}), for the monomial model $v (r) = v_{0} r^{p}$ (\textit{left}), for the $R^{2}$-type model and the power-law model (\textit{right}). 
        In all the models, the overall factor in the potentials is fixed to be $v_{0} = 10^{-4}$. 
    }
    \label{fig:bd}
\end{figure*}

The competition between the classical drift force and the noise-induced centrifugal force can be observed more intuitively by introducing the effective potential, 
\begin{equation}
    v_{\rm E} (r) 
    = 
    \ln \qty( v_{0} \frac{d-1}{2 p} r^{p} ) 
    - v_{0} \frac{d-1}{2 p} r^{p} 
    + 1 
    \,\, . 
    \label{eq:demo_monq_vef}
\end{equation}
From Eq.~(\ref{eq:demo_monq_vef}), the effective force terms in Eq.~(\ref{eq:d_mean_geno_lan}) are derived according to $f_{\rm E} (r) = - \dd v_{\rm E} (r) / \dd r$, which is displayed in the top-left panel in Figure~\ref{fig:fneff}. 
The effective potential (\ref{eq:demo_monq_vef}) shown in the top-right panel in Figure~\ref{fig:fneff} is normalised in such a way that 
the maximum becomes zero, \textit{i.e.}, 
$v_{\rm E} (\widetilde{r}) = 0$, where 
\begin{equation}
    \widetilde{r} 
    = \qty[ 
        \frac{2 p}{ v_{0} (d-1) } 
    ]^{1/p} 
    \label{eq:demo_monq_rtil}
\end{equation}
satisfies $f_{\rm E} (\widetilde{r}) = 0$, \textit{i.e.}, it is the location where the two forces are balanced. 
It should be mentioned, for the monomial model (\ref{eq:m1_pot}), that $f_{\rm E} (r)$ is a monotonically increasing function, and $\widetilde{r}$ is thus determined uniquely for the monomial model. 
When $r > \widetilde{r}$, the noise-induced force dominates over the gradient force, so that $f_{\rm E} (r) > 0$. 
This means that $r$ on average ascends the potential, resulting in eternal inflation deterministically. 
For given $v_{0}$ and $p$, the critical location $\widetilde{r}$ becomes smaller and smaller when more and more fields are prevailing, as can be seen from Eq.~(\ref{eq:demo_monq_rtil}). 

The discussion can be generalised to a class of $O (d)$-symmetric models. 
For this purpose, let us introduce the effective force function defined through Eq.~(\ref{eq:pre_lan_r}), by 
\begin{equation}
	f_{\rm E} (r) 
	= - \dv{v_{\rm E} (r)}{r} 
	\equiv - \frac{\dd v (r) / \dd r}{v (r)} + \frac{v (r)}{2} \frac{d-1}{r} 
    \,\, . 
    \label{eq:d_mean_veff}
\end{equation}
With Eq.~(\ref{eq:d_mean_veff}), the classification (\ref{eq:dm_mon_class}) is generalised to $f_{\rm E} (r) \gtrless 0$. 
When $f_{\rm E} (r) < 0$, the inflatons roll down the potential on average, resulting in a finite number of $e$-folds. 
On the other hand, the continuously ascending motion is realised if $f_{\rm E} (r) > 0$, leading to eternal inflation. 
The situations can therefore be summarised as 
\begin{subequations}
\label{eq:d_ninf_class}
\begin{align}
		&\expval*{ \mathcal{N} } (r) 
		< \infty \,\, , &~\text{if}\quad f_{\rm E} (r) < 0 
		\,\, ; 
		\\[2.0ex] 
		&\expval*{ \mathcal{N} } (r) 
		= \infty \,\, , &~\text{if}\quad f_{\rm E} (r) \geq 0 
		\,\, , 
\end{align}
\end{subequations}
for a given initial location $r$. 
This implies that, the more fields are prevailing, the more finely tuned initial location is required for a finite number of $e$-folds to be realised. 
The classification (\ref{eq:d_ninf_class}) can equivalently be stated in terms of the number of fields $d$, for a set of the given parameters including $r$ once an $O (d)$-symmetric model of inflation is fixed. 
If $v (r)$ is a function solely of $r$ and does not explicitly depend on $d$, the critical number of fields is given by 
\begin{equation}
    d_{\star} 
    = 1 + 2 r \, \frac{ \dd v (r) / \dd r }{ [ v (r) ]^{2} } 
    = 1 - 2 r \, \dv{r} \qty[ \frac{1}{v (r)} ] 
    \,\, . 
    \label{eq:critd_gen}
\end{equation}
This is the purely theoretical bound on the number of fields $d$ from above, as illustrated in Ref.~\cite{Takahashi:2025hqt}. 

The left panel in Figure~\ref{fig:bd} shows the bound on the number of fields, Eq.~(\ref{eq:critd_gen}), for the monomial model (\ref{eq:m1_pot}). 
The quadratic monomial model corresponds to the black curve, while other curves to different powers $p$. 
As the field location $r$ becomes large, the allowed number of fields $d_{\star}$ monotonically becomes smaller, since for $r \gg 1$ the gradient force proportional to $1/r$ is small, and thus the noise-induced centrifugal force can easily dominate over the gradient force. 
For each $r$, the upper bound on $d_{\star}$ to obtain a finite number of $e$-folds becomes smaller as the potential becomes steeper, that is, the power $p$ becomes larger. 
This is because the noise-induced centrifugal force is proportional to $v (r)$, see Eq.~(\ref{eq:critd_gen}), so that a larger $p$ can more easily result in the domination of the noise-induced force. 

In summary, the existence of the radius of convergence originates from the fact that, when there are more than one fields, not only the classical drift force but also the noise-induced centrifugal force affect the evolution of $r$. 
In particular, the latter force either partially reduces the former, or dominates over the former. 
In addition to the small-noise assumption $v \ll 1$, the validity of the perturbative expansion requires $v d / 2 p \ll 1$, which is equivalent to the condition that the noise-induced force never dominates over the classical drift force. 
In such a regime, the inflaton fields still roll down the potential on average, although more slowly than classically expected due to the weakened dragging force, and the finite number of $e$-folds may thus be realised. 
On the other hand, the inflaton fields deterministically ascend the potential in the opposite regime. 
The critical number of fields resides at the boundary of the two regimes.\footnote{
    One finds that our critical number of fields $d_{\star}$ is obtained solely from the effective potential with the centrifugal-force term, and has nothing to do with the presence or absence of the reflective boundary. 
    This is a fundamentally different quantity from the one obtained in Ref.~\cite{Vennin:2016wnk}, which is controlled 
    by the procedure of taking $r_+$ to infinity. 
}

\subsection{Quadratic monomial model}
\label{subsec:demo_q}

To further illustrate the multifield dynamics, let us consider the model that consists solely of the mass terms, 
\begin{equation}
    V (\vb*{\phi}) 
    = \sum_{i = 1}^{d} \frac{m_{i}^{2}}{2} \phi_{i}^{2} 
    \,\, , 
    \qquad 
    m_{1} = \cdots = m_{d} 
    \,\, , 
    \label{eq:pre_eqmasspot}
\end{equation}
from which the nondimensionalised potential is given by $v (r) = v_{0} r^{2}$ with $v_{0} = (m_{1} / M_{\rm P})^{2} / 24 \pi^{2}$. 
It corresponds to a $p = 2$ case in Section~\ref{subsec:demo_mon}, so all the concrete formulae summarised there can be directly used. 
The reason why the model (\ref{eq:pre_eqmasspot}) is specifically considered here is that it enables us to observe the field-space evolution analytically, together with the deterministic part of the number of $e$-folds and its divergence at the critical $d$, in the presence of the noise-induced centrifugal force. 

When the stochastic effects are absent (\textit{i.e.}, if both the stochastic noise and the noise-induced centrifugal force for $d > 1$ are neglected), the evolution of $\vb*{r}$ is governed by the standard classical and slow-roll equation of motion, 
\begin{equation}
	\dv{\vb*{r}}{N} = - \frac{2 \vb*{r}}{r^{2}} 
	\,\, . 
	\label{eq:pre_ex_cleom}
\end{equation}
The direction of $\vb*{r}$ does not change in time, and $\vb*{r}$ is absorbed towards the bottom of the potential without detour until the slow-roll condition is violated at $r_{-} = \sqrt{2}$. 
In particular, the motion of $\vb*{r}$ deterministically ends in a finite number of $e$-folds estimated by Eq.~(\ref{eq:rvd_mean0_mon}). 

The solution to Eq.~(\ref{eq:pre_ex_cleom}) gives the classical trajectory, 
\begin{equation}
	r (N) = \sqrt{ r_{0}^{2} - 4 (N - N_{0}) } 
	\,\, . 
	\label{eq:pre_ex_trajc}
\end{equation}
In Eq.~(\ref{eq:pre_ex_trajc}), $r_{0}$ is the initial condition at $N_{0}$ that satisfies $r_{-} \leq r (N) \leq r_{0}$ during the evolution, due to the absence of the stochastic effects. 

Let us now take into account the stochastic effect. 
For a general case with $d \geq 1$, the equation of motion for $r$ is given by Eq.~(\ref{eq:pre_lan_r}), which reduces to Eq.~(\ref{eq:d_mean_geno_lan}) with $p = 2$ for the quadratic monomial model. 
The effective force is given by $f_{\rm E} (r) = - 2 / r + [ v_{0} (d - 1) / 2 ] r$, see also Eq.~(\ref{eq:d_mean_veff}). 

The mean trajectory of $r$ in this situation follows from Eq.~(\ref{eq:d_mean_geno_lan}), with the noise term still being neglected. 
For $p = 2$, the time evolution of $r$ can be chased analytically, 
\begin{align}
	r (N) 
	&= 
	\left\{ 
		\frac{4}{v_{0} (d-1)} 
		+ \qty[ r_{0}^{2} - \frac{4}{v_{0} (d-1)} ] 
	\right. 
	\notag \\[0.5ex] 
	&\quad \quad~\, \left. 
		\times \exp \qty[ \frac{v_{0} (d-1)}{2} (N - N_{0}) ] 
	\right\}^{1/2} 
	\,\, . 
    \label{eq:p2_dtraj}
\end{align}
The deterministic number of $e$-folds in the presence of the noise-induced centrifugal force is then given by~\cite{Takahashi:2025hqt} 
\begin{equation}
	N_{d} (r) 
	\equiv 
	\frac{2}{v_{0} (d-1)} \ln \qty[ 
		\frac{ \displaystyle 1 - \frac{v_{0} r^{2} (d-1)}{4} \qty( \frac{\sqrt{2}}{r} )^{2} }{ \displaystyle 1 - \frac{v_{0} r^{2} (d-1)}{4} \vphantom{ \qty( \frac{\sqrt{2}}{r} )^{2} } } 
	] 
	\,\, . 
	\label{eq:demo_nd}
\end{equation}

\begin{figure*}
    \centering
    \includegraphics[width=0.995\columnwidth]{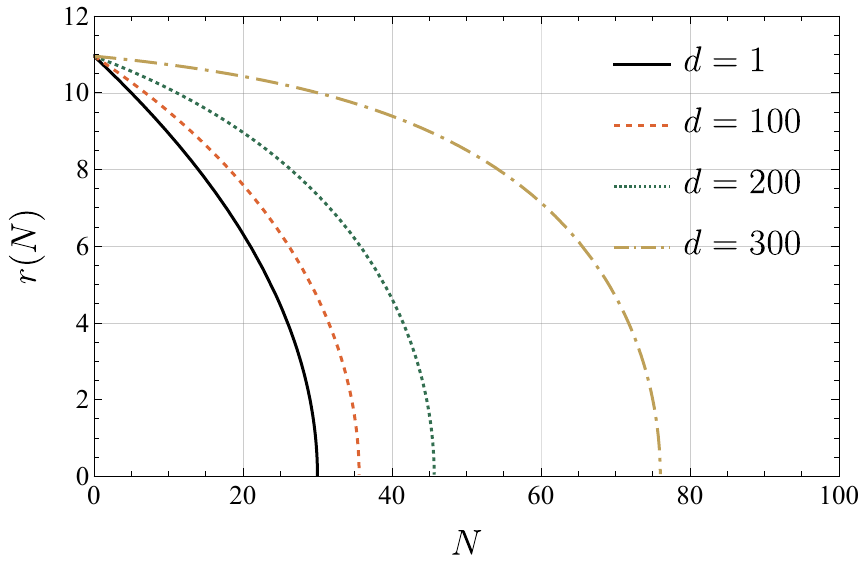} 
    \hfil
    \includegraphics[width=0.995\columnwidth]{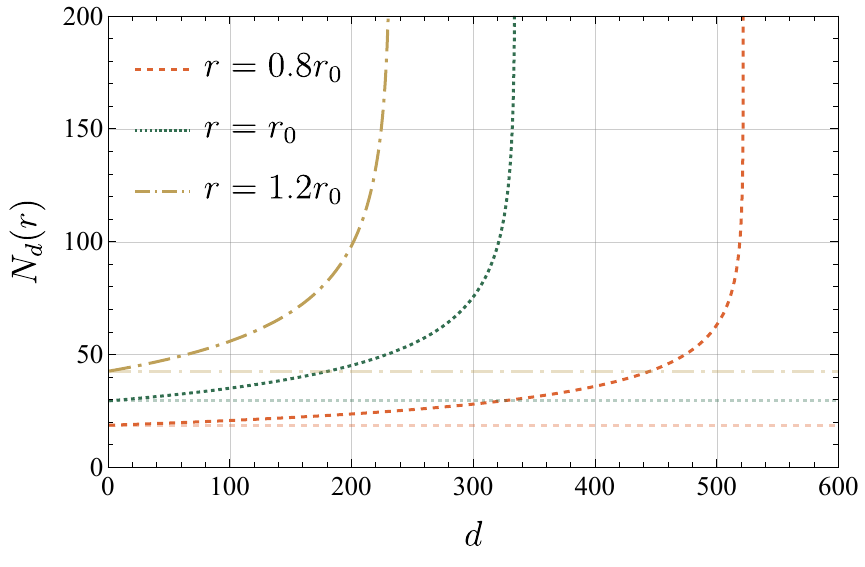} 
    \hfill 
    \caption{
        (\textit{Left}) 
        The deterministic evolution of $r$ for several $d$'s, Eq.~(\ref{eq:p2_dtraj}). 
        When more and more fields are present, the velocity of $r$ decreases. 
        (\textit{Right}) 
        The deterministic number of $e$-folds as a function of $d$, Eq.~(\ref{eq:demo_nd}), for several initial conditions. 
        In both figures, the fiducial initial location is fixed to be $r_{0} = \sqrt{4 \times 30} \approx 11$ for the quadratic monomial model, as well as $v_{0} = 10^{-4}$. 
    }
    \label{fig:fsas}
\end{figure*}

The left panel in Figure~\ref{fig:fsas} shows the evolution of $r$ for several $d$, as a function of time. 
The black solid curve is the classical trajectory, Eq.~(\ref{eq:pre_ex_trajc}). 
This is the case where the stochastic effects, \textit{i.e.}, both the stochastic noise and the noise-induced centrifugal force, are absent for an arbitrary $d$, or equivalently, the stochastic noise is switched off when only a single field is present. 
On the other hand, three non-solid curves illustrate the average trajectories for $d = 100$ (red dashed), $200$ (green dotted), and $300$ (mustard dotted-dashed). 
For those three curves, the noise-induced centrifugal force becomes relevant in the sense that it partially compensates the gradient force, by which the velocity of $r$ is decreased. 
The number of $e$-folds, Eq.~(\ref{eq:demo_nd}), is thus extended, as illustrated in the green dotted curve in the right panel in Figure~\ref{fig:fsas}. 
This effect becomes more prominent when more fields are prevalent. 
When the number of fields $d$ reaches its critical value, $r$ on average becomes constant in time, depicting the horizontal line $r (N) = r_{0}$ in the left panel. 
It corresponds to the divergent behaviour of $N_{d}$ in the right panel, at which the denominator inside the logarithm in Eq.~(\ref{eq:demo_nd}) vanishes~\cite{Takahashi:2025hqt}. 

The right panel in Figure~\ref{fig:fsas} not only shows such a behaviour of $N_{d}$ at the critical number of fields, but also demonstrates the relevance of the initial location. 
The choice $r_{0} = \sqrt{4 \times 30} \approx 11$ is referred to as our benchmark, following Ref.~\cite{Takahashi:2025hqt}. 
The three horizontal lines show the classically realised number of $e$-folds, which do not depend on the number of fields and are given by Eq.~(\ref{eq:rvd_mean0_mon}). 
The three curves, on the other hand, are the deterministically realised number of $e$-folds that account for the noise-induced centrifugal force, namely Eq.~(\ref{eq:demo_nd}). 
For a given initial condition $r$ at the critical $d$, the classical drift and centrifugal force are balanced to give rise to $f_{\rm E} (r) = 0$, by which $r$ cannot move without the stochastic noise. 
This gives rise to $N_{d} (r) = \infty$. 

The critical number of fields for the model is given by 
\begin{equation}
	d_{\star} 
	= 1 + \frac{4}{v_{0} r^{2}} 
	\,\, , 
\end{equation}
at which, again, $N_{d} (r) = \infty$ in the right panel in Figure~\ref{fig:fsas}. 
It depends on $v_{0}$, $r$, and actually $p = 2$, and is generalised to $d_{\star} = 1 + 2 p / v_{0} r^{p}$ for the monomial potential, $v (r) = v_{0} r^{p}$. 
Those are the special cases of Eq.~(\ref{eq:critd_gen}) for the models. 

The essence regarding the relation between the mean number of $e$-folds $\expval*{\mathcal{N}}$ and the number of fields $d$ has been observed through the monomial model. 
However, even if a class of monotonically increasing function for which $\dd v (r) / \dd r > 0$ is focused on, 
there is a variety of the functional forms of the effective force and the effective potential, as well as the field-value dependence of the critical number of fields. 
Henceforth, other two models are further studied separately in the remainder of this section. 

\subsection{\texorpdfstring{$R^{2}$}{R2}-type model}

Let us now turn our attention to the situation where the radial degree of freedom $r$ has the following potential, 
\begin{equation}
    v (r) 
    = v_{0} \qty[ 
        1 - \exp \qty( - \frac{\sqrt{6}}{3} \, r ) 
    ]^{2} 
    \,\, . 
    \label{eq:demo_sov_v}
\end{equation}
It should be noted that, rather than assuming that each individual field has the $R^{2}$-type potential, an $O (d)$-symmetric situation (\ref{eq:demo_sov_v}) is considered here. 
For $V (\vb*{\phi}) = (3 M_{\rm P}^{2} m^{2} / 4) [ 1 - \exp \qty( - \sqrt{6} \, \norm{ \vb*{\phi} } / 3 M_{\rm P} ) ]^{2}$, the prefactor is determined as $v_{0} = (m / M_{\rm P})^{2} / 16 \pi^{2}$. 

The classical mean number of $e$-folds and the next-order stochastic correction are estimated respectively by 
\begin{subequations}
    \begin{align}
        \dv{ \expval*{ \mathcal{N} }^{(0)} (x) }{ x } 
        &= \frac{\sqrt{6}}{4} \exp \qty( \frac{ \sqrt{6} }{3} \, x ) \qty[ 
            1 - \exp \qty( - \frac{ \sqrt{6} }{3} \, x ) 
        ] 
        \,\, , 
        \\ 
        \dv{ \expval*{ \mathcal{N} }^{(1)} (x) }{ x } 
        &= \frac{v_{0}}{16 x} 
        \exp \qty( - \frac{4}{ \sqrt{6} } \, x ) 
        \qty[ \exp \qty( \frac{2}{ \sqrt{6} } \, x ) - 1 ]^{3} 
        \notag \\[0.5ex] 
        &\quad \times 
        \qty{ 
            3 (d - 1) \qty[ \exp \qty( \frac{2}{\sqrt{6}} \, x ) - 1 ] 
            + \sqrt{6} \, x 
        } 
        \,\, . 
    \end{align}
\end{subequations}
Though the higher-order corrections can also be obtained by virtue of the results in Section~\ref{sec:statistical_moments}, they are omitted here since the derivation is straightforward but the expressions are complicated. 

For the effective force defined by Eq.~(\ref{eq:d_mean_veff}), illustrated in the middle-left panel in Figure~\ref{fig:fneff}, the critical location $\widetilde{r}$ can be obtained numerically. 
The normalised effective potential in this model then reads 
\begin{align}
    v_{\rm E} (r) 
    &= 2 \ln \qty[ 
        \frac{ 
            \displaystyle 
            1 - \exp \qty( - \frac{2}{\sqrt{6}} r ) 
        }{ 
            \displaystyle 
            1 - \exp \qty( - \frac{2}{\sqrt{6}} \widetilde{r} ) 
        }
    ] 
    - v_{0} \frac{d-1}{2} \Bigg[ 
        \ln \qty( \frac{r}{\widetilde{r}} ) 
    \Bigg. 
    \notag \\[0.5ex] 
    &\quad~ \left. 
        + \, 
        \int_{4 \widetilde{r} / \sqrt{6}}^{4 r / \sqrt{6}} \dd t \, \frac{e^{-t}}{t} 
        - 2 
        \int_{2 \widetilde{r} / \sqrt{6}}^{2 r / \sqrt{6}} \dd t \, \frac{e^{-t}}{t} 
    \right] 
    \,\, . 
\end{align}
The middle-right panel in Figure~\ref{fig:fneff} shows the effective potential. 
Both $f_{\rm E} (r)$ and $v_{\rm E} (r)$ in this model approach a constant as $r \to \infty$ contrary to the monomial model. 

The critical number of fields (\ref{eq:critd_gen}) for the model under consideration is given by 
\begin{equation}
    d_{\star} 
    = 1 
    + \frac{r}{\sqrt{6} \, v_{0}} 
    \qty[ 1 + \coth \qty( \frac{r}{\sqrt{6}} ) ] \mathrm{cosech}^{2} \qty( \frac{r}{\sqrt{6}} ) 
    \,\, . 
\end{equation}
It is shown in the right panel in Figure~\ref{fig:bd}. 
Since $d_{\star} = \mathcal{O} (1 / v_{0})$ neglecting $r$, the critical number of fields does not change much compared to the monomial model. 

\subsection{Power-law model}

Finally, let us consider the power-law model, 
\begin{equation}
    v (r) 
    = v_{0} \exp \qty( \lambda r ) 
    \,\, , 
    \qquad 
    \lambda > 0 
    \,\, . 
    \label{eq:m3_pl}
\end{equation}
As in the $R^{2}$-type model, the model in which the radial variable $r$ enters the exponent is considered. 
The classical and higher-order terms of (the derivative of) the mean number of $e$-folds are given by 
\begin{subequations}
    \begin{align}
        \dv{ \expval*{ \mathcal{N} }^{(0)} (x) }{x} 
	    &= \frac{ 1 }{ \lambda }
	    \,\, , 
        \label{eq:rvd_mean0_mon_lambda}
        \\[0.5ex] 
        \dv{ \expval*{ \mathcal{N} }^{(1)} (x) }{x} 
	    &= v_{0} \frac{ d - 1 }{ 2 \lambda^{2} } \frac{ e^{\lambda x} }{x}  
    	\,\, , 
	    \label{eq:rvd_mean1_mon_lambda}
        \\[0.5ex] 
        \dv{ \expval*{ \mathcal{N} }^{(2)} (x) }{x} 
	    &= v_{0}^{2} \frac{ (d - 1) (d - 2 + \lambda x) }{ 4 \lambda^{3} } \frac{ e^{2 \lambda x} }{x^{2}} 
        \label{eq:rvd_mean2_mon_lambda} 
	    \,\, . 
    \end{align}
\end{subequations}

A notable difference from the previous models is that the stochastic corrections vanish at $d = 1$, at least up to second order. 
This statement can generically be confirmed, by virtue of umbral calculus~\cite{roman2019umbral}. 
When $d = 1$, and for the power-law model (\ref{eq:m3_pl}), the general formula for the $k$-th order of $\expval*{ \mathcal{N} }$, Eq.~(\ref{eq:rvd_mean_k_c1}), reduces to 
\begin{align}
    \dv{ \expval*{ \mathcal{N} }^{(k)} (x) }{x} 
    &= 
    \frac{1}{\lambda} 
    \qty[ \frac{v (x)}{2 \lambda} ]^{ k } 
    \sum_{i = 0}^{k} 
    \tikz[baseline={(0, -0.1)}]{
				\fill (0, 0) circle (2.5pt);
  				\node at (0, -0.3) {\scriptsize {$k-i$}};
			} 
    \notag \\[0.5ex] 
    &\quad \times 
    \sum_{j = 0}^{i} 
    \frac{ (k+j)!  }{j! \lambda^{j}} 
    D_{i, \, j} (x) 
    \,\, . 
    \label{eq:mex_mgen}
\end{align}
For $v (r) = v_{0} e^{\lambda r}$, Eq.~(\ref{eq:sdot}) simplifies to $z_{\ell} \equiv \ell ! D_{\ell, \, 1} (x) = (-)^{\ell + 1} \lambda^{\ell + 1} / ( \ell + 1 )$. 
Therefore, it follows from Eq.~(\ref{eq:fnD_Bell}) that $D_{i, \, j} (x) = ( j! / i! ) B_{i, \, j} (z_{1}, \, z_{2}, \, \dots, \, z_{i-j+1})$. 
The summation over $j$ can then be reduced to a closed form, 
\begin{equation}
    \sum_{j = 0}^{i} 
    \frac{ (k+j)!  }{j! \lambda^{j}} 
    D_{i, \, j} (x) 
    = k! \lambda^{i} B_{i}^{(k+1)} (k + 1) 
    \,\, , 
    \label{eq:mex_mid}
\end{equation}
in terms of the generalised Bernoulli polynomial $B_{i}^{(\alpha)} (z)$, defined through its generating function, 
\begin{equation}
    \sum_{i = 0}^{\infty} B_{i}^{(\alpha)} (z) \frac{t^{i}}{i!} 
    = \qty( \frac{t}{e^{t} - 1} )^{\alpha} e^{z t} 
    \,\, . 
\end{equation}
Substitution of Eq.~(\ref{eq:mex_mid}) into Eq.~(\ref{eq:mex_mgen}) gives, for the generalised Bernoulli numbers $B_{i}^{(\alpha)} \equiv B_{i}^{(\alpha)} (z = 0)$, the desired confirmation can be obtained for single-field situations as 
\begin{align}
    \dv{ \expval*{ \mathcal{N} }^{(k)} (x) }{x} 
    &= \frac{1}{\lambda} 
    \qty[ \frac{ v (x) }{2} ]^{k} 
    \sum_{i = 0}^{k} (-)^{k} C (k, \, i) B_{i}^{(k+1)} 
    \notag \\[0.5ex] 
    &= 
    \begin{cases}
        \displaystyle 
        \frac{1}{\lambda} 
        &\quad (k = 0) \,\, , 
        \\[2.0ex] 
        \, 0 
        &\quad (k \geq 1) \,\, . 
    \end{cases}
\end{align}

On the other hand, the stochastic correction terms in the variance does not have such a characteristic. 
Those are given, for instance, by 
\begin{subequations}
    \begin{align}
        \dv{ [ \delta \mathcal{N}^{2} ]^{(1)} (x) }{ x } 
        &\simeq 
        v_{0} \frac{ e^{\lambda x} }{ \lambda^{3} } 
        \,\, , 
        \\[0.5ex] 
        \dv{ [ \delta \mathcal{N}^{2} ]^{(2)} (x) }{ x } 
        &\simeq 
        v_0^{2} \frac{ 3 (d-1) + \lambda x }{ 2 \lambda^{4} } \frac{e^{2 \lambda x}}{x} 
        \,\, , 
    \end{align}
\end{subequations}
As a consequence of those results, the spectral tilt $n_{\rm S}$ experiences the stochastic effect that is present even when $d = 1$, as the previous two models. 
At the leading and next-to-leading orders, the correction terms are given by 
\begin{subequations}
    \begin{align}
        ( n_{\rm S} - 1 )^{(0)} 
        &\simeq - \lambda^{2} 
        \,\, , 
        \\[0.5ex] 
        ( n_{\rm S} - 1 )^{(1)} 
        &\simeq - v_{0} 
        \frac{ (d - 1) ( \lambda x - 2 ) + (\lambda x)^{2} }{2 x^{2}} e^{ \lambda x} 
        \,\, . 
    \end{align}
\end{subequations}
The higher-order correction terms can also be derived. 
Different from the results of the monomial model, see Eqs.~(\ref{eq:mon_nS}), the first-order correction can be either positive or negative depending on the value of $\lambda x$. 

For the model (\ref{eq:m3_pl}), the effective force reads 
\begin{equation}
    f_{\rm E} (r) 
    = - \lambda \qty[ 
        1 - \frac{ v_{0} }{2} (d-1) \frac{ e^{\lambda r} }{ \lambda r } 
    ] 
    \,\, . 
    \label{eq:mSov_fE}
\end{equation}
The bottom-left panel in Figure~\ref{fig:fneff} shows $f_{\rm E} (r) / \lambda$, as a function of $\lambda r$. 
A difference from the previous two models can be found, that it diverges positively as $r \to 0$. 
This is because $- v' (r) / v (r) = - \lambda = \mathrm{const.}$ in the power-law model, so the behaviour for small $r$ is determined by the second term in Eq.~(\ref{eq:d_mean_veff}). 
This feature is in contrast to the previous two models, where the logarithmic derivative of the potential is proportional to $- 1 / r$, diverging negatively as $r \to 0$. 
However, other terms in the potential would often be assumed in the power-law model to terminate inflation since violation of the slow-roll condition may not be realised otherwise. 
Therefore, the diverging behaviour of $f_{\rm E} (r)$ at $r \to 0$ does not play a relevant role in most practical situations. 
Then, the function $f_{\rm E} (r)$ divides the potential into two, in which the inflaton field is driven towards smaller-field direction if $f_{\rm E} (r) < 0$, while it ascends the potential if $f_{\rm E} (r) > 0$. 
In this sense, the situation is similar to the previous two models. 

While the function (\ref{eq:mSov_fE}) has two roots as can be seen in the figure, the larger one is identified with $\widetilde{r}$ for our purposes. 
The effective potential is then given by 
\begin{equation}
    v_{\rm E} (r) 
    = (\lambda r - \lambda \widetilde{r}) 
    - \frac{v_{0}}{2} (d-1) 
    \int_{\lambda \widetilde{r}}^{\lambda r} \dd t \, \frac{e^{t}}{t} 
    \,\, . 
\end{equation}
This is displayed in the bottom-right panel in Figure~\ref{fig:fneff}. 
A common feature with the previous two models is that for a large $\lambda r$ it climbs the potential. 
It is also common that, as long as $v_{0} d \lesssim 1$, there exists a region, for $\lambda r$ smaller than $\widetilde{r}$, where it rolls down. 
However, for much smaller $\lambda r$, the noise-induced centrifugal force dominates over the classical drift, as already seen in Eq.~(\ref{eq:mSov_fE}). 
It should be noted that, in the nearly flat region in the bottom right panel, the derivative of $v_{\rm E} (r)$ is actually negative. 
This behaviour is associated with the feature of $f_{\rm E} (r)$ that it diverges positively as $r \to 0$. 

The above-mentioned second common feature ceases to exist once the condition that $v_{0} d \lesssim 1$ is violated. 
This can be confirmed by observing the minimum of Eq.~(\ref{eq:mSov_fE}), which is given by $- [ 1 - (v_{0} / 2) (d-1) ]$. 
The necessary condition for a successful end of inflation reads 
\begin{equation}
    d < 1 + \frac{2}{v_{0}} 
    \,\, . 
    \label{eq:mp_d0}
\end{equation}
However, one realises that $f_{\rm E} (r) > 0$ for all $r > 0$ in the opposite situation where $d > 1 + 2 / v_{0}$. 
In the latter case, the inflaton field climbs the potential everywhere, due to which a successful termination of inflation cannot be realised unless rare stochastic fluctuations occur by chance. 

Even if the condition (\ref{eq:mp_d0}) is satisfied, termination of inflation is not necessarily guaranteed. 
However, the situation is similar to the previous two models in the sense that the region on the potential is divided into two with $f_{\rm E} (r) \gtrless 0$, as already mentioned above. 
For each given $r$, a successful termination of inflation can be realised when the number of the fields is smaller than the critical value, 
\begin{equation}
    d_{\star} 
    = 1 + \frac{2 \lambda r}{v_{0}} e^{- \lambda r} 
    \,\, . 
    \label{eq:m2_dcrit}
\end{equation}
The right panel in Figure~\ref{fig:bd} shows Eq.~(\ref{eq:m2_dcrit}), as a function of $r$. 
Different from the previous two models,  $d_{\star}$ is not a monotonically decreasing function with respect to $r$. 
This is because, the critical number of fields is given by $d_{\star} - 1 = [ 2 r / v (r) ] \cdot [ \dd v (r) / \dd r ] / v (r) = [ v (r) / 2 r ]^{-1} \cdot \dd \ln v (r) / \dd r$, which consists of a non-monotonic function for the power-law model, whereas it is a multiplication of two monotonically decreasing functions for the previous two models. 
Therefore, the critical number of fields $d_{\star}$ for the monomial model and the $R^{2}$-type model decreases as the field value $r$ gets large, while the maximum number of fields is allowed for $r = 1 / \lambda$ for the power-law model. 

In summary, the number of inflaton fields $d$ is irrelevant when the dynamics is analysed at the classical level for $O (d)$-symmetric models. 
Once the stochastic effect is taken into account, however, the statistical moments of the number of $e$-folds are substantially affected, and in particular there exists the critical number of fields that distinguishes a successful end of inflation or eternally inflating evolution. 

\section{Conclusion}
\label{sec:conclusion}

Small-scale quantum fluctuations 
stochastically affect the evolution of the large-scale field configurations, whose dynamics may be altered due to those stochastic effects especially in multifield models. 
This article is in particular interested in the extension of the mean number of $e$-folds that explicitly depends on the number of fields, as well as influences on the higher-order statistical moments that are directly related to the observable quantities. 
While a variety of multifield models has been proposed so far, this article restricts itself to a class of $O (d)$-symmetric ones in order to focus on the essential features and to maintain technical simplicity. 
The stochastic-inflation formalism is fully exploited for our purpose, deriving the perturbative results that are valid at arbitrary order from the non-perturbative solution to the governing equation. 

Within the small-noise regime where the stochastic effect does not dominate over the classical evolution, in Section~\ref{sec:smex}, the all-order perturbative expansion (\ref{eq:sngen_main}) has been derived. 
It reveals the recursive structure of the statistical moments of the number of $e$-folds, keeping the dependence on the number of fields analytical. 
Despite its form involving multiple sums, the relevant terms at a given perturbative order can be collected recursively and systematically, 
as is demonstrated in Section~\ref{sec:statistical_moments}. 
A concrete functional form of the $O (d)$-symmetric potential has not been specified in those derived results, meaning that they can immediately be applied to concrete models. 
The derived perturbative results clarify how the number of fields arises in the statistical moments. 

The emergence of the explicit $d$-dependence in those quantities can be intuitively understood by observing the effective Langevin equation (\ref{eq:pre_lan_r}) for the radial degree of freedom. 
There, the noise-induced centrifugal force is one-order higher than the classical drift force, and is thus absent at the classical level. 
When the former is taken into account, however, it partly compensates the latter collective force and the net force that the radial degree of freedom receives is then weakened, resulting an extension of the elapsed mean number of $e$-folds. 
With this viewpoint, the regime in which the perturbative expansion remains valid can be reinterpreted as the condition that the classical drift force is dominant. 
This, in turn, is equivalent to the condition that inflation should on average be terminated, giving rise to our observable Universe. 
In order to realise such a successful termination of inflation, a theoretical bound on the number of fields is derived in Section~\ref{sec:implications}. 

What is extended due to the stochastic effect is not restricted to the mean number of $e$-folds, which is also demonstrated in Section~\ref{sec:statistical_moments}. 
For instance, the variance is also increased as the mean becomes larger, again depending on the number of fields. 
Though the correction terms in the higher-order statistical moments are in general very complicated, the all-order mean number of $e$-folds were obtained for the monomial model in Section~\ref{sec:implications}, and enables us to find the radius of convergence for
the perturbative expansion. 
The existence of the radius of convergence can again be attributed to the balance condition between the gradient force and the noise-induced centrifugal force. 
The latter collective force is due to stochastic effects, and to the prevalence of a number of fields. 
Even if the condition that one is inside the small-noise regime (that is, $v_{0} \ll 1$ for the monomial potential), accumulation of the fields can give rise to the domination of the collective force. 
In such cases, the inflaton fields continue to ascend the potential to results in eternal inflation with $\expval*{ \mathcal{N} } = \infty$. 
It should be emphasised that this infinite number of $e$-folds is fundamentally different from those known in the literature. 

An interesting direction includes scrutinising the effects of the number of fields $d$ on the tail, relevant for the formation of primordial black holes. 
Since the gradient force is partly compensated by the noise-induced centrifugal force, the velocity of the inflaton fields is decreased, giving rise to a heavier tail and an enhancement of fluctuations. 
Another direction consists of extensions of our analysis to a more general class of models, in particular without $O(d)$ symmetry. 
In such cases, for instance, several mass spectra can be implemented 
(see \textit{e.g.}~Refs.~\cite{Easther:2005zr, Kim:2007bc, McAllister:2012am}). 
Direct numerical simulations of the stochastic equations would be required there, which may also be an interesting avenue to explore.\\ 

\begin{acknowledgments}
    This work was supported by 
    JSPS KAKENHI Grant Numbers 25K01004 (TT) and 24K22877 (KT), 
    MEXT KAKENHI Grant Number 23H04515 (TT) and 25H01543 (TT), 
    and JSPS Overseas Research Fellowships (KT). 
\end{acknowledgments}

\onecolumngrid

\appendix 

\renewcommand{\thesection}{}
\section{Relevant terms to be included in perturbative calculations}

A bunch of terms must be collected in the derived formula, Eq.~(\ref{eq:rvd_mean_master}) in Section~\ref{subsec:mean} and Eq.~(\ref{eq:gen_main}) in Section~\ref{subsec:var} for the first and second moment respectively. 
This Appendix provides, at each order in $v$, the combinations of dummy indices in those formulas. 

\renewcommand{\theequation}{A\arabic{equation}}

\subsection{Mean}
\label{appx:cb_n1}

The relevant terms for the calculation of the mean number of $e$-folds are listed below. 

\begin{itemize}
    \item 
        $N_{1} (0) = 1$ term in $\mathcal{O} (v^{0})$. 
        \begin{fleqn}[40pt]
            \begin{align}
                (k_{1}, \, k_{2}, \, k_{3}, \, k_{4}) 
                = 
                (0, \, 0, \, 0, \, 0) 
                . 
            \end{align}
        \end{fleqn}

    \item 
        $N_{1} (1) = 3$ terms in $\mathcal{O} (v^{1})$. 
        \begin{fleqn}[40pt]
            \begin{align}
                (k_{1}, \, k_{2}, \, k_{3}, \, k_{4}) 
                = 
                (0, \, 0, \, 0, \, 1)
                , \,\, 
                (0, \, 1, \, 1, \, 2)
                , \,\, 
                (1, \, 0, \, 0, \, 1) 
                . 
            \end{align}
        \end{fleqn}

    \item 
        $N_{1} (2) = 7$ terms in $\mathcal{O} (v^{2})$. 
        \begin{fleqn}[40pt]
            \begin{align}
                (k_{1}, \, k_{2}, \, k_{3}, \, k_{4}) 
                &= 
                (0, \, 0, \, 0, \, 2) 
                , \,\, 
                (0, \, 1, \, 1, \, 3) 
                , \,\, 
                (0, \, 2, \, 1, \, 3) 
                , \,\, 
                (0, \, 2, \, 2, \, 4) 
                , \,\, 
                (1, \, 0, \, 0, \, 2) 
                , \,\, 
                (1, \, 1, \, 1, \, 3) 
                , \,\, 
                \notag \\ 
                &\phantom{{}={}} 
                (2, \, 0, \, 0, \, 2) 
                . 
            \end{align}
        \end{fleqn}

    \item 
        $N_{1} (3) = 14$ terms in $\mathcal{O} (v^{3})$. 
        \begin{fleqn}[40pt]
            \begin{align}
                (k_{1}, \, k_{2}, \, k_{3}, \, k_{4}) 
                &= 
                (0, \, 0, \, 0, \, 3) 
                , \,\, 
                (0, \, 1, \, 1, \, 4) 
                , \,\, 
                (0, \, 2, \, 1, \, 4) 
                , \,\, 
                (0, \, 2, \, 2, \, 5) 
                , \,\, 
                (0, \, 3, \, 1, \, 4) 
                , \,\, 
                (0, \, 3, \, 2, \, 5) 
                , \,\, 
                \notag \\ 
                &\phantom{{}={}} 
                (0, \, 3, \, 3, \, 6) 
                , \,\, 
                (1, \, 0, \, 0, \, 3) 
                , \,\, 
                (1, \, 1, \, 1, \, 4) 
                , \,\, 
                (1, \, 2, \, 1, \, 4) 
                , \,\, 
                (1, \, 2, \, 2, \, 5) 
                , \,\, 
                (2, \, 0, \, 0, \, 3) 
                , \,\, 
                \notag \\ 
                &\phantom{{}={}} 
                (2, \, 1, \, 1, \, 4) 
                , \,\, 
                (3, \, 0, \, 0, \, 3)                
                . 
            \end{align}
        \end{fleqn}
\end{itemize}

\subsection{Variance} 
\label{appx:cb_n2}

The relevant terms for the calculation of the second moment of $\mathcal{N}$ are listed below. 
Collecting those terms enables one to obtain the perturbative expansion of $\expval*{ \mathcal{N}^{2} }$, from which the variance $\delta \mathcal{N}^{2}$ can be derived. 
Though not listed here, a similar counting can also be performed for the formula (\ref{eq:var_ex_gen}), which directly derives the perturbative expansion of the variance. 

\begin{itemize}
    \item 
        $N_{2} (0) = 1$ term in $\mathcal{O} (v^{0})$. 
        \begin{fleqn}[40pt]
            \begin{align}
                (k_{1}, \, k_{2}, \, k_{3}, \, k_{4}, \, k_{5}, \, k_{6}) 
                = 
                (0, \, 0, \, 0, \, 0, \, 0, \, 0) 
                . 
            \end{align}
        \end{fleqn}

    \item 
        $N_{2} (1) = 5$ terms in $\mathcal{O} (v^{1})$. 
        \begin{fleqn}[40pt]
            \begin{align}
                (k_{1}, \, k_{2}, \, k_{3}, \, k_{4}, \, k_{5}, \, k_{6}) 
                &= 
                (0, \, 0, \, 0, \, 0, \, 0, \, 1) 
                , \,\, 
                (0, \, 0, \, 0, \, 0, \, 1, \, 0) 
                , \,\, 
                (0, \, 0, \, 0, \, 1, \, 0, \, 1) 
                , \,\, 
                (0, \, 1, \, 1, \, 0, \, 0, \, 2) 
                , \,\, 
                \notag \\ 
                &\phantom{{}={}} 
                (1, \, 0, \, 0, \, 0, \, 0, \, 1) 
                . 
            \end{align}
        \end{fleqn}

    \item 
        $N_{2} (2) = 16$ terms in $\mathcal{O} (v^{2})$. 
        \begin{fleqn}[40pt]
            \begin{align}
                (k_{1}, \, k_{2}, \, k_{3}, \, k_{4}, \, k_{5}, \, k_{6}) 
                &= 
                (0, \, 0, \, 0, \, 0, \, 0, \, 2) 
                , \,\, 
                (0, \, 0, \, 0, \, 0, \, 1, \, 1) 
                , \,\, 
                (0, \, 0, \, 0, \, 0, \, 2, \, 0) 
                , \,\, 
                (0, \, 0, \, 0, \, 1, \, 0, \, 2) 
                , \,\, 
                \notag \\ 
                &\phantom{{}={}} 
                (0, \, 0, \, 0, \, 1, \, 1, \, 1) 
                , \,\, 
                (0, \, 0, \, 0, \, 2, \, 0, \, 2) 
                , \,\, 
                (0, \, 1, \, 1, \, 0, \, 0, \, 3) 
                , \,\, 
                (0, \, 1, \, 1, \, 0, \, 1, \, 2) 
                , \,\, 
                \notag \\ 
                &\phantom{{}={}} 
                (0, \, 1, \, 1, \, 1, \, 0, \, 3) 
                , \,\, 
                (0, \, 2, \, 1, \, 0, \, 0, \, 3) 
                , \,\, 
                (0, \, 2, \, 2, \, 0, \, 0, \, 4) 
                , \,\, 
                (1, \, 0, \, 0, \, 0, \, 0, \, 2) 
                , \,\, 
                \notag \\ 
                &\phantom{{}={}} 
                (1, \, 0, \, 0, \, 0, \, 1, \, 1) 
                , \,\, 
                (1, \, 0, \, 0, \, 1, \, 0, \, 2) 
                , \,\, 
                (1, \, 1, \, 1, \, 0, \, 0, \, 3) 
                , \,\, 
                (2, \, 0, \, 0, \, 0, \, 0, \, 2) 
                . 
            \end{align}
        \end{fleqn}
\end{itemize}
The $N_{2} (3) = 41$ terms are to be included in the third-order calculation that derives Eq.~(\ref{eq:rvd_var_o3}), however they are omitted here to avoid clutter. 

\twocolumngrid
\bibliography{apssamp}

\begin{thebibliography}{93}%
\makeatletter
\providecommand \@ifxundefined [1]{%
 \@ifx{#1\undefined}
}%
\providecommand \@ifnum [1]{%
 \ifnum #1\expandafter \@firstoftwo
 \else \expandafter \@secondoftwo
 \fi
}%
\providecommand \@ifx [1]{%
 \ifx #1\expandafter \@firstoftwo
 \else \expandafter \@secondoftwo
 \fi
}%
\providecommand \natexlab [1]{#1}%
\providecommand \enquote  [1]{``#1''}%
\providecommand \bibnamefont  [1]{#1}%
\providecommand \bibfnamefont [1]{#1}%
\providecommand \citenamefont [1]{#1}%
\providecommand \href@noop [0]{\@secondoftwo}%
\providecommand \href [0]{\begingroup \@sanitize@url \@href}%
\providecommand \@href[1]{\@@startlink{#1}\@@href}%
\providecommand \@@href[1]{\endgroup#1\@@endlink}%
\providecommand \@sanitize@url [0]{\catcode `\\12\catcode `\$12\catcode
  `\&12\catcode `\#12\catcode `\^12\catcode `\_12\catcode `\%12\relax}%
\providecommand \@@startlink[1]{}%
\providecommand \@@endlink[0]{}%
\providecommand \url  [0]{\begingroup\@sanitize@url \@url }%
\providecommand \@url [1]{\endgroup\@href {#1}{\urlprefix }}%
\providecommand \urlprefix  [0]{URL }%
\providecommand \Eprint [0]{\href }%
\providecommand \doibase [0]{https://doi.org/}%
\providecommand \selectlanguage [0]{\@gobble}%
\providecommand \bibinfo  [0]{\@secondoftwo}%
\providecommand \bibfield  [0]{\@secondoftwo}%
\providecommand \translation [1]{[#1]}%
\providecommand \BibitemOpen [0]{}%
\providecommand \bibitemStop [0]{}%
\providecommand \bibitemNoStop [0]{.\EOS\space}%
\providecommand \EOS [0]{\spacefactor3000\relax}%
\providecommand \BibitemShut  [1]{\csname bibitem#1\endcsname}%
\let\auto@bib@innerbib\@empty
\bibitem [{\citenamefont {Starobinsky}(1980)}]{Starobinsky:1980te}%
  \BibitemOpen
  \bibfield  {author} {\bibinfo {author} {\bibfnamefont {A.~A.}\ \bibnamefont
  {Starobinsky}},\ }\bibfield  {title} {\bibinfo {title} {{A New Type of
  Isotropic Cosmological Models Without Singularity}},\ }\href
  {https://doi.org/10.1016/0370-2693(80)90670-X} {\bibfield  {journal}
  {\bibinfo  {journal} {Phys. Lett. B}\ }\textbf {\bibinfo {volume} {91}},\
  \bibinfo {pages} {99} (\bibinfo {year} {1980})}\BibitemShut {NoStop}%
\bibitem [{\citenamefont {Sato}(1981)}]{Sato:1980yn}%
  \BibitemOpen
  \bibfield  {author} {\bibinfo {author} {\bibfnamefont {K.}~\bibnamefont
  {Sato}},\ }\bibfield  {title} {\bibinfo {title} {{First Order Phase
  Transition of a Vacuum and Expansion of the Universe}},\ }\href@noop {}
  {\bibfield  {journal} {\bibinfo  {journal} {Mon. Not. Roy. Astron. Soc.}\
  }\textbf {\bibinfo {volume} {195}},\ \bibinfo {pages} {467} (\bibinfo {year}
  {1981})}\BibitemShut {NoStop}%
\bibitem [{\citenamefont {Guth}(1981)}]{Guth:1980zm}%
  \BibitemOpen
  \bibfield  {author} {\bibinfo {author} {\bibfnamefont {A.~H.}\ \bibnamefont
  {Guth}},\ }\bibfield  {title} {\bibinfo {title} {{The Inflationary Universe:
  A Possible Solution to the Horizon and Flatness Problems}},\ }\href
  {https://doi.org/10.1103/PhysRevD.23.347} {\bibfield  {journal} {\bibinfo
  {journal} {Phys. Rev. D}\ }\textbf {\bibinfo {volume} {23}},\ \bibinfo
  {pages} {347} (\bibinfo {year} {1981})}\BibitemShut {NoStop}%
\bibitem [{\citenamefont {Linde}(1982{\natexlab{a}})}]{Linde:1981mu}%
  \BibitemOpen
  \bibfield  {author} {\bibinfo {author} {\bibfnamefont {A.~D.}\ \bibnamefont
  {Linde}},\ }\bibfield  {title} {\bibinfo {title} {{A New Inflationary
  Universe Scenario: A Possible Solution of the Horizon, Flatness, Homogeneity,
  Isotropy and Primordial Monopole Problems}},\ }\href
  {https://doi.org/10.1016/0370-2693(82)91219-9} {\bibfield  {journal}
  {\bibinfo  {journal} {Phys. Lett. B}\ }\textbf {\bibinfo {volume} {108}},\
  \bibinfo {pages} {389} (\bibinfo {year} {1982}{\natexlab{a}})}\BibitemShut
  {NoStop}%
\bibitem [{\citenamefont {Albrecht}\ and\ \citenamefont
  {Steinhardt}(1982)}]{Albrecht:1982wi}%
  \BibitemOpen
  \bibfield  {author} {\bibinfo {author} {\bibfnamefont {A.}~\bibnamefont
  {Albrecht}}\ and\ \bibinfo {author} {\bibfnamefont {P.~J.}\ \bibnamefont
  {Steinhardt}},\ }\bibfield  {title} {\bibinfo {title} {{Cosmology for Grand
  Unified Theories with Radiatively Induced Symmetry Breaking}},\ }\href
  {https://doi.org/10.1103/PhysRevLett.48.1220} {\bibfield  {journal} {\bibinfo
   {journal} {Phys. Rev. Lett.}\ }\textbf {\bibinfo {volume} {48}},\ \bibinfo
  {pages} {1220} (\bibinfo {year} {1982})}\BibitemShut {NoStop}%
\bibitem [{\citenamefont {Linde}(1983)}]{Linde:1983gd}%
  \BibitemOpen
  \bibfield  {author} {\bibinfo {author} {\bibfnamefont {A.~D.}\ \bibnamefont
  {Linde}},\ }\bibfield  {title} {\bibinfo {title} {{Chaotic Inflation}},\
  }\href {https://doi.org/10.1016/0370-2693(83)90837-7} {\bibfield  {journal}
  {\bibinfo  {journal} {Phys. Lett. B}\ }\textbf {\bibinfo {volume} {129}},\
  \bibinfo {pages} {177} (\bibinfo {year} {1983})}\BibitemShut {NoStop}%
\bibitem [{\citenamefont {Mukhanov}\ and\ \citenamefont
  {Chibisov}(1981)}]{Mukhanov:1981xt}%
  \BibitemOpen
  \bibfield  {author} {\bibinfo {author} {\bibfnamefont {V.~F.}\ \bibnamefont
  {Mukhanov}}\ and\ \bibinfo {author} {\bibfnamefont {G.~V.}\ \bibnamefont
  {Chibisov}},\ }\bibfield  {title} {\bibinfo {title} {{Quantum Fluctuations
  and a Nonsingular Universe}},\ }\href@noop {} {\bibfield  {journal} {\bibinfo
   {journal} {JETP Lett.}\ }\textbf {\bibinfo {volume} {33}},\ \bibinfo {pages}
  {532} (\bibinfo {year} {1981})}\BibitemShut {NoStop}%
\bibitem [{\citenamefont {Starobinsky}(1982)}]{Starobinsky:1982ee}%
  \BibitemOpen
  \bibfield  {author} {\bibinfo {author} {\bibfnamefont {A.~A.}\ \bibnamefont
  {Starobinsky}},\ }\bibfield  {title} {\bibinfo {title} {{Dynamics of Phase
  Transition in the New Inflationary Universe Scenario and Generation of
  Perturbations}},\ }\href {https://doi.org/10.1016/0370-2693(82)90541-X}
  {\bibfield  {journal} {\bibinfo  {journal} {Phys. Lett. B}\ }\textbf
  {\bibinfo {volume} {117}},\ \bibinfo {pages} {175} (\bibinfo {year}
  {1982})}\BibitemShut {NoStop}%
\bibitem [{\citenamefont {Guth}\ and\ \citenamefont {Pi}(1982)}]{Guth:1982ec}%
  \BibitemOpen
  \bibfield  {author} {\bibinfo {author} {\bibfnamefont {A.~H.}\ \bibnamefont
  {Guth}}\ and\ \bibinfo {author} {\bibfnamefont {S.~Y.}\ \bibnamefont {Pi}},\
  }\bibfield  {title} {\bibinfo {title} {{Fluctuations in the New Inflationary
  Universe}},\ }\href {https://doi.org/10.1103/PhysRevLett.49.1110} {\bibfield
  {journal} {\bibinfo  {journal} {Phys. Rev. Lett.}\ }\textbf {\bibinfo
  {volume} {49}},\ \bibinfo {pages} {1110} (\bibinfo {year}
  {1982})}\BibitemShut {NoStop}%
\bibitem [{\citenamefont {Bardeen}\ \emph {et~al.}(1983)\citenamefont
  {Bardeen}, \citenamefont {Steinhardt},\ and\ \citenamefont
  {Turner}}]{Bardeen:1983qw}%
  \BibitemOpen
  \bibfield  {author} {\bibinfo {author} {\bibfnamefont {J.~M.}\ \bibnamefont
  {Bardeen}}, \bibinfo {author} {\bibfnamefont {P.~J.}\ \bibnamefont
  {Steinhardt}},\ and\ \bibinfo {author} {\bibfnamefont {M.~S.}\ \bibnamefont
  {Turner}},\ }\bibfield  {title} {\bibinfo {title} {{Spontaneous Creation of
  Almost Scale - Free Density Perturbations in an Inflationary Universe}},\
  }\href {https://doi.org/10.1103/PhysRevD.28.679} {\bibfield  {journal}
  {\bibinfo  {journal} {Phys. Rev. D}\ }\textbf {\bibinfo {volume} {28}},\
  \bibinfo {pages} {679} (\bibinfo {year} {1983})}\BibitemShut {NoStop}%
\bibitem [{\citenamefont {Starobinsky}(1979)}]{Starobinsky:1979ty}%
  \BibitemOpen
  \bibfield  {author} {\bibinfo {author} {\bibfnamefont {A.~A.}\ \bibnamefont
  {Starobinsky}},\ }\bibfield  {title} {\bibinfo {title} {{Spectrum of relict
  gravitational radiation and the early state of the universe}},\ }\href@noop
  {} {\bibfield  {journal} {\bibinfo  {journal} {JETP Lett.}\ }\textbf
  {\bibinfo {volume} {30}},\ \bibinfo {pages} {682} (\bibinfo {year}
  {1979})}\BibitemShut {NoStop}%
\bibitem [{\citenamefont {Polarski}\ and\ \citenamefont
  {Starobinsky}(1996)}]{Polarski:1995jg}%
  \BibitemOpen
  \bibfield  {author} {\bibinfo {author} {\bibfnamefont {D.}~\bibnamefont
  {Polarski}}\ and\ \bibinfo {author} {\bibfnamefont {A.~A.}\ \bibnamefont
  {Starobinsky}},\ }\bibfield  {title} {\bibinfo {title} {{Semiclassicality and
  decoherence of cosmological perturbations}},\ }\href
  {https://doi.org/10.1088/0264-9381/13/3/006} {\bibfield  {journal} {\bibinfo
  {journal} {Class. Quant. Grav.}\ }\textbf {\bibinfo {volume} {13}},\ \bibinfo
  {pages} {377} (\bibinfo {year} {1996})},\ \Eprint
  {https://arxiv.org/abs/gr-qc/9504030} {arXiv:gr-qc/9504030} \BibitemShut
  {NoStop}%
\bibitem [{\citenamefont {Lesgourgues}\ \emph {et~al.}(1997)\citenamefont
  {Lesgourgues}, \citenamefont {Polarski},\ and\ \citenamefont
  {Starobinsky}}]{Lesgourgues:1996jc}%
  \BibitemOpen
  \bibfield  {author} {\bibinfo {author} {\bibfnamefont {J.}~\bibnamefont
  {Lesgourgues}}, \bibinfo {author} {\bibfnamefont {D.}~\bibnamefont
  {Polarski}},\ and\ \bibinfo {author} {\bibfnamefont {A.~A.}\ \bibnamefont
  {Starobinsky}},\ }\bibfield  {title} {\bibinfo {title} {{Quantum to classical
  transition of cosmological perturbations for nonvacuum initial states}},\
  }\href {https://doi.org/10.1016/S0550-3213(97)00224-1} {\bibfield  {journal}
  {\bibinfo  {journal} {Nucl. Phys. B}\ }\textbf {\bibinfo {volume} {497}},\
  \bibinfo {pages} {479} (\bibinfo {year} {1997})},\ \Eprint
  {https://arxiv.org/abs/gr-qc/9611019} {arXiv:gr-qc/9611019} \BibitemShut
  {NoStop}%
\bibitem [{\citenamefont {Akrami}\ \emph
  {et~al.}(2020{\natexlab{a}})\citenamefont {Akrami} \emph
  {et~al.}}]{Planck:2018jri}%
  \BibitemOpen
  \bibfield  {author} {\bibinfo {author} {\bibfnamefont {Y.}~\bibnamefont
  {Akrami}} \emph {et~al.} (\bibinfo {collaboration} {Planck}),\ }\bibfield
  {title} {\bibinfo {title} {{Planck 2018 results. X. Constraints on
  inflation}},\ }\href {https://doi.org/10.1051/0004-6361/201833887} {\bibfield
   {journal} {\bibinfo  {journal} {Astron. Astrophys.}\ }\textbf {\bibinfo
  {volume} {641}},\ \bibinfo {pages} {A10} (\bibinfo {year}
  {2020}{\natexlab{a}})},\ \Eprint {https://arxiv.org/abs/1807.06211}
  {arXiv:1807.06211 [astro-ph.CO]} \BibitemShut {NoStop}%
\bibitem [{\citenamefont {Akrami}\ \emph
  {et~al.}(2020{\natexlab{b}})\citenamefont {Akrami} \emph
  {et~al.}}]{Planck:2019kim}%
  \BibitemOpen
  \bibfield  {author} {\bibinfo {author} {\bibfnamefont {Y.}~\bibnamefont
  {Akrami}} \emph {et~al.} (\bibinfo {collaboration} {Planck}),\ }\bibfield
  {title} {\bibinfo {title} {{Planck 2018 results. IX. Constraints on
  primordial non-Gaussianity}},\ }\href
  {https://doi.org/10.1051/0004-6361/201935891} {\bibfield  {journal} {\bibinfo
   {journal} {Astron. Astrophys.}\ }\textbf {\bibinfo {volume} {641}},\
  \bibinfo {pages} {A9} (\bibinfo {year} {2020}{\natexlab{b}})},\ \Eprint
  {https://arxiv.org/abs/1905.05697} {arXiv:1905.05697 [astro-ph.CO]}
  \BibitemShut {NoStop}%
\bibitem [{\citenamefont {Enqvist}\ and\ \citenamefont
  {Sloth}(2002)}]{Enqvist:2001zp}%
  \BibitemOpen
  \bibfield  {author} {\bibinfo {author} {\bibfnamefont {K.}~\bibnamefont
  {Enqvist}}\ and\ \bibinfo {author} {\bibfnamefont {M.~S.}\ \bibnamefont
  {Sloth}},\ }\bibfield  {title} {\bibinfo {title} {{Adiabatic CMB
  perturbations in pre - big bang string cosmology}},\ }\href
  {https://doi.org/10.1016/S0550-3213(02)00043-3} {\bibfield  {journal}
  {\bibinfo  {journal} {Nucl. Phys. B}\ }\textbf {\bibinfo {volume} {626}},\
  \bibinfo {pages} {395} (\bibinfo {year} {2002})},\ \Eprint
  {https://arxiv.org/abs/hep-ph/0109214} {arXiv:hep-ph/0109214} \BibitemShut
  {NoStop}%
\bibitem [{\citenamefont {Lyth}\ and\ \citenamefont
  {Wands}(2002)}]{Lyth:2001nq}%
  \BibitemOpen
  \bibfield  {author} {\bibinfo {author} {\bibfnamefont {D.~H.}\ \bibnamefont
  {Lyth}}\ and\ \bibinfo {author} {\bibfnamefont {D.}~\bibnamefont {Wands}},\
  }\bibfield  {title} {\bibinfo {title} {{Generating the curvature perturbation
  without an inflaton}},\ }\href
  {https://doi.org/10.1016/S0370-2693(01)01366-1} {\bibfield  {journal}
  {\bibinfo  {journal} {Phys. Lett. B}\ }\textbf {\bibinfo {volume} {524}},\
  \bibinfo {pages} {5} (\bibinfo {year} {2002})},\ \Eprint
  {https://arxiv.org/abs/hep-ph/0110002} {arXiv:hep-ph/0110002} \BibitemShut
  {NoStop}%
\bibitem [{\citenamefont {Moroi}\ and\ \citenamefont
  {Takahashi}(2001)}]{Moroi:2001ct}%
  \BibitemOpen
  \bibfield  {author} {\bibinfo {author} {\bibfnamefont {T.}~\bibnamefont
  {Moroi}}\ and\ \bibinfo {author} {\bibfnamefont {T.}~\bibnamefont
  {Takahashi}},\ }\bibfield  {title} {\bibinfo {title} {{Effects of
  cosmological moduli fields on cosmic microwave background}},\ }\href
  {https://doi.org/10.1016/S0370-2693(01)01295-3} {\bibfield  {journal}
  {\bibinfo  {journal} {Phys. Lett. B}\ }\textbf {\bibinfo {volume} {522}},\
  \bibinfo {pages} {215} (\bibinfo {year} {2001})},\ \bibinfo {note} {[Erratum:
  Phys.Lett.B 539, 303--303 (2002)]},\ \Eprint
  {https://arxiv.org/abs/hep-ph/0110096} {arXiv:hep-ph/0110096} \BibitemShut
  {NoStop}%
\bibitem [{\citenamefont {Wands}(2008)}]{Wands:2007bd}%
  \BibitemOpen
  \bibfield  {author} {\bibinfo {author} {\bibfnamefont {D.}~\bibnamefont
  {Wands}},\ }\bibfield  {title} {\bibinfo {title} {{Multiple field
  inflation}},\ }\href {https://doi.org/10.1007/978-3-540-74353-8_8} {\bibfield
   {journal} {\bibinfo  {journal} {Lect. Notes Phys.}\ }\textbf {\bibinfo
  {volume} {738}},\ \bibinfo {pages} {275} (\bibinfo {year} {2008})},\ \Eprint
  {https://arxiv.org/abs/astro-ph/0702187} {arXiv:astro-ph/0702187}
  \BibitemShut {NoStop}%
\bibitem [{\citenamefont {Baumann}\ and\ \citenamefont
  {McAllister}(2015)}]{Baumann:2014nda}%
  \BibitemOpen
  \bibfield  {author} {\bibinfo {author} {\bibfnamefont {D.}~\bibnamefont
  {Baumann}}\ and\ \bibinfo {author} {\bibfnamefont {L.}~\bibnamefont
  {McAllister}},\ }\href {https://doi.org/10.1017/CBO9781316105733} {\emph
  {\bibinfo {title} {{Inflation and String Theory}}}},\ Cambridge Monographs on
  Mathematical Physics\ (\bibinfo  {publisher} {Cambridge University Press},\
  \bibinfo {year} {2015})\ \Eprint {https://arxiv.org/abs/1404.2601}
  {arXiv:1404.2601 [hep-th]} \BibitemShut {NoStop}%
\bibitem [{\citenamefont {Gordon}\ \emph {et~al.}(2000)\citenamefont {Gordon},
  \citenamefont {Wands}, \citenamefont {Bassett},\ and\ \citenamefont
  {Maartens}}]{Gordon:2000hv}%
  \BibitemOpen
  \bibfield  {author} {\bibinfo {author} {\bibfnamefont {C.}~\bibnamefont
  {Gordon}}, \bibinfo {author} {\bibfnamefont {D.}~\bibnamefont {Wands}},
  \bibinfo {author} {\bibfnamefont {B.~A.}\ \bibnamefont {Bassett}},\ and\
  \bibinfo {author} {\bibfnamefont {R.}~\bibnamefont {Maartens}},\ }\bibfield
  {title} {\bibinfo {title} {{Adiabatic and entropy perturbations from
  inflation}},\ }\href {https://doi.org/10.1103/PhysRevD.63.023506} {\bibfield
  {journal} {\bibinfo  {journal} {Phys. Rev. D}\ }\textbf {\bibinfo {volume}
  {63}},\ \bibinfo {pages} {023506} (\bibinfo {year} {2000})},\ \Eprint
  {https://arxiv.org/abs/astro-ph/0009131} {arXiv:astro-ph/0009131}
  \BibitemShut {NoStop}%
\bibitem [{\citenamefont {Wands}\ \emph {et~al.}(2002)\citenamefont {Wands},
  \citenamefont {Bartolo}, \citenamefont {Matarrese},\ and\ \citenamefont
  {Riotto}}]{Wands:2002bn}%
  \BibitemOpen
  \bibfield  {author} {\bibinfo {author} {\bibfnamefont {D.}~\bibnamefont
  {Wands}}, \bibinfo {author} {\bibfnamefont {N.}~\bibnamefont {Bartolo}},
  \bibinfo {author} {\bibfnamefont {S.}~\bibnamefont {Matarrese}},\ and\
  \bibinfo {author} {\bibfnamefont {A.}~\bibnamefont {Riotto}},\ }\bibfield
  {title} {\bibinfo {title} {{An Observational test of two-field inflation}},\
  }\href {https://doi.org/10.1103/PhysRevD.66.043520} {\bibfield  {journal}
  {\bibinfo  {journal} {Phys. Rev. D}\ }\textbf {\bibinfo {volume} {66}},\
  \bibinfo {pages} {043520} (\bibinfo {year} {2002})},\ \Eprint
  {https://arxiv.org/abs/astro-ph/0205253} {arXiv:astro-ph/0205253}
  \BibitemShut {NoStop}%
\bibitem [{\citenamefont {Bassett}\ \emph {et~al.}(2006)\citenamefont
  {Bassett}, \citenamefont {Tsujikawa},\ and\ \citenamefont
  {Wands}}]{Bassett:2005xm}%
  \BibitemOpen
  \bibfield  {author} {\bibinfo {author} {\bibfnamefont {B.~A.}\ \bibnamefont
  {Bassett}}, \bibinfo {author} {\bibfnamefont {S.}~\bibnamefont {Tsujikawa}},\
  and\ \bibinfo {author} {\bibfnamefont {D.}~\bibnamefont {Wands}},\ }\bibfield
   {title} {\bibinfo {title} {{Inflation dynamics and reheating}},\ }\href
  {https://doi.org/10.1103/RevModPhys.78.537} {\bibfield  {journal} {\bibinfo
  {journal} {Rev. Mod. Phys.}\ }\textbf {\bibinfo {volume} {78}},\ \bibinfo
  {pages} {537} (\bibinfo {year} {2006})},\ \Eprint
  {https://arxiv.org/abs/astro-ph/0507632} {arXiv:astro-ph/0507632}
  \BibitemShut {NoStop}%
\bibitem [{\citenamefont {Peterson}\ and\ \citenamefont
  {Tegmark}(2011)}]{Peterson:2010np}%
  \BibitemOpen
  \bibfield  {author} {\bibinfo {author} {\bibfnamefont {C.~M.}\ \bibnamefont
  {Peterson}}\ and\ \bibinfo {author} {\bibfnamefont {M.}~\bibnamefont
  {Tegmark}},\ }\bibfield  {title} {\bibinfo {title} {{Testing Two-Field
  Inflation}},\ }\href {https://doi.org/10.1103/PhysRevD.83.023522} {\bibfield
  {journal} {\bibinfo  {journal} {Phys. Rev. D}\ }\textbf {\bibinfo {volume}
  {83}},\ \bibinfo {pages} {023522} (\bibinfo {year} {2011})},\ \Eprint
  {https://arxiv.org/abs/1005.4056} {arXiv:1005.4056 [astro-ph.CO]}
  \BibitemShut {NoStop}%
\bibitem [{\citenamefont {Peterson}\ and\ \citenamefont
  {Tegmark}(2013)}]{Peterson:2011yt}%
  \BibitemOpen
  \bibfield  {author} {\bibinfo {author} {\bibfnamefont {C.~M.}\ \bibnamefont
  {Peterson}}\ and\ \bibinfo {author} {\bibfnamefont {M.}~\bibnamefont
  {Tegmark}},\ }\bibfield  {title} {\bibinfo {title} {{Testing multifield
  inflation: A geometric approach}},\ }\href
  {https://doi.org/10.1103/PhysRevD.87.103507} {\bibfield  {journal} {\bibinfo
  {journal} {Phys. Rev. D}\ }\textbf {\bibinfo {volume} {87}},\ \bibinfo
  {pages} {103507} (\bibinfo {year} {2013})},\ \Eprint
  {https://arxiv.org/abs/1111.0927} {arXiv:1111.0927 [astro-ph.CO]}
  \BibitemShut {NoStop}%
\bibitem [{\citenamefont {Pattison}\ \emph {et~al.}(2017)\citenamefont
  {Pattison}, \citenamefont {Vennin}, \citenamefont {Assadullahi},\ and\
  \citenamefont {Wands}}]{Pattison:2017mbe}%
  \BibitemOpen
  \bibfield  {author} {\bibinfo {author} {\bibfnamefont {C.}~\bibnamefont
  {Pattison}}, \bibinfo {author} {\bibfnamefont {V.}~\bibnamefont {Vennin}},
  \bibinfo {author} {\bibfnamefont {H.}~\bibnamefont {Assadullahi}},\ and\
  \bibinfo {author} {\bibfnamefont {D.}~\bibnamefont {Wands}},\ }\bibfield
  {title} {\bibinfo {title} {{Quantum diffusion during inflation and primordial
  black holes}},\ }\href {https://doi.org/10.1088/1475-7516/2017/10/046}
  {\bibfield  {journal} {\bibinfo  {journal} {JCAP}\ }\textbf {\bibinfo
  {volume} {10}}\bibfield  {number} {\bibinfo  {number} { (2017)},\ \bibinfo
  {pages} {046}},\ }\Eprint {https://arxiv.org/abs/1707.00537}
  {arXiv:1707.00537 [hep-th]} \BibitemShut {NoStop}%
\bibitem [{\citenamefont {Prokopec}\ and\ \citenamefont
  {Rigopoulos}(2021)}]{Prokopec:2019srf}%
  \BibitemOpen
  \bibfield  {author} {\bibinfo {author} {\bibfnamefont {T.}~\bibnamefont
  {Prokopec}}\ and\ \bibinfo {author} {\bibfnamefont {G.}~\bibnamefont
  {Rigopoulos}},\ }\bibfield  {title} {\bibinfo {title} {{{\ensuremath{\Delta
  N}} and the stochastic conveyor belt of ultra slow-roll inflation}},\ }\href
  {https://doi.org/10.1103/PhysRevD.104.083505} {\bibfield  {journal} {\bibinfo
   {journal} {Phys. Rev. D}\ }\textbf {\bibinfo {volume} {104}},\ \bibinfo
  {pages} {083505} (\bibinfo {year} {2021})},\ \Eprint
  {https://arxiv.org/abs/1910.08487} {arXiv:1910.08487 [gr-qc]} \BibitemShut
  {NoStop}%
\bibitem [{\citenamefont {Ezquiaga}\ \emph {et~al.}(2020)\citenamefont
  {Ezquiaga}, \citenamefont {Garc{\'i}a-Bellido},\ and\ \citenamefont
  {Vennin}}]{Ezquiaga:2019ftu}%
  \BibitemOpen
  \bibfield  {author} {\bibinfo {author} {\bibfnamefont {J.~M.}\ \bibnamefont
  {Ezquiaga}}, \bibinfo {author} {\bibfnamefont {J.}~\bibnamefont
  {Garc{\'i}a-Bellido}},\ and\ \bibinfo {author} {\bibfnamefont
  {V.}~\bibnamefont {Vennin}},\ }\bibfield  {title} {\bibinfo {title} {{The
  exponential tail of inflationary fluctuations: consequences for primordial
  black holes}},\ }\href {https://doi.org/10.1088/1475-7516/2020/03/029}
  {\bibfield  {journal} {\bibinfo  {journal} {JCAP}\ }\textbf {\bibinfo
  {volume} {03}}\bibfield  {number} {\bibinfo  {number} { (2020)},\ \bibinfo
  {pages} {029}},\ }\Eprint {https://arxiv.org/abs/1912.05399}
  {arXiv:1912.05399 [astro-ph.CO]} \BibitemShut {NoStop}%
\bibitem [{\citenamefont {Pattison}\ \emph {et~al.}(2021)\citenamefont
  {Pattison}, \citenamefont {Vennin}, \citenamefont {Wands},\ and\
  \citenamefont {Assadullahi}}]{Pattison:2021oen}%
  \BibitemOpen
  \bibfield  {author} {\bibinfo {author} {\bibfnamefont {C.}~\bibnamefont
  {Pattison}}, \bibinfo {author} {\bibfnamefont {V.}~\bibnamefont {Vennin}},
  \bibinfo {author} {\bibfnamefont {D.}~\bibnamefont {Wands}},\ and\ \bibinfo
  {author} {\bibfnamefont {H.}~\bibnamefont {Assadullahi}},\ }\bibfield
  {title} {\bibinfo {title} {{Ultra-slow-roll inflation with quantum
  diffusion}},\ }\href {https://doi.org/10.1088/1475-7516/2021/04/080}
  {\bibfield  {journal} {\bibinfo  {journal} {JCAP}\ }\textbf {\bibinfo
  {volume} {04}}\bibfield  {number} {\bibinfo  {number} { (2021)},\ \bibinfo
  {pages} {080}},\ }\Eprint {https://arxiv.org/abs/2101.05741}
  {arXiv:2101.05741 [astro-ph.CO]} \BibitemShut {NoStop}%
\bibitem [{\citenamefont {Rigopoulos}\ and\ \citenamefont
  {Wilkins}(2021)}]{Rigopoulos:2021nhv}%
  \BibitemOpen
  \bibfield  {author} {\bibinfo {author} {\bibfnamefont {G.}~\bibnamefont
  {Rigopoulos}}\ and\ \bibinfo {author} {\bibfnamefont {A.}~\bibnamefont
  {Wilkins}},\ }\bibfield  {title} {\bibinfo {title} {{Inflation is always
  semi-classical: diffusion domination overproduces Primordial Black Holes}},\
  }\href {https://doi.org/10.1088/1475-7516/2021/12/027} {\bibfield  {journal}
  {\bibinfo  {journal} {JCAP}\ }\textbf {\bibinfo {volume} {12}}\bibfield
  {number} {\bibinfo  {number} { (2021)},\ \bibinfo {pages} {027}},\ }\Eprint
  {https://arxiv.org/abs/2107.05317} {arXiv:2107.05317 [astro-ph.CO]}
  \BibitemShut {NoStop}%
\bibitem [{\citenamefont {Tada}\ and\ \citenamefont
  {Vennin}(2022)}]{Tada:2021zzj}%
  \BibitemOpen
  \bibfield  {author} {\bibinfo {author} {\bibfnamefont {Y.}~\bibnamefont
  {Tada}}\ and\ \bibinfo {author} {\bibfnamefont {V.}~\bibnamefont {Vennin}},\
  }\bibfield  {title} {\bibinfo {title} {{Statistics of coarse-grained
  cosmological fields in stochastic inflation}},\ }\href
  {https://doi.org/10.1088/1475-7516/2022/02/021} {\bibfield  {journal}
  {\bibinfo  {journal} {JCAP}\ }\textbf {\bibinfo {volume} {02}}\bibfield
  {number} {\bibinfo  {number} { (2022)},\ \bibinfo {pages} {021}},\ }\Eprint
  {https://arxiv.org/abs/2111.15280} {arXiv:2111.15280 [astro-ph.CO]}
  \BibitemShut {NoStop}%
\bibitem [{\citenamefont {Hooshangi}\ \emph {et~al.}(2022)\citenamefont
  {Hooshangi}, \citenamefont {Namjoo},\ and\ \citenamefont
  {Noorbala}}]{Hooshangi:2021ubn}%
  \BibitemOpen
  \bibfield  {author} {\bibinfo {author} {\bibfnamefont {S.}~\bibnamefont
  {Hooshangi}}, \bibinfo {author} {\bibfnamefont {M.~H.}\ \bibnamefont
  {Namjoo}},\ and\ \bibinfo {author} {\bibfnamefont {M.}~\bibnamefont
  {Noorbala}},\ }\bibfield  {title} {\bibinfo {title} {{Rare events are
  nonperturbative: Primordial black holes from heavy-tailed distributions}},\
  }\href {https://doi.org/10.1016/j.physletb.2022.137400} {\bibfield  {journal}
  {\bibinfo  {journal} {Phys. Lett. B}\ }\textbf {\bibinfo {volume} {834}},\
  \bibinfo {pages} {137400} (\bibinfo {year} {2022})},\ \Eprint
  {https://arxiv.org/abs/2112.04520} {arXiv:2112.04520 [astro-ph.CO]}
  \BibitemShut {NoStop}%
\bibitem [{\citenamefont {Achucarro}\ \emph {et~al.}(2022)\citenamefont
  {Achucarro}, \citenamefont {Cespedes}, \citenamefont {Davis},\ and\
  \citenamefont {Palma}}]{Achucarro:2021pdh}%
  \BibitemOpen
  \bibfield  {author} {\bibinfo {author} {\bibfnamefont {A.}~\bibnamefont
  {Achucarro}}, \bibinfo {author} {\bibfnamefont {S.}~\bibnamefont {Cespedes}},
  \bibinfo {author} {\bibfnamefont {A.-C.}\ \bibnamefont {Davis}},\ and\
  \bibinfo {author} {\bibfnamefont {G.~A.}\ \bibnamefont {Palma}},\ }\bibfield
  {title} {\bibinfo {title} {{The hand-made tail: non-perturbative tails from
  multifield inflation}},\ }\href {https://doi.org/10.1007/JHEP05(2022)052}
  {\bibfield  {journal} {\bibinfo  {journal} {JHEP}\ }\textbf {\bibinfo
  {volume} {05}}\bibfield  {number} {\bibinfo  {number} { (2022)},\ \bibinfo
  {pages} {052}},\ }\Eprint {https://arxiv.org/abs/2112.14712}
  {arXiv:2112.14712 [hep-th]} \BibitemShut {NoStop}%
\bibitem [{\citenamefont {Animali}\ and\ \citenamefont
  {Vennin}(2023)}]{Animali:2022otk}%
  \BibitemOpen
  \bibfield  {author} {\bibinfo {author} {\bibfnamefont {C.}~\bibnamefont
  {Animali}}\ and\ \bibinfo {author} {\bibfnamefont {V.}~\bibnamefont
  {Vennin}},\ }\bibfield  {title} {\bibinfo {title} {{Primordial black holes
  from stochastic tunnelling}},\ }\href
  {https://doi.org/10.1088/1475-7516/2023/02/043} {\bibfield  {journal}
  {\bibinfo  {journal} {JCAP}\ }\textbf {\bibinfo {volume} {02}}\bibfield
  {number} {\bibinfo  {number} { (02)},\ \bibinfo {pages} {043}},\ }\Eprint
  {https://arxiv.org/abs/2210.03812} {arXiv:2210.03812 [astro-ph.CO]}
  \BibitemShut {NoStop}%
\bibitem [{\citenamefont {Tomberg}(2023)}]{Tomberg:2023kli}%
  \BibitemOpen
  \bibfield  {author} {\bibinfo {author} {\bibfnamefont {E.}~\bibnamefont
  {Tomberg}},\ }\bibfield  {title} {\bibinfo {title} {{Stochastic constant-roll
  inflation and primordial black holes}},\ }\href
  {https://doi.org/10.1103/PhysRevD.108.043502} {\bibfield  {journal} {\bibinfo
   {journal} {Phys. Rev. D}\ }\textbf {\bibinfo {volume} {108}},\ \bibinfo
  {pages} {043502} (\bibinfo {year} {2023})},\ \Eprint
  {https://arxiv.org/abs/2304.10903} {arXiv:2304.10903 [astro-ph.CO]}
  \BibitemShut {NoStop}%
\bibitem [{\citenamefont {Vennin}\ and\ \citenamefont
  {Wands}(2025)}]{Vennin:2024yzl}%
  \BibitemOpen
  \bibfield  {author} {\bibinfo {author} {\bibfnamefont {V.}~\bibnamefont
  {Vennin}}\ and\ \bibinfo {author} {\bibfnamefont {D.}~\bibnamefont {Wands}},\
  }\bibinfo {title} {Quantum diffusion and large primordial perturbations from
  inflation},\ in\ \href {https://doi.org/10.1007/978-981-97-8887-3_8} {\emph
  {\bibinfo {booktitle} {Primordial Black Holes}}},\ \bibinfo {editor} {edited
  by\ \bibinfo {editor} {\bibfnamefont {C.}~\bibnamefont {Byrnes}}, \bibinfo
  {editor} {\bibfnamefont {G.}~\bibnamefont {Franciolini}}, \bibinfo {editor}
  {\bibfnamefont {T.}~\bibnamefont {Harada}}, \bibinfo {editor} {\bibfnamefont
  {P.}~\bibnamefont {Pani}},\ and\ \bibinfo {editor} {\bibfnamefont
  {M.}~\bibnamefont {Sasaki}}}\ (\bibinfo  {publisher} {Springer Nature
  Singapore},\ \bibinfo {address} {Singapore},\ \bibinfo {year} {2025})\ pp.\
  \bibinfo {pages} {201--227}\BibitemShut {NoStop}%
\bibitem [{\citenamefont {Murata}\ and\ \citenamefont
  {Tada}(2026)}]{Murata:2025onc}%
  \BibitemOpen
  \bibfield  {author} {\bibinfo {author} {\bibfnamefont {T.}~\bibnamefont
  {Murata}}\ and\ \bibinfo {author} {\bibfnamefont {Y.}~\bibnamefont {Tada}},\
  }\bibfield  {title} {\bibinfo {title} {{Stochastic tail of the curvature
  perturbation in hybrid inflation}},\ }\href
  {https://doi.org/10.1103/h2j9-p3y8} {\bibfield  {journal} {\bibinfo
  {journal} {Phys. Rev. D}\ }\textbf {\bibinfo {volume} {113}},\ \bibinfo
  {pages} {023552} (\bibinfo {year} {2026})},\ \Eprint
  {https://arxiv.org/abs/2507.22439} {arXiv:2507.22439 [astro-ph.CO]}
  \BibitemShut {NoStop}%
\bibitem [{\citenamefont {Zel'dovich}\ and\ \citenamefont
  {Novikov}(1967)}]{Zeldovich:1967lct}%
  \BibitemOpen
  \bibfield  {author} {\bibinfo {author} {\bibfnamefont {Y.~B.}\ \bibnamefont
  {Zel'dovich}}\ and\ \bibinfo {author} {\bibfnamefont {I.~D.}\ \bibnamefont
  {Novikov}},\ }\bibfield  {title} {\bibinfo {title} {{The Hypothesis of Cores
  Retarded during Expansion and the Hot Cosmological Model}},\ }\href@noop {}
  {\bibfield  {journal} {\bibinfo  {journal} {Soviet Astron. AJ (Engl. Transl.
  ),}\ }\textbf {\bibinfo {volume} {10}},\ \bibinfo {pages} {602} (\bibinfo
  {year} {1967})}\BibitemShut {NoStop}%
\bibitem [{\citenamefont {Hawking}(1971)}]{Hawking:1971ei}%
  \BibitemOpen
  \bibfield  {author} {\bibinfo {author} {\bibfnamefont {S.}~\bibnamefont
  {Hawking}},\ }\bibfield  {title} {\bibinfo {title} {{Gravitationally
  collapsed objects of very low mass}},\ }\href
  {https://doi.org/10.1093/mnras/152.1.75} {\bibfield  {journal} {\bibinfo
  {journal} {Mon. Not. Roy. Astron. Soc.}\ }\textbf {\bibinfo {volume} {152}},\
  \bibinfo {pages} {75} (\bibinfo {year} {1971})}\BibitemShut {NoStop}%
\bibitem [{\citenamefont {Carr}\ and\ \citenamefont
  {Hawking}(1974)}]{Carr:1974nx}%
  \BibitemOpen
  \bibfield  {author} {\bibinfo {author} {\bibfnamefont {B.~J.}\ \bibnamefont
  {Carr}}\ and\ \bibinfo {author} {\bibfnamefont {S.~W.}\ \bibnamefont
  {Hawking}},\ }\bibfield  {title} {\bibinfo {title} {{Black holes in the early
  Universe}},\ }\href {https://doi.org/10.1093/mnras/168.2.399} {\bibfield
  {journal} {\bibinfo  {journal} {Mon. Not. Roy. Astron. Soc.}\ }\textbf
  {\bibinfo {volume} {168}},\ \bibinfo {pages} {399} (\bibinfo {year}
  {1974})}\BibitemShut {NoStop}%
\bibitem [{\citenamefont {Carr}(1975)}]{Carr:1975qj}%
  \BibitemOpen
  \bibfield  {author} {\bibinfo {author} {\bibfnamefont {B.~J.}\ \bibnamefont
  {Carr}},\ }\bibfield  {title} {\bibinfo {title} {{The Primordial black hole
  mass spectrum}},\ }\href {https://doi.org/10.1086/153853} {\bibfield
  {journal} {\bibinfo  {journal} {Astrophys. J.}\ }\textbf {\bibinfo {volume}
  {201}},\ \bibinfo {pages} {1} (\bibinfo {year} {1975})}\BibitemShut {NoStop}%
\bibitem [{\citenamefont {Vennin}\ and\ \citenamefont
  {Starobinsky}(2015)}]{Vennin:2015hra}%
  \BibitemOpen
  \bibfield  {author} {\bibinfo {author} {\bibfnamefont {V.}~\bibnamefont
  {Vennin}}\ and\ \bibinfo {author} {\bibfnamefont {A.~A.}\ \bibnamefont
  {Starobinsky}},\ }\bibfield  {title} {\bibinfo {title} {{Correlation
  Functions in Stochastic Inflation}},\ }\href
  {https://doi.org/10.1140/epjc/s10052-015-3643-y} {\bibfield  {journal}
  {\bibinfo  {journal} {Eur. Phys. J. C}\ }\textbf {\bibinfo {volume} {75}},\
  \bibinfo {pages} {413} (\bibinfo {year} {2015})},\ \Eprint
  {https://arxiv.org/abs/1506.04732} {arXiv:1506.04732 [hep-th]} \BibitemShut
  {NoStop}%
\bibitem [{\citenamefont {Linde}(1982{\natexlab{b}})}]{Linde:1982ur}%
  \BibitemOpen
  \bibfield  {author} {\bibinfo {author} {\bibfnamefont {A.~D.}\ \bibnamefont
  {Linde}},\ }\href@noop {} {\bibinfo {title} {{Nonsingular Regenerating
  Inflationary Universe}}} (\bibinfo {year} {1982}{\natexlab{b}}),\ \bibinfo
  {note} {{Cambridge University preprint}}\BibitemShut {NoStop}%
\bibitem [{\citenamefont {Vilenkin}(1983)}]{Vilenkin:1983xq}%
  \BibitemOpen
  \bibfield  {author} {\bibinfo {author} {\bibfnamefont {A.}~\bibnamefont
  {Vilenkin}},\ }\bibfield  {title} {\bibinfo {title} {{The Birth of
  Inflationary Universes}},\ }\href {https://doi.org/10.1103/PhysRevD.27.2848}
  {\bibfield  {journal} {\bibinfo  {journal} {Phys. Rev. D}\ }\textbf {\bibinfo
  {volume} {27}},\ \bibinfo {pages} {2848} (\bibinfo {year}
  {1983})}\BibitemShut {NoStop}%
\bibitem [{\citenamefont {Winitzki}(2009)}]{Winitzki:2008zz}%
  \BibitemOpen
  \bibfield  {author} {\bibinfo {author} {\bibfnamefont {S.}~\bibnamefont
  {Winitzki}},\ }\href {https://doi.org/10.1142/6923} {\emph {\bibinfo {title}
  {{Eternal Inflation}}}}\ (\bibinfo  {publisher} {World Scientific},\ \bibinfo
  {year} {2009})\BibitemShut {NoStop}%
\bibitem [{\citenamefont {Takahashi}\ and\ \citenamefont
  {Tokeshi}(2025)}]{Takahashi:2025hqt}%
  \BibitemOpen
  \bibfield  {author} {\bibinfo {author} {\bibfnamefont {T.}~\bibnamefont
  {Takahashi}}\ and\ \bibinfo {author} {\bibfnamefont {K.}~\bibnamefont
  {Tokeshi}},\ }\href@noop {} {\bibinfo {title} {{More fields are different:
  Stochastic view of multi-field inflationary scenario}}} (\bibinfo {year}
  {2025}),\ \Eprint {https://arxiv.org/abs/2504.11158} {arXiv:2504.11158
  [astro-ph.CO]} \BibitemShut {NoStop}%
\bibitem [{\citenamefont {Salopek}\ and\ \citenamefont
  {Bond}(1990)}]{Salopek:1990jq}%
  \BibitemOpen
  \bibfield  {author} {\bibinfo {author} {\bibfnamefont {D.~S.}\ \bibnamefont
  {Salopek}}\ and\ \bibinfo {author} {\bibfnamefont {J.~R.}\ \bibnamefont
  {Bond}},\ }\bibfield  {title} {\bibinfo {title} {{Nonlinear evolution of long
  wavelength metric fluctuations in inflationary models}},\ }\href
  {https://doi.org/10.1103/PhysRevD.42.3936} {\bibfield  {journal} {\bibinfo
  {journal} {Phys. Rev. D}\ }\textbf {\bibinfo {volume} {42}},\ \bibinfo
  {pages} {3936} (\bibinfo {year} {1990})}\BibitemShut {NoStop}%
\bibitem [{\citenamefont {Sasaki}\ and\ \citenamefont
  {Stewart}(1996)}]{Sasaki:1995aw}%
  \BibitemOpen
  \bibfield  {author} {\bibinfo {author} {\bibfnamefont {M.}~\bibnamefont
  {Sasaki}}\ and\ \bibinfo {author} {\bibfnamefont {E.~D.}\ \bibnamefont
  {Stewart}},\ }\bibfield  {title} {\bibinfo {title} {{A General analytic
  formula for the spectral index of the density perturbations produced during
  inflation}},\ }\href {https://doi.org/10.1143/PTP.95.71} {\bibfield
  {journal} {\bibinfo  {journal} {Prog. Theor. Phys.}\ }\textbf {\bibinfo
  {volume} {95}},\ \bibinfo {pages} {71} (\bibinfo {year} {1996})},\ \Eprint
  {https://arxiv.org/abs/astro-ph/9507001} {arXiv:astro-ph/9507001}
  \BibitemShut {NoStop}%
\bibitem [{\citenamefont {Sasaki}\ and\ \citenamefont
  {Tanaka}(1998)}]{Sasaki:1998ug}%
  \BibitemOpen
  \bibfield  {author} {\bibinfo {author} {\bibfnamefont {M.}~\bibnamefont
  {Sasaki}}\ and\ \bibinfo {author} {\bibfnamefont {T.}~\bibnamefont
  {Tanaka}},\ }\bibfield  {title} {\bibinfo {title} {{Superhorizon scale
  dynamics of multiscalar inflation}},\ }\href
  {https://doi.org/10.1143/PTP.99.763} {\bibfield  {journal} {\bibinfo
  {journal} {Prog. Theor. Phys.}\ }\textbf {\bibinfo {volume} {99}},\ \bibinfo
  {pages} {763} (\bibinfo {year} {1998})},\ \Eprint
  {https://arxiv.org/abs/gr-qc/9801017} {arXiv:gr-qc/9801017} \BibitemShut
  {NoStop}%
\bibitem [{\citenamefont {Lyth}\ \emph {et~al.}(2005)\citenamefont {Lyth},
  \citenamefont {Malik},\ and\ \citenamefont {Sasaki}}]{Lyth:2004gb}%
  \BibitemOpen
  \bibfield  {author} {\bibinfo {author} {\bibfnamefont {D.~H.}\ \bibnamefont
  {Lyth}}, \bibinfo {author} {\bibfnamefont {K.~A.}\ \bibnamefont {Malik}},\
  and\ \bibinfo {author} {\bibfnamefont {M.}~\bibnamefont {Sasaki}},\
  }\bibfield  {title} {\bibinfo {title} {{A General proof of the conservation
  of the curvature perturbation}},\ }\href
  {https://doi.org/10.1088/1475-7516/2005/05/004} {\bibfield  {journal}
  {\bibinfo  {journal} {JCAP}\ }\textbf {\bibinfo {volume} {05}}\bibfield
  {number} {\bibinfo  {number} { (2005)},\ \bibinfo {pages} {004}},\ }\Eprint
  {https://arxiv.org/abs/astro-ph/0411220} {arXiv:astro-ph/0411220}
  \BibitemShut {NoStop}%
\bibitem [{\citenamefont {Abolhasani}\ \emph {et~al.}(2019)\citenamefont
  {Abolhasani}, \citenamefont {Firouzjahi}, \citenamefont {Naruko},\ and\
  \citenamefont {Sasaki}}]{Abolhasani:2019cqw}%
  \BibitemOpen
  \bibfield  {author} {\bibinfo {author} {\bibfnamefont {A.~A.}\ \bibnamefont
  {Abolhasani}}, \bibinfo {author} {\bibfnamefont {H.}~\bibnamefont
  {Firouzjahi}}, \bibinfo {author} {\bibfnamefont {A.}~\bibnamefont {Naruko}},\
  and\ \bibinfo {author} {\bibfnamefont {M.}~\bibnamefont {Sasaki}},\ }\href
  {https://doi.org/10.1142/10953} {\emph {\bibinfo {title} {{Delta N Formalism
  in Cosmological Perturbation Theory}}}}\ (\bibinfo  {publisher} {WSP},\
  \bibinfo {year} {2019})\BibitemShut {NoStop}%
\bibitem [{\citenamefont {Fujita}\ \emph {et~al.}(2013)\citenamefont {Fujita},
  \citenamefont {Kawasaki}, \citenamefont {Tada},\ and\ \citenamefont
  {Takesako}}]{Fujita:2013cna}%
  \BibitemOpen
  \bibfield  {author} {\bibinfo {author} {\bibfnamefont {T.}~\bibnamefont
  {Fujita}}, \bibinfo {author} {\bibfnamefont {M.}~\bibnamefont {Kawasaki}},
  \bibinfo {author} {\bibfnamefont {Y.}~\bibnamefont {Tada}},\ and\ \bibinfo
  {author} {\bibfnamefont {T.}~\bibnamefont {Takesako}},\ }\bibfield  {title}
  {\bibinfo {title} {{A new algorithm for calculating the curvature
  perturbations in stochastic inflation}},\ }\href
  {https://doi.org/10.1088/1475-7516/2013/12/036} {\bibfield  {journal}
  {\bibinfo  {journal} {JCAP}\ }\textbf {\bibinfo {volume} {12}}\bibfield
  {number} {\bibinfo  {number} { (2013)},\ \bibinfo {pages} {036}},\ }\Eprint
  {https://arxiv.org/abs/1308.4754} {arXiv:1308.4754 [astro-ph.CO]}
  \BibitemShut {NoStop}%
\bibitem [{\citenamefont {Fujita}\ \emph {et~al.}(2014)\citenamefont {Fujita},
  \citenamefont {Kawasaki},\ and\ \citenamefont {Tada}}]{Fujita:2014tja}%
  \BibitemOpen
  \bibfield  {author} {\bibinfo {author} {\bibfnamefont {T.}~\bibnamefont
  {Fujita}}, \bibinfo {author} {\bibfnamefont {M.}~\bibnamefont {Kawasaki}},\
  and\ \bibinfo {author} {\bibfnamefont {Y.}~\bibnamefont {Tada}},\ }\bibfield
  {title} {\bibinfo {title} {{Non-perturbative approach for curvature
  perturbations in stochastic $\delta N$ formalism}},\ }\href
  {https://doi.org/10.1088/1475-7516/2014/10/030} {\bibfield  {journal}
  {\bibinfo  {journal} {JCAP}\ }\textbf {\bibinfo {volume} {10}}\bibfield
  {number} {\bibinfo  {number} { (10)},\ \bibinfo {pages} {030}},\ }\Eprint
  {https://arxiv.org/abs/1405.2187} {arXiv:1405.2187 [astro-ph.CO]}
  \BibitemShut {NoStop}%
\bibitem [{\citenamefont {Ando}\ and\ \citenamefont
  {Vennin}(2021)}]{Ando:2020fjm}%
  \BibitemOpen
  \bibfield  {author} {\bibinfo {author} {\bibfnamefont {K.}~\bibnamefont
  {Ando}}\ and\ \bibinfo {author} {\bibfnamefont {V.}~\bibnamefont {Vennin}},\
  }\bibfield  {title} {\bibinfo {title} {{Power spectrum in stochastic
  inflation}},\ }\href {https://doi.org/10.1088/1475-7516/2021/04/057}
  {\bibfield  {journal} {\bibinfo  {journal} {JCAP}\ }\textbf {\bibinfo
  {volume} {04}}\bibfield  {number} {\bibinfo  {number} { (04)},\ \bibinfo
  {pages} {057}},\ }\Eprint {https://arxiv.org/abs/2012.02031}
  {arXiv:2012.02031 [astro-ph.CO]} \BibitemShut {NoStop}%
\bibitem [{\citenamefont {Starobinsky}(1986)}]{Starobinsky:1986fx}%
  \BibitemOpen
  \bibfield  {author} {\bibinfo {author} {\bibfnamefont {A.~A.}\ \bibnamefont
  {Starobinsky}},\ }\bibfield  {title} {\bibinfo {title} {{Stochastic de Siter
  (inflationary) stage in the early Universe}},\ }\href
  {https://doi.org/10.1007/3-540-16452-9_6} {\bibfield  {journal} {\bibinfo
  {journal} {Lect. Notes Phys.}\ }\textbf {\bibinfo {volume} {246}},\ \bibinfo
  {pages} {107} (\bibinfo {year} {1986})}\BibitemShut {NoStop}%
\bibitem [{\citenamefont {Starobinsky}\ and\ \citenamefont
  {Yokoyama}(1994)}]{Starobinsky:1994bd}%
  \BibitemOpen
  \bibfield  {author} {\bibinfo {author} {\bibfnamefont {A.~A.}\ \bibnamefont
  {Starobinsky}}\ and\ \bibinfo {author} {\bibfnamefont {J.}~\bibnamefont
  {Yokoyama}},\ }\bibfield  {title} {\bibinfo {title} {{Equilibrium state of a
  selfinteracting scalar field in the De Sitter background}},\ }\href
  {https://doi.org/10.1103/PhysRevD.50.6357} {\bibfield  {journal} {\bibinfo
  {journal} {Phys. Rev. D}\ }\textbf {\bibinfo {volume} {50}},\ \bibinfo
  {pages} {6357} (\bibinfo {year} {1994})},\ \Eprint
  {https://arxiv.org/abs/astro-ph/9407016} {arXiv:astro-ph/9407016}
  \BibitemShut {NoStop}%
\bibitem [{\citenamefont {Cruces}(2022)}]{Cruces:2022imf}%
  \BibitemOpen
  \bibfield  {author} {\bibinfo {author} {\bibfnamefont {D.}~\bibnamefont
  {Cruces}},\ }\bibfield  {title} {\bibinfo {title} {{Review on Stochastic
  Approach to Inflation}},\ }\href {https://doi.org/10.3390/universe8060334}
  {\bibfield  {journal} {\bibinfo  {journal} {Universe}\ }\textbf {\bibinfo
  {volume} {8}},\ \bibinfo {pages} {334} (\bibinfo {year} {2022})},\ \Eprint
  {https://arxiv.org/abs/2203.13852} {arXiv:2203.13852 [gr-qc]} \BibitemShut
  {NoStop}%
\bibitem [{\citenamefont {Mollerach}\ \emph {et~al.}(1991)\citenamefont
  {Mollerach}, \citenamefont {Matarrese}, \citenamefont {Ortolan},\ and\
  \citenamefont {Lucchin}}]{Mollerach:1990zf}%
  \BibitemOpen
  \bibfield  {author} {\bibinfo {author} {\bibfnamefont {S.}~\bibnamefont
  {Mollerach}}, \bibinfo {author} {\bibfnamefont {S.}~\bibnamefont
  {Matarrese}}, \bibinfo {author} {\bibfnamefont {A.}~\bibnamefont {Ortolan}},\
  and\ \bibinfo {author} {\bibfnamefont {F.}~\bibnamefont {Lucchin}},\
  }\bibfield  {title} {\bibinfo {title} {{Stochastic inflation in a simple two
  field model}},\ }\href {https://doi.org/10.1103/PhysRevD.44.1670} {\bibfield
  {journal} {\bibinfo  {journal} {Phys. Rev. D}\ }\textbf {\bibinfo {volume}
  {44}},\ \bibinfo {pages} {1670} (\bibinfo {year} {1991})}\BibitemShut
  {NoStop}%
\bibitem [{\citenamefont {Assadullahi}\ \emph {et~al.}(2016)\citenamefont
  {Assadullahi}, \citenamefont {Firouzjahi}, \citenamefont {Noorbala},
  \citenamefont {Vennin},\ and\ \citenamefont {Wands}}]{Assadullahi:2016gkk}%
  \BibitemOpen
  \bibfield  {author} {\bibinfo {author} {\bibfnamefont {H.}~\bibnamefont
  {Assadullahi}}, \bibinfo {author} {\bibfnamefont {H.}~\bibnamefont
  {Firouzjahi}}, \bibinfo {author} {\bibfnamefont {M.}~\bibnamefont
  {Noorbala}}, \bibinfo {author} {\bibfnamefont {V.}~\bibnamefont {Vennin}},\
  and\ \bibinfo {author} {\bibfnamefont {D.}~\bibnamefont {Wands}},\ }\bibfield
   {title} {\bibinfo {title} {{Multiple Fields in Stochastic Inflation}},\
  }\href {https://doi.org/10.1088/1475-7516/2016/06/043} {\bibfield  {journal}
  {\bibinfo  {journal} {JCAP}\ }\textbf {\bibinfo {volume} {06}}\bibfield
  {number} {\bibinfo  {number} { (2016)},\ \bibinfo {pages} {043}},\ }\Eprint
  {https://arxiv.org/abs/1604.04502} {arXiv:1604.04502 [hep-th]} \BibitemShut
  {NoStop}%
\bibitem [{\citenamefont {Vennin}\ \emph {et~al.}(2017)\citenamefont {Vennin},
  \citenamefont {Assadullahi}, \citenamefont {Firouzjahi}, \citenamefont
  {Noorbala},\ and\ \citenamefont {Wands}}]{Vennin:2016wnk}%
  \BibitemOpen
  \bibfield  {author} {\bibinfo {author} {\bibfnamefont {V.}~\bibnamefont
  {Vennin}}, \bibinfo {author} {\bibfnamefont {H.}~\bibnamefont {Assadullahi}},
  \bibinfo {author} {\bibfnamefont {H.}~\bibnamefont {Firouzjahi}}, \bibinfo
  {author} {\bibfnamefont {M.}~\bibnamefont {Noorbala}},\ and\ \bibinfo
  {author} {\bibfnamefont {D.}~\bibnamefont {Wands}},\ }\bibfield  {title}
  {\bibinfo {title} {{Critical Number of Fields in Stochastic Inflation}},\
  }\href {https://doi.org/10.1103/PhysRevLett.118.031301} {\bibfield  {journal}
  {\bibinfo  {journal} {Phys. Rev. Lett.}\ }\textbf {\bibinfo {volume} {118}},\
  \bibinfo {pages} {031301} (\bibinfo {year} {2017})},\ \Eprint
  {https://arxiv.org/abs/1604.06017} {arXiv:1604.06017 [astro-ph.CO]}
  \BibitemShut {NoStop}%
\bibitem [{\citenamefont {Finelli}\ \emph {et~al.}(2009)\citenamefont
  {Finelli}, \citenamefont {Marozzi}, \citenamefont {Starobinsky},
  \citenamefont {Vacca},\ and\ \citenamefont {Venturi}}]{Finelli:2008zg}%
  \BibitemOpen
  \bibfield  {author} {\bibinfo {author} {\bibfnamefont {F.}~\bibnamefont
  {Finelli}}, \bibinfo {author} {\bibfnamefont {G.}~\bibnamefont {Marozzi}},
  \bibinfo {author} {\bibfnamefont {A.~A.}\ \bibnamefont {Starobinsky}},
  \bibinfo {author} {\bibfnamefont {G.~P.}\ \bibnamefont {Vacca}},\ and\
  \bibinfo {author} {\bibfnamefont {G.}~\bibnamefont {Venturi}},\ }\bibfield
  {title} {\bibinfo {title} {{Generation of fluctuations during inflation:
  Comparison of stochastic and field-theoretic approaches}},\ }\href
  {https://doi.org/10.1103/PhysRevD.79.044007} {\bibfield  {journal} {\bibinfo
  {journal} {Phys. Rev. D}\ }\textbf {\bibinfo {volume} {79}},\ \bibinfo
  {pages} {044007} (\bibinfo {year} {2009})},\ \Eprint
  {https://arxiv.org/abs/0808.1786} {arXiv:0808.1786 [hep-th]} \BibitemShut
  {NoStop}%
\bibitem [{\citenamefont {Finelli}\ \emph {et~al.}(2010)\citenamefont
  {Finelli}, \citenamefont {Marozzi}, \citenamefont {Starobinsky},
  \citenamefont {Vacca},\ and\ \citenamefont {Venturi}}]{Finelli:2010sh}%
  \BibitemOpen
  \bibfield  {author} {\bibinfo {author} {\bibfnamefont {F.}~\bibnamefont
  {Finelli}}, \bibinfo {author} {\bibfnamefont {G.}~\bibnamefont {Marozzi}},
  \bibinfo {author} {\bibfnamefont {A.~A.}\ \bibnamefont {Starobinsky}},
  \bibinfo {author} {\bibfnamefont {G.~P.}\ \bibnamefont {Vacca}},\ and\
  \bibinfo {author} {\bibfnamefont {G.}~\bibnamefont {Venturi}},\ }\bibfield
  {title} {\bibinfo {title} {{Stochastic growth of quantum fluctuations during
  slow-roll inflation}},\ }\href {https://doi.org/10.1103/PhysRevD.82.064020}
  {\bibfield  {journal} {\bibinfo  {journal} {Phys. Rev. D}\ }\textbf {\bibinfo
  {volume} {82}},\ \bibinfo {pages} {064020} (\bibinfo {year} {2010})},\
  \Eprint {https://arxiv.org/abs/1003.1327} {arXiv:1003.1327 [hep-th]}
  \BibitemShut {NoStop}%
\bibitem [{\citenamefont {Pattison}\ \emph {et~al.}(2019)\citenamefont
  {Pattison}, \citenamefont {Vennin}, \citenamefont {Assadullahi},\ and\
  \citenamefont {Wands}}]{Pattison:2019hef}%
  \BibitemOpen
  \bibfield  {author} {\bibinfo {author} {\bibfnamefont {C.}~\bibnamefont
  {Pattison}}, \bibinfo {author} {\bibfnamefont {V.}~\bibnamefont {Vennin}},
  \bibinfo {author} {\bibfnamefont {H.}~\bibnamefont {Assadullahi}},\ and\
  \bibinfo {author} {\bibfnamefont {D.}~\bibnamefont {Wands}},\ }\bibfield
  {title} {\bibinfo {title} {{Stochastic inflation beyond slow roll}},\ }\href
  {https://doi.org/10.1088/1475-7516/2019/07/031} {\bibfield  {journal}
  {\bibinfo  {journal} {JCAP}\ }\textbf {\bibinfo {volume} {07}}\bibfield
  {number} {\bibinfo  {number} { (2019)},\ \bibinfo {pages} {031}},\ }\Eprint
  {https://arxiv.org/abs/1905.06300} {arXiv:1905.06300 [astro-ph.CO]}
  \BibitemShut {NoStop}%
\bibitem [{\citenamefont {Casini}\ \emph {et~al.}(1999)\citenamefont {Casini},
  \citenamefont {Montemayor},\ and\ \citenamefont {Sisterna}}]{Casini:1998wr}%
  \BibitemOpen
  \bibfield  {author} {\bibinfo {author} {\bibfnamefont {H.}~\bibnamefont
  {Casini}}, \bibinfo {author} {\bibfnamefont {R.}~\bibnamefont {Montemayor}},\
  and\ \bibinfo {author} {\bibfnamefont {P.}~\bibnamefont {Sisterna}},\
  }\bibfield  {title} {\bibinfo {title} {{Stochastic approach to inflation. 2.
  Classicality, coarse graining and noises}},\ }\href
  {https://doi.org/10.1103/PhysRevD.59.063512} {\bibfield  {journal} {\bibinfo
  {journal} {Phys. Rev. D}\ }\textbf {\bibinfo {volume} {59}},\ \bibinfo
  {pages} {063512} (\bibinfo {year} {1999})},\ \Eprint
  {https://arxiv.org/abs/gr-qc/9811083} {arXiv:gr-qc/9811083} \BibitemShut
  {NoStop}%
\bibitem [{\citenamefont {Winitzki}\ and\ \citenamefont
  {Vilenkin}(2000)}]{Winitzki:1999ve}%
  \BibitemOpen
  \bibfield  {author} {\bibinfo {author} {\bibfnamefont {S.}~\bibnamefont
  {Winitzki}}\ and\ \bibinfo {author} {\bibfnamefont {A.}~\bibnamefont
  {Vilenkin}},\ }\bibfield  {title} {\bibinfo {title} {{Effective noise in
  stochastic description of inflation}},\ }\href
  {https://doi.org/10.1103/PhysRevD.61.084008} {\bibfield  {journal} {\bibinfo
  {journal} {Phys. Rev. D}\ }\textbf {\bibinfo {volume} {61}},\ \bibinfo
  {pages} {084008} (\bibinfo {year} {2000})},\ \Eprint
  {https://arxiv.org/abs/gr-qc/9911029} {arXiv:gr-qc/9911029} \BibitemShut
  {NoStop}%
\bibitem [{\citenamefont {Matarrese}\ \emph {et~al.}(2004)\citenamefont
  {Matarrese}, \citenamefont {Musso},\ and\ \citenamefont
  {Riotto}}]{Matarrese:2003ye}%
  \BibitemOpen
  \bibfield  {author} {\bibinfo {author} {\bibfnamefont {S.}~\bibnamefont
  {Matarrese}}, \bibinfo {author} {\bibfnamefont {M.~A.}\ \bibnamefont
  {Musso}},\ and\ \bibinfo {author} {\bibfnamefont {A.}~\bibnamefont
  {Riotto}},\ }\bibfield  {title} {\bibinfo {title} {{Influence of superhorizon
  scales on cosmological observables generated during inflation}},\ }\href
  {https://doi.org/10.1088/1475-7516/2004/05/008} {\bibfield  {journal}
  {\bibinfo  {journal} {JCAP}\ }\textbf {\bibinfo {volume} {05}}\bibfield
  {number} {\bibinfo  {number} { (2004)},\ \bibinfo {pages} {008}},\ }\Eprint
  {https://arxiv.org/abs/hep-th/0311059} {arXiv:hep-th/0311059} \BibitemShut
  {NoStop}%
\bibitem [{\citenamefont {Liguori}\ \emph {et~al.}(2004)\citenamefont
  {Liguori}, \citenamefont {Matarrese}, \citenamefont {Musso},\ and\
  \citenamefont {Riotto}}]{Liguori:2004fa}%
  \BibitemOpen
  \bibfield  {author} {\bibinfo {author} {\bibfnamefont {M.}~\bibnamefont
  {Liguori}}, \bibinfo {author} {\bibfnamefont {S.}~\bibnamefont {Matarrese}},
  \bibinfo {author} {\bibfnamefont {M.}~\bibnamefont {Musso}},\ and\ \bibinfo
  {author} {\bibfnamefont {A.}~\bibnamefont {Riotto}},\ }\bibfield  {title}
  {\bibinfo {title} {{Stochastic inflation and the lower multipoles in the CMB
  anisotropies}},\ }\href {https://doi.org/10.1088/1475-7516/2004/08/011}
  {\bibfield  {journal} {\bibinfo  {journal} {JCAP}\ }\textbf {\bibinfo
  {volume} {08}}\bibfield  {number} {\bibinfo  {number} { (2004)},\ \bibinfo
  {pages} {011}},\ }\Eprint {https://arxiv.org/abs/astro-ph/0405544}
  {arXiv:astro-ph/0405544} \BibitemShut {NoStop}%
\bibitem [{\citenamefont {Andersen}\ \emph {et~al.}(2022)\citenamefont
  {Andersen}, \citenamefont {Eriksson},\ and\ \citenamefont
  {Tranberg}}]{Andersen:2021lii}%
  \BibitemOpen
  \bibfield  {author} {\bibinfo {author} {\bibfnamefont {J.~O.}\ \bibnamefont
  {Andersen}}, \bibinfo {author} {\bibfnamefont {M.}~\bibnamefont {Eriksson}},\
  and\ \bibinfo {author} {\bibfnamefont {A.}~\bibnamefont {Tranberg}},\
  }\bibfield  {title} {\bibinfo {title} {{Stochastic inflation from quantum
  field theory and the parametric dependence of the effective noise
  amplitude}},\ }\href {https://doi.org/10.1007/JHEP02(2022)121} {\bibfield
  {journal} {\bibinfo  {journal} {JHEP}\ }\textbf {\bibinfo {volume}
  {02}}\bibfield  {number} {\bibinfo  {number} { (2022)},\ \bibinfo {pages}
  {121}},\ }\Eprint {https://arxiv.org/abs/2111.14503} {arXiv:2111.14503
  [hep-ph]} \BibitemShut {NoStop}%
\bibitem [{\citenamefont {Mahbub}\ and\ \citenamefont
  {De}(2022)}]{Mahbub:2022osb}%
  \BibitemOpen
  \bibfield  {author} {\bibinfo {author} {\bibfnamefont {R.}~\bibnamefont
  {Mahbub}}\ and\ \bibinfo {author} {\bibfnamefont {A.}~\bibnamefont {De}},\
  }\bibfield  {title} {\bibinfo {title} {{Smooth coarse-graining and colored
  noise dynamics in stochastic inflation}},\ }\href
  {https://doi.org/10.1088/1475-7516/2022/09/045} {\bibfield  {journal}
  {\bibinfo  {journal} {JCAP}\ }\textbf {\bibinfo {volume} {09}}\bibfield
  {number} {\bibinfo  {number} { (2022)},\ \bibinfo {pages} {045}},\ }\Eprint
  {https://arxiv.org/abs/2204.03859} {arXiv:2204.03859 [astro-ph.CO]}
  \BibitemShut {NoStop}%
\bibitem [{\citenamefont {Brahma}\ \emph {et~al.}(2025)\citenamefont {Brahma},
  \citenamefont {Calder{\'o}n-Figueroa},\ and\ \citenamefont
  {Luo}}]{Brahma:2024yor}%
  \BibitemOpen
  \bibfield  {author} {\bibinfo {author} {\bibfnamefont {S.}~\bibnamefont
  {Brahma}}, \bibinfo {author} {\bibfnamefont {J.}~\bibnamefont
  {Calder{\'o}n-Figueroa}},\ and\ \bibinfo {author} {\bibfnamefont
  {X.}~\bibnamefont {Luo}},\ }\bibfield  {title} {\bibinfo {title}
  {{Time-convolutionless cosmological master equations: late-time resummations
  and decoherence for non-local kernels}},\ }\href
  {https://doi.org/10.1088/1475-7516/2025/08/019} {\bibfield  {journal}
  {\bibinfo  {journal} {JCAP}\ }\textbf {\bibinfo {volume} {08}}\bibfield
  {number} {\bibinfo  {number} { (2025)},\ \bibinfo {pages} {019}},\ }\Eprint
  {https://arxiv.org/abs/2407.12091} {arXiv:2407.12091 [hep-th]} \BibitemShut
  {NoStop}%
\bibitem [{\citenamefont {Figueroa}\ \emph {et~al.}(2021)\citenamefont
  {Figueroa}, \citenamefont {Raatikainen}, \citenamefont {Rasanen},\ and\
  \citenamefont {Tomberg}}]{Figueroa:2020jkf}%
  \BibitemOpen
  \bibfield  {author} {\bibinfo {author} {\bibfnamefont {D.~G.}\ \bibnamefont
  {Figueroa}}, \bibinfo {author} {\bibfnamefont {S.}~\bibnamefont
  {Raatikainen}}, \bibinfo {author} {\bibfnamefont {S.}~\bibnamefont
  {Rasanen}},\ and\ \bibinfo {author} {\bibfnamefont {E.}~\bibnamefont
  {Tomberg}},\ }\bibfield  {title} {\bibinfo {title} {{Non-Gaussian Tail of the
  Curvature Perturbation in Stochastic Ultraslow-Roll Inflation: Implications
  for Primordial Black Hole Production}},\ }\href
  {https://doi.org/10.1103/PhysRevLett.127.101302} {\bibfield  {journal}
  {\bibinfo  {journal} {Phys. Rev. Lett.}\ }\textbf {\bibinfo {volume} {127}},\
  \bibinfo {pages} {101302} (\bibinfo {year} {2021})},\ \Eprint
  {https://arxiv.org/abs/2012.06551} {arXiv:2012.06551 [astro-ph.CO]}
  \BibitemShut {NoStop}%
\bibitem [{\citenamefont {Cruces}\ \emph {et~al.}(2025)\citenamefont {Cruces},
  \citenamefont {Germani}, \citenamefont {Nassiri-Rad},\ and\ \citenamefont
  {Yamaguchi}}]{Cruces:2024pni}%
  \BibitemOpen
  \bibfield  {author} {\bibinfo {author} {\bibfnamefont {D.}~\bibnamefont
  {Cruces}}, \bibinfo {author} {\bibfnamefont {C.}~\bibnamefont {Germani}},
  \bibinfo {author} {\bibfnamefont {A.}~\bibnamefont {Nassiri-Rad}},\ and\
  \bibinfo {author} {\bibfnamefont {M.}~\bibnamefont {Yamaguchi}},\ }\bibfield
  {title} {\bibinfo {title} {{Small noise expansion of stochastic inflation}},\
  }\href {https://doi.org/10.1088/1475-7516/2025/04/090} {\bibfield  {journal}
  {\bibinfo  {journal} {JCAP}\ }\textbf {\bibinfo {volume} {04}}\bibfield
  {number} {\bibinfo  {number} { (2025)},\ \bibinfo {pages} {090}},\ }\Eprint
  {https://arxiv.org/abs/2410.17987} {arXiv:2410.17987 [astro-ph.CO]}
  \BibitemShut {NoStop}%
\bibitem [{\citenamefont {Ahmadi}\ and\ \citenamefont
  {Noorbala}(2025)}]{Ahmadi:2025oon}%
  \BibitemOpen
  \bibfield  {author} {\bibinfo {author} {\bibfnamefont {Z.}~\bibnamefont
  {Ahmadi}}\ and\ \bibinfo {author} {\bibfnamefont {M.}~\bibnamefont
  {Noorbala}},\ }\href@noop {} {\bibinfo {title} {{Deviations from Gaussian
  White Noise in Stochastic Inflation}}} (\bibinfo {year} {2025}),\ \Eprint
  {https://arxiv.org/abs/2512.17070} {arXiv:2512.17070 [gr-qc]} \BibitemShut
  {NoStop}%
\bibitem [{\citenamefont {Kawasaki}\ and\ \citenamefont
  {Kuroda}(2026)}]{Kawasaki:2026hnx}%
  \BibitemOpen
  \bibfield  {author} {\bibinfo {author} {\bibfnamefont {M.}~\bibnamefont
  {Kawasaki}}\ and\ \bibinfo {author} {\bibfnamefont {T.}~\bibnamefont
  {Kuroda}},\ }\href@noop {} {\bibinfo {title} {{Numerical simulation of the
  stochastic formalism including non-Markovianity}}} (\bibinfo {year} {2026}),\
  \Eprint {https://arxiv.org/abs/2602.11652} {arXiv:2602.11652 [astro-ph.CO]}
  \BibitemShut {NoStop}%
\bibitem [{\citenamefont {Schwarz}\ \emph {et~al.}(2001)\citenamefont
  {Schwarz}, \citenamefont {Terrero-Escalante},\ and\ \citenamefont
  {Garcia}}]{Schwarz:2001vv}%
  \BibitemOpen
  \bibfield  {author} {\bibinfo {author} {\bibfnamefont {D.~J.}\ \bibnamefont
  {Schwarz}}, \bibinfo {author} {\bibfnamefont {C.~A.}\ \bibnamefont
  {Terrero-Escalante}},\ and\ \bibinfo {author} {\bibfnamefont {A.~A.}\
  \bibnamefont {Garcia}},\ }\bibfield  {title} {\bibinfo {title} {{Higher order
  corrections to primordial spectra from cosmological inflation}},\ }\href
  {https://doi.org/10.1016/S0370-2693(01)01036-X} {\bibfield  {journal}
  {\bibinfo  {journal} {Phys. Lett. B}\ }\textbf {\bibinfo {volume} {517}},\
  \bibinfo {pages} {243} (\bibinfo {year} {2001})},\ \Eprint
  {https://arxiv.org/abs/astro-ph/0106020} {arXiv:astro-ph/0106020}
  \BibitemShut {NoStop}%
\bibitem [{\citenamefont {Schwarz}\ and\ \citenamefont
  {Terrero-Escalante}(2004)}]{Schwarz:2004tz}%
  \BibitemOpen
  \bibfield  {author} {\bibinfo {author} {\bibfnamefont {D.~J.}\ \bibnamefont
  {Schwarz}}\ and\ \bibinfo {author} {\bibfnamefont {C.~A.}\ \bibnamefont
  {Terrero-Escalante}},\ }\bibfield  {title} {\bibinfo {title} {{Primordial
  fluctuations and cosmological inflation after WMAP 1.0}},\ }\href
  {https://doi.org/10.1088/1475-7516/2004/08/003} {\bibfield  {journal}
  {\bibinfo  {journal} {JCAP}\ }\textbf {\bibinfo {volume} {08}}\bibfield
  {number} {\bibinfo  {number} { (2004)},\ \bibinfo {pages} {003}},\ }\Eprint
  {https://arxiv.org/abs/hep-ph/0403129} {arXiv:hep-ph/0403129} \BibitemShut
  {NoStop}%
\bibitem [{\citenamefont {Tokeshi}\ and\ \citenamefont
  {Vennin}(2024)}]{Tokeshi:2023swe}%
  \BibitemOpen
  \bibfield  {author} {\bibinfo {author} {\bibfnamefont {K.}~\bibnamefont
  {Tokeshi}}\ and\ \bibinfo {author} {\bibfnamefont {V.}~\bibnamefont
  {Vennin}},\ }\bibfield  {title} {\bibinfo {title} {{Why Does Inflation Look
  Single Field to Us?}},\ }\href
  {https://doi.org/10.1103/PhysRevLett.132.251001} {\bibfield  {journal}
  {\bibinfo  {journal} {Phys. Rev. Lett.}\ }\textbf {\bibinfo {volume} {132}},\
  \bibinfo {pages} {251001} (\bibinfo {year} {2024})},\ \Eprint
  {https://arxiv.org/abs/2310.16649} {arXiv:2310.16649 [astro-ph.CO]}
  \BibitemShut {NoStop}%
\bibitem [{\citenamefont {Adshead}\ \emph {et~al.}(2020)\citenamefont
  {Adshead}, \citenamefont {Pearce}, \citenamefont {Shelton},\ and\
  \citenamefont {Weiner}}]{Adshead:2020ijf}%
  \BibitemOpen
  \bibfield  {author} {\bibinfo {author} {\bibfnamefont {P.}~\bibnamefont
  {Adshead}}, \bibinfo {author} {\bibfnamefont {L.}~\bibnamefont {Pearce}},
  \bibinfo {author} {\bibfnamefont {J.}~\bibnamefont {Shelton}},\ and\ \bibinfo
  {author} {\bibfnamefont {Z.~J.}\ \bibnamefont {Weiner}},\ }\bibfield  {title}
  {\bibinfo {title} {{Stochastic evolution of scalar fields with continuous
  symmetries during inflation}},\ }\href
  {https://doi.org/10.1103/PhysRevD.102.023526} {\bibfield  {journal} {\bibinfo
   {journal} {Phys. Rev. D}\ }\textbf {\bibinfo {volume} {102}},\ \bibinfo
  {pages} {2} (\bibinfo {year} {2020})},\ \Eprint
  {https://arxiv.org/abs/2002.07201} {arXiv:2002.07201 [hep-ph]} \BibitemShut
  {NoStop}%
\bibitem [{\citenamefont {Tada}\ and\ \citenamefont
  {Yamada}(2023)}]{Tada:2023fvd}%
  \BibitemOpen
  \bibfield  {author} {\bibinfo {author} {\bibfnamefont {Y.}~\bibnamefont
  {Tada}}\ and\ \bibinfo {author} {\bibfnamefont {M.}~\bibnamefont {Yamada}},\
  }\bibfield  {title} {\bibinfo {title} {{Stochastic dynamics of
  multi-waterfall hybrid inflation and formation of primordial black holes}},\
  }\href {https://doi.org/10.1088/1475-7516/2023/11/089} {\bibfield  {journal}
  {\bibinfo  {journal} {JCAP}\ }\textbf {\bibinfo {volume} {11}}\bibfield
  {number} {\bibinfo  {number} { (2023)},\ \bibinfo {pages} {089}},\ }\Eprint
  {https://arxiv.org/abs/2306.07324} {arXiv:2306.07324 [astro-ph.CO]}
  \BibitemShut {NoStop}%
\bibitem [{\citenamefont {Gardiner}(2009)}]{gardiner2009stochastic}%
  \BibitemOpen
  \bibfield  {author} {\bibinfo {author} {\bibfnamefont {C.}~\bibnamefont
  {Gardiner}},\ }\href@noop {} {\emph {\bibinfo {title} {Stochastic Methods: A
  Handbook for the Natural and Social Sciences}}},\ Springer Series in
  Synergetics\ (\bibinfo  {publisher} {Springer Berlin Heidelberg},\ \bibinfo
  {year} {2009})\BibitemShut {NoStop}%
\bibitem [{\citenamefont {Noorbala}\ and\ \citenamefont
  {Firouzjahi}(2019)}]{Noorbala:2019kdd}%
  \BibitemOpen
  \bibfield  {author} {\bibinfo {author} {\bibfnamefont {M.}~\bibnamefont
  {Noorbala}}\ and\ \bibinfo {author} {\bibfnamefont {H.}~\bibnamefont
  {Firouzjahi}},\ }\bibfield  {title} {\bibinfo {title} {{Boundary crossing in
  stochastic inflation with a critical number of fields}},\ }\href
  {https://doi.org/10.1103/PhysRevD.100.083510} {\bibfield  {journal} {\bibinfo
   {journal} {Phys. Rev. D}\ }\textbf {\bibinfo {volume} {100}},\ \bibinfo
  {pages} {083510} (\bibinfo {year} {2019})},\ \Eprint
  {https://arxiv.org/abs/1907.13149} {arXiv:1907.13149 [hep-th]} \BibitemShut
  {NoStop}%
\bibitem [{\citenamefont {Pinol}\ \emph {et~al.}(2021)\citenamefont {Pinol},
  \citenamefont {Renaux-Petel},\ and\ \citenamefont {Tada}}]{Pinol:2020cdp}%
  \BibitemOpen
  \bibfield  {author} {\bibinfo {author} {\bibfnamefont {L.}~\bibnamefont
  {Pinol}}, \bibinfo {author} {\bibfnamefont {S.}~\bibnamefont
  {Renaux-Petel}},\ and\ \bibinfo {author} {\bibfnamefont {Y.}~\bibnamefont
  {Tada}},\ }\bibfield  {title} {\bibinfo {title} {{A manifestly covariant
  theory of multifield stochastic inflation in phase space: solving the
  discretisation ambiguity in stochastic inflation}},\ }\href
  {https://doi.org/10.1088/1475-7516/2021/04/048} {\bibfield  {journal}
  {\bibinfo  {journal} {JCAP}\ }\textbf {\bibinfo {volume} {04}}\bibfield
  {number} {\bibinfo  {number} { (2021)},\ \bibinfo {pages} {048}},\ }\Eprint
  {https://arxiv.org/abs/2008.07497} {arXiv:2008.07497 [astro-ph.CO]}
  \BibitemShut {NoStop}%
\bibitem [{\citenamefont {Tokuda}\ and\ \citenamefont
  {Tanaka}(2018{\natexlab{a}})}]{Tokuda:2017fdh}%
  \BibitemOpen
  \bibfield  {author} {\bibinfo {author} {\bibfnamefont {J.}~\bibnamefont
  {Tokuda}}\ and\ \bibinfo {author} {\bibfnamefont {T.}~\bibnamefont
  {Tanaka}},\ }\bibfield  {title} {\bibinfo {title} {{Statistical nature of
  infrared dynamics on de Sitter background}},\ }\href
  {https://doi.org/10.1088/1475-7516/2018/02/014} {\bibfield  {journal}
  {\bibinfo  {journal} {JCAP}\ }\textbf {\bibinfo {volume} {02}}\bibfield
  {number} {\bibinfo  {number} { (2018)},\ \bibinfo {pages} {014}},\ }\Eprint
  {https://arxiv.org/abs/1708.01734} {arXiv:1708.01734 [gr-qc]} \BibitemShut
  {NoStop}%
\bibitem [{\citenamefont {Tokuda}\ and\ \citenamefont
  {Tanaka}(2018{\natexlab{b}})}]{Tokuda:2018eqs}%
  \BibitemOpen
  \bibfield  {author} {\bibinfo {author} {\bibfnamefont {J.}~\bibnamefont
  {Tokuda}}\ and\ \bibinfo {author} {\bibfnamefont {T.}~\bibnamefont
  {Tanaka}},\ }\bibfield  {title} {\bibinfo {title} {{Can all the infrared
  secular growth really be understood as increase of classical statistical
  variance?}},\ }\href {https://doi.org/10.1088/1475-7516/2018/11/022}
  {\bibfield  {journal} {\bibinfo  {journal} {JCAP}\ }\textbf {\bibinfo
  {volume} {11}}\bibfield  {number} {\bibinfo  {number} { (2018)},\ \bibinfo
  {pages} {022}},\ }\Eprint {https://arxiv.org/abs/1806.03262}
  {arXiv:1806.03262 [hep-th]} \BibitemShut {NoStop}%
\bibitem [{\citenamefont {Roman}(1980)}]{faa}%
  \BibitemOpen
  \bibfield  {author} {\bibinfo {author} {\bibfnamefont {S.}~\bibnamefont
  {Roman}},\ }\bibfield  {title} {\bibinfo {title} {The formula of faa di
  bruno},\ }\href@noop {} {\bibfield  {journal} {\bibinfo  {journal} {Amer.
  Math. Monthly}\ }\textbf {\bibinfo {volume} {87}},\ \bibinfo {pages} {805}
  (\bibinfo {year} {1980})}\BibitemShut {NoStop}%
\bibitem [{\citenamefont {Bell}(1927)}]{5ff4f149-c808-3dd9-a33e-1cf21fe301dc}%
  \BibitemOpen
  \bibfield  {author} {\bibinfo {author} {\bibfnamefont {E.~T.}\ \bibnamefont
  {Bell}},\ }\bibfield  {title} {\bibinfo {title} {Partition polynomials},\
  }\href@noop {} {\bibfield  {journal} {\bibinfo  {journal} {Annals of
  Mathematics}\ }\textbf {\bibinfo {volume} {29}},\ \bibinfo {pages} {38}
  (\bibinfo {year} {1927})}\BibitemShut {NoStop}%
\bibitem [{\citenamefont {Comtet}(1974)}]{Louis:1974}%
  \BibitemOpen
  \bibfield  {author} {\bibinfo {author} {\bibfnamefont {L.}~\bibnamefont
  {Comtet}},\ }\href@noop {} {\emph {\bibinfo {title} {{Advanced Combinatorics:
  The Art of Finite and Infinite Expansions}}}}\ (\bibinfo  {publisher}
  {Springer Dordrecht},\ \bibinfo {year} {1974})\BibitemShut {NoStop}%
\bibitem [{{\relax DLMF}()}]{NIST:DLMF}%
  \BibitemOpen
  {\relax DLMF},\ \href {https://dlmf.nist.gov/} {\bibinfo {title} {{\it NIST
  Digital Library of Mathematical Functions}}},\ \bibinfo {howpublished}
  {\url{https://dlmf.nist.gov/}, Release 1.1.12 of 2023-12-15} (\bibinfo {year}
  {2023}),\ \bibinfo {note} {f.~W.~J. Olver, A.~B. {Olde Daalhuis}, D.~W.
  Lozier, B.~I. Schneider, R.~F. Boisvert, C.~W. Clark, B.~R. Miller, B.~V.
  Saunders, H.~S. Cohl, and M.~A. McClain, eds.}\BibitemShut {Stop}%
\bibitem [{\citenamefont {Gauss}(1813)}]{gauss1813disquisitiones}%
  \BibitemOpen
  \bibfield  {author} {\bibinfo {author} {\bibfnamefont {C.}~\bibnamefont
  {Gauss}},\ }\href@noop {} {\emph {\bibinfo {title} {{Disquisitiones generales
  circa seriem infinitam}}}}\ (\bibinfo  {publisher} {Apud H. Dieterich},\
  \bibinfo {year} {1813})\BibitemShut {NoStop}%
\bibitem [{\citenamefont {Roman}(2019)}]{roman2019umbral}%
  \BibitemOpen
  \bibfield  {author} {\bibinfo {author} {\bibfnamefont {S.}~\bibnamefont
  {Roman}},\ }\href@noop {} {\emph {\bibinfo {title} {Umbral Calculus}}},\
  Dover Books on Mathematics\ (\bibinfo  {publisher} {Dover Publications},\
  \bibinfo {year} {2019})\BibitemShut {NoStop}%
\bibitem [{\citenamefont {Easther}\ and\ \citenamefont
  {McAllister}(2006)}]{Easther:2005zr}%
  \BibitemOpen
  \bibfield  {author} {\bibinfo {author} {\bibfnamefont {R.}~\bibnamefont
  {Easther}}\ and\ \bibinfo {author} {\bibfnamefont {L.}~\bibnamefont
  {McAllister}},\ }\bibfield  {title} {\bibinfo {title} {{Random matrices and
  the spectrum of N-flation}},\ }\href
  {https://doi.org/10.1088/1475-7516/2006/05/018} {\bibfield  {journal}
  {\bibinfo  {journal} {JCAP}\ }\textbf {\bibinfo {volume} {05}}\bibfield
  {number} {\bibinfo  {number} { (2006)},\ \bibinfo {pages} {018}},\ }\Eprint
  {https://arxiv.org/abs/hep-th/0512102} {arXiv:hep-th/0512102} \BibitemShut
  {NoStop}%
\bibitem [{\citenamefont {Kim}\ and\ \citenamefont
  {Liddle}(2007)}]{Kim:2007bc}%
  \BibitemOpen
  \bibfield  {author} {\bibinfo {author} {\bibfnamefont {S.~A.}\ \bibnamefont
  {Kim}}\ and\ \bibinfo {author} {\bibfnamefont {A.~R.}\ \bibnamefont
  {Liddle}},\ }\bibfield  {title} {\bibinfo {title} {{Nflation: observable
  predictions from the random matrix mass spectrum}},\ }\href
  {https://doi.org/10.1103/PhysRevD.76.063515} {\bibfield  {journal} {\bibinfo
  {journal} {Phys. Rev. D}\ }\textbf {\bibinfo {volume} {76}},\ \bibinfo
  {pages} {063515} (\bibinfo {year} {2007})},\ \Eprint
  {https://arxiv.org/abs/0707.1982} {arXiv:0707.1982 [astro-ph]} \BibitemShut
  {NoStop}%
\bibitem [{\citenamefont {McAllister}\ \emph {et~al.}(2012)\citenamefont
  {McAllister}, \citenamefont {Renaux-Petel},\ and\ \citenamefont
  {Xu}}]{McAllister:2012am}%
  \BibitemOpen
  \bibfield  {author} {\bibinfo {author} {\bibfnamefont {L.}~\bibnamefont
  {McAllister}}, \bibinfo {author} {\bibfnamefont {S.}~\bibnamefont
  {Renaux-Petel}},\ and\ \bibinfo {author} {\bibfnamefont {G.}~\bibnamefont
  {Xu}},\ }\bibfield  {title} {\bibinfo {title} {{A Statistical Approach to
  Multifield Inflation: Many-field Perturbations Beyond Slow Roll}},\ }\href
  {https://doi.org/10.1088/1475-7516/2012/10/046} {\bibfield  {journal}
  {\bibinfo  {journal} {JCAP}\ }\textbf {\bibinfo {volume} {10}}\bibfield
  {number} {\bibinfo  {number} { (2012)},\ \bibinfo {pages} {046}},\ }\Eprint
  {https://arxiv.org/abs/1207.0317} {arXiv:1207.0317 [astro-ph.CO]}
  \BibitemShut {NoStop}%
\end{thebibliography}%

\end{document}